\documentclass{ws-rv975x65}
\usepackage{ws-rv-van}     
\usepackage{graphicx}
\usepackage{psfig}
\usepackage{enumitem}

\makeindex

\begin{document}

\chapter[Event-based simulation of quantum physics experiments]{Event-based simulation of quantum physics experiments}\label{ra_ch1}

\author[K. Michielsen and H. De Raedt]{K. Michielsen$^{(1)}$\footnote{Corresponding author\\
Int. J. Mod. Phys. C Vol. 25, No. 8 (2014) 1430003} 
and H. De Raedt$^{(2)}$}

\address{$^{(1)}$Institute for Advanced Simulation, J\"ulich Supercomputing Centre, \\Forschungszentrum J\"ulich,\\
D-52425 J\"ulich, Germany and\\
RWTH Aachen University, D-52056 Aachen, Germany\\
k.michielsen@fz-juelich.de
}

\address{$^{(2)}$Department of Applied Physics, Zernike Institute for Advanced Materials, \\University of Groningen, Nijenborgh 4,\\
NL-9747 AG Groningen, The Netherlands\\
h.a.de.raedt@rug.nl}

\begin{abstract}
We review an event-based simulation approach which
reproduces the statistical distributions of wave theory
not by requiring the knowledge of the solution of the wave equation of the whole system
but by generating detection events one-by-one according to an unknown distribution.
We illustrate its applicability to various single photon and single neutron interferometry experiments
and to two Bell-test experiments, a single-photon  Einstein-Podolsky-Rosen experiment
employing post-selection for photon pair identification and a single-neutron Bell test
interferometry experiment with nearly 100\% detection efficiency.

\end{abstract}

Keywords: {computational techniques; discrete event simulation; quantum theory}

{PACS: 02.70.-c, 03.65.-w, 03.65.Ud}

\body

\section{Introduction}
The statistical properties of a vast number
of laboratory experiments with individual entities such as electrons, atoms,
molecules, photons, $\ldots$ can be extremely well described by quantum theory.
The mathematical framework of quantum theory allows for a straightforward
calculation of numbers which can be compared with experimental
data as long as these numbers refer to statistical
averages of measured quantities, such as for example an interference pattern, the specific heat and magnetic susceptibility.

However, as soon as an experiment records individual
clicks of a detector which contribute to the statistical average of a quantity then a fundamental problem appears.
Quantum theory provides a recipe to compute
the frequencies for observing events but it does not account
for the observation of the individual events themselves, a
manifestation of the quantum measurement problem.~\cite{HOME97,BALL03}
Examples of such experiments are single-particle interference experiments in which the interference
pattern is built up by successive discrete detection events and Bell-test experiments in which
two-particle correlations are computed as averages of pairs of individual detection events recorded at two different
detectors and seen to take values which correspond to those of the singlet state in the quantum theoretical description.

An intriguing question to be answered is why individual entities which do not interact with each other can exhibit the collective behavior
that gives rise to the observed interference pattern and why two particles, which only interacted in the past, after individual
local manipulation and detection can show correlations corresponding to those of the singlet state.
Since quantum theory postulates that it is fundamentally impossible to go beyond the description in terms of probability distributions,
an answer in terms of a cause-and-effect description of the observed phenomena cannot be given within the framework of quantum
theory.

We provide an answer by constructing an event-based simulation model that reproduces the statistical distributions of
quantum (and Maxwell's) theory without solving a wave equation but by modeling physical phenomena as a chronological sequence of events
whereby events can be actions of an experimenter, particle emissions by a source, signal generations by a detector,
interactions of a particle with a material and so on.~\cite{MICH11a,RAED12a,RAED12b}
The underlying assumption of the event-based simulation approach is that current scientific knowledge derives from the discrete events which
are observed in laboratory experiments and from relations between those events.
Hence, the event-based simulation approach concerns what we can say about these experiments but not what ``really'' happens in Nature.
This underlying assumption strongly differs from the premise that the observed discrete events are signatures
of an underlying objective reality which is mathematical in nature.

The general idea of the event-based simulation method is that simple rules define discrete-event processes which may lead to the
behavior that is observed in experiments. The basic strategy in designing these rules is to carefully examine
the experimental procedure and to devise rules such that they produce the same kind of data as those
recorded in experiment, while avoiding the trap of simulating thought experiments that are difficult to realize in
the laboratory. Evidently, mainly because of insufficient knowledge, the rules are not unique.
Hence, the simplest rules can be used until a new experiment indicates otherwise.
On the one hand one may
consider the method being entirely classical since it only uses concepts of the macroscopic world,
but on the other hand one could consider the method being nonclassical because some of the rules are not those
of classical Newtonian dynamics.

Obviously, using trial and error to find discrete-event rules that reproduce experimental results
is unlikely to be successful.
Instead, we started our search for useful rules by asking ourselves the question ``by
what kind of discrete-event rule should a beam splitter
operate in order to mimic the build-up, event-by-event, of the interference pattern
observed in the single-photon Mach-Zehnder experiments performed by Grangier {\sl et al.}~\cite{GRAN86}?''
The simplest rule (discussed below) that performs this task seems to be rather generic in the sense
that it can be used to construct discrete-event processes that reproduce the results of many interference experiments.
Of course, for some experiments, the simple rule is ``too simple''
and more sophisticated, backwards compatible variants are required.
However, the guiding principle for designing the latter is the same as for the simple rule.

The event-based approach has successfully been used for discrete-event simulations of the single beam splitter and
Mach-Zehnder interferometer experiments
of Grangier {\sl et al.}~\cite{GRAN86} (see Refs.~\citen{RAED05d,RAED05b,MICH11a}),
Wheeler's delayed choice experiment of Jacques {\sl et al.}~\cite{JACQ07}
(see Refs.~\citen{ZHAO08b,MICH10a,MICH11a}),
the quantum eraser experiment of Schwindt {\sl et al.}~\cite{SCHW99} (see Refs.~\citen{JIN10c,MICH11a,RAED12e}),
two-beam single-photon interference experiments and the single-photon interference experiment with
a Fresnel biprism of Jacques {\sl et al.}~\cite{JACQ05} (see Refs.~\citen{JIN10b,MICH11a,RAED12a}),
quantum cryptography protocols (see Ref.~\citen{ZHAO08a}),
the Hanbury Brown-Twiss experiment of Agafonov {\sl et al.}~\cite{AGAF08} (see Refs.~\citen{JIN10a,MICH11a,MICH12b}),
universal quantum computation (see Refs.~\citen{RAED05c,MICH05}),
Einstein-Podolsky-Rosen-Bohm (EPRB)-type of experiments of Aspect {\sl et al.}~\cite{ASPE82a,ASPE82b}
and Weihs {\sl et al.}~\cite{WEIH98} (see Refs.~\citen{RAED06c,RAED07a,RAED07b,RAED07c,RAED07d,ZHAO08,MICH11a,RAED12a}),
the propagation of electromagnetic plane waves through homogeneous thin films and stratified media (see Refs.~\citen{TRIE11,MICH11a}),
and neutron interferometry experiments (see Refs.~\citen{RAED12a,RAED12b}).

In this paper, we review the applicability of the event-based simulation method to various single-photon and single-neutron
interferometry experiments and to Bell-test experiments.
The paper is organized as follows.
Section 2 is devoted to the single-particle two-slit experiment, one of the most fundamental experiments in quantum physics.
We first discuss Feynman's thought experiment, demonstrating single-electron interference, and briefly review its laboratory realizations.
We then describe the two-beam experiment with single-photons, a variant of Young's double slit experiment.
It is seen that for these single-particle interference experiments quantum theory gives a recipe to compute the observed interference pattern after many detection events are registered,
but quantum theory does not account for the one-by-one build-up process of the pattern in terms of the individual detection events.
Hence, as formulated in section 3, the challenge is to come up with a set of rules which allow to produce detection events with frequencies which agree with a given distribution
(in this particular case a two-slit interference pattern) without these rules referring, in any way, to the distribution itself.
The event-based simulation method solves this challenging problem by modeling various physical phenomena as a chronological sequence of different events, such as
actions of the experimenter, particles emitted by a source, signals generated by a detector and so on.
In section 4 we explain the basis of the event-based simulation method by specifying rules which allow to reproduce the results of the quantum theoretical description of
the idealized Stern-Gerlach experiment and of a single-photon experiment with a linearly birefringent crystal demonstrating Malus' law, without making any use of quantum theoretical concepts.
In this section, we also discuss the efficiency of two types of single-particle detectors used in the event-based simulation method.
In section 5 we show that a similar set of rules can be used to simulate single-particle interference.
We demonstrate this on the basis of the single-photon two-beam experiment thereby also exactly simulating Feynman's thought experiment, the Mach-Zehnder interferometer experiment,
Wheeler's delayed choice experiment and a single-neutron interferometry experiment with a Mach-Zehnder type of interferometer.
We explain why the event-based simulation method can produce interference without solving a wave problem.
Section 6 is devoted to the event-based simulation of EPRB-type of experiments with correlated photon pairs and with neutrons with correlated spatial and spin degrees of freedom.
Since both experiments are Bell-test experiments testing whether or not a Bell-CHSH (Clauser-Horne-Shimony-Holt) inequality can be violated, we also elaborate on
the conclusions that can be drawn from such a violation.
For both experiments we explain why the event-based model, a classical causal model, can produce the results of quantum theory.
A discussion is given in section 7.

\section{Two-slit and two-beam experiments}
One of the most fundamental experiments in quantum physics is the single-particle double-slit experiment.
Feynman stated that the phenomenon of electron diffraction by a double-slit structure is ``impossible,
{\sl absolutely} impossible, to explain in any classical way, and has in it the heart of quantum mechanics.
In reality it contains the only mystery.''~\cite{FEYN65}
While Young's original double-slit experiment helped
establish the wave theory of light,~\cite{YOUN02} variants of the experiment over the years with electrons (see below), single photons (see below),
neutrons,~\cite{ZEIL88,RAUC00} atoms~\cite{KEIT91,CARN91} and molecules~\cite{ARND99,BREZ02,JUFF12}
helped the development of ideas on concepts such as wave-particle duality in quantum theory.~\cite{BALL03}

Two prevailing variants of the double-slit experiments can be recognized, one consists of a source $S$
and a screen with two apertures and another one consists of a source $S$ and a biprism.
The first one is a real two-slit experiment in which the two slits can be regarded as two virtual sources $S_1$ and $S_2$,
the latter one is a two-beam experiment which can also be replaced by a system with two virtual sources $S_1$ and $S_2$.~\cite{BORN64}
In contrast to the two-slit experiment in which diffraction or scattering and interference phenomena play a role, the phenomenon of diffraction or scattering
is absent in the two-beam experiment, except for the diffraction or scattering at the sources themselves.

A brief note on the difference in usage of the words diffraction, scattering and interference is here in place.
Feynman mentioned in his lecture notes that
``no-one has ever been able to define the difference between interference and diffraction satisfactorily.
It is just a question of usage, and there is no specific, important physical difference between them.''~\cite{FEYN63}
In classical optics, diffraction is the effect of a wave bending as it passes through
an opening or goes around an object. The amount of bending depends on the relative dimensions of the object or opening compared to the wavelength of the wave.
Interference is the superposition of two or more waves resulting in a new wave pattern.
Therefore a double-slit, as well as a single-slit structure illuminated by (classical) light yields an interference (or diffraction) pattern due to diffraction {\sl and} interference.
In principle, diffraction and interference are phenomena observed only with waves. However, an interference pattern identical in form to that of classical optics can be
observed by collecting many detector spots or clicks which are the result of electrons, photons, neutrons, atoms or molecules travelling one-by-one through a double-slit structure.
In these experiments the so-called interference pattern is the statistical distribution of the detection events (spots at or clicks of the detector).
Hence in these particle-like experiments, only the correlations between detection events reveal interference.
Misleadingly this interference pattern is often called a diffraction pattern in analogy with classical optics where both the phenomena of diffraction and interference
are responsible for the resulting pattern. In the particle-like experiment it would be better to replace the word diffraction by scattering because scattering refers
to the spreading of a beam of particles (or a beam of rays) over a range of directions as a result of collisions with other particles or objects.
In what follows we use the term interference pattern for the statistical distribution of detection events.

\subsection{Two-slit experiment with electrons}
In 1964 Feynman described a thought experiment consisting of an electron gun emitting individual electrons in the
direction of a thin metal plate with two slits in it behind which is placed a movable detector.~\cite{FEYN65}
Feynman made the following observations:

\begin{itemize}
\item{Sharp identical ``clicks'' which are distributed erratically, are heard from the detector.}
\item{The probability $P_1(x)$ or $P_2(x)$ of arrival, through one slit with the other slit closed,
at position $x$ is a symmetric curve with its maximum located at
the centre position of the open slit.}
\item{The probability $P_{12}(x)$ of arrival through both slits looks like the intensity of water waves which propagated
through two holes thereby forming a so-called ``interference pattern'' and looks completely different from the
curve $P_1(x)+P_2(x)$, a curve that would be obtained by repeating the experiment with bullets.}
\end{itemize}
which lead him to the conclusions:
\begin{itemize}
\item{Electrons arrive at the detector in identical ``lumps'', like particles.}
\item{The probability of arrival of these lumps is distributed like the distribution of intensity
of a wave propagating through both holes.}
\item{It is in this sense that an electron behaves``sometimes like a particle and sometimes like a wave''.}
\end{itemize}
Note that Feynman made his reasoning with probabilities $P_1(x)$, $P_2(x)$, $P_{12}(x)$, which he said to be proportional
to the average rate of clicks $N_1(x)$, $N_2(x)$, $N_{12}(x)$. However, one cannot simply add $P_1(x)$ and $P_2(x)$
and compare the result with $P_{12}(x)$ because these are probabilities for different conditions (different ``contexts''), namely only slit 1 open,
only slit 2 open and both slits 1 and 2 open, respectively.~\cite{BALL03}
Hence, no conclusions can be drawn from making the comparison between $P_{12}(x)$ and $P_1(x)+P_2(x)$.

Although Feynman wrote ``you should not try to set up this experiment'' because ``the apparatus would
have to be made on an impossibly small scale to show the effects we are interested in'', advances in
(nano)technology made possible various laboratory implementations of his fundamental thought experiment.
The first electron interference pattern obtained with an electron-biprism, the analog of a Fresnel biprism in optics,
was reported in 1955.~\cite{MOEL55, MOEL56}
In 1961 J\"onsson performed
the first electron interference experiment with multiple (up to five) slits in the
micrometer range.~\cite{JONS61} However, these were not
single-electron interference experiments since there was not just one electron in the apparatus at any one time.
The first real single-electron interference experiments that were conducted were electron-biprism experiments
(for a review see Refs.~\citen{HASS10,ROSA12})
in which single electrons either pass to the left or to the right of a conducting wire
(there are no real slits in this type of experiments).~\cite{DONA73,MERL76,TONO89} In these experiments the interference pattern is
built up from many independent detection events. Electron-electron interaction plays no role in the
interference process since the electrons pass the wire one-by-one.
More recently, single-electron interference experiments have been demonstrated
with one-, two-, three and four slits fabricated by focused ion beam milling.~\cite{FRAB08,FRAB10,FRAB11}
However, in these experiments only the final recorded electron intensity is shown.
In a follow-up single-electron two-slit experiment a fast-readout pixel
detector was used which allows the measurement of the distribution of the electron arrival times and the observation
of the build-up of the interference pattern by individual detection events.~\cite{FRAB12}
Hence, this experiment comes very close to Feynman's thought experiment except that the two electron distributions for
one slit open and the other one closed are not measured. Note that one of these distributions was measured in
Ref.~\citen{FRAB08} by a non-reversible process of closing one slit and without using the fast-readout pixel detector.
Very recently, it has been reported that a full realization of Feynman's thought experiment has been performed.~\cite{BACH13}
In this experiment a movable mask is placed behind the double-slit structure to open/close the slits.
Unfortunately, the mask is positioned behind the slits and not in front of them, so that all electrons always encounter a double-slit
structure and are filtered afterwards by the mask. Hence, one could say that anno 2014
Feynman's thought experiment has yet to be performed.

\subsection{Two-beam experiment with photons}
Another interesting variant of Young's double slit experiment involves a very dim light source so that on average only one photon
is emitted by the source at any time.
Inspired by Thomson's idea that light consists of indivisible units that are more widely separated
when the intensity of light is reduced,~\cite{THOM08} in 1909 Taylor conducted an experiment with a light source varying in strength and illuminating a needle thereby demonstrating
that the diffraction pattern observed with a feeble light source (exposure time of three months) was as sharp
as the one obtained with an intense source and a shorter exposure time.~\cite{TAYL09}
In 1985, a double-slit experiment was performed with a low-pressure mercury lamp and neutral density filters to realize a very low-light level.~\cite{TSUC85}
It was shown that at the start of the measurement bright dots appeared at random positions on the detection screen and that after a couple of minutes an
interference pattern appeared.
Demonstration versions of double-slit experiments illuminated by strongly attenuated lasers are reported in Refs.~\citen{PARK71,WEIS03} and in figure 1 of Ref.~\citen{DIMI08}.
However, attenuated laser sources are imperfect single-photon sources.
Light from these sources attenuated to the single-photon level never antibunches, which means that the anticorrelation parameter $\alpha\ge 1$.
For a real single-photon source $0<\alpha<1$.
In 2005, a variation of Young's experiment was performed with a Fresnel biprism and a single-photon source based on the pulsed, optically excited photoluminescence
of a single N-V colour centre in a diamond nanocrystal.~\cite{JACQ05}
In this two-beam experiment there is always only one photon between the source and the detection plane.
Is was observed that the interference pattern gradually builds up starting from a couple of dots spread over the screen for small exposure times.
A time-resolved two-beam experiment has been reported in Refs.~\citen{SAVE02,GARC02}.
Recently, a temporally and spatially resolved two-beam experiment has been performed with entangled photons, providing
insight in the dynamics of the build-up process of the interference pattern.~\cite{KOLE13}

\subsection{The experimental observations and their quantum theoretical description}
The common observation in these single-particle interference experiments, where ``single particle'' can be read as electron, photon, neutron, atom or molecule,
is that individual detection events gradually build up an interference pattern
and that the final interference pattern can be described by wave theory.
In trying to give a pictorial (cause-and-effect) view of what is going on in these experiments, it is commonly assumed that
there is a one-to-one correspondence between an emission event, ``the departure of a single particle from the source'' and a detection event,
``the arrival of the single particle at the detector''. This assumption might be wrong.
The only conclusion that can be drawn from the experiments is that there is some relation between the
emission and detection events.

In view of the quantum measurement problem,~\cite{HOME97,BALL03,NIEU13} a cause-and-effect description of the observed phenomena is unlikely to be found in the framework of quantum theory.
Quantum theory provides a recipe to compute the frequencies for observing events and thus to compute the final interference pattern which is observed after the experiment is finished.
However, it does not account for the observation of the individual detection events building up the interference pattern.
In fact quantum theory postulates that it is fundamentally impossible to go beyond the description in terms of probability distributions.
Of course, one could simply use pseudo-random numbers to generate events according to the probability distribution that is obtained by
solving the time-dependent Schr\"odinger equation. However, that is not the problem one has to solve as it assumes that the probability distribution
of the quantum mechanical problem is known, which is exactly the knowledge that one has to generate without making reference to quantum theory.
If we would like to produce, event-by-event, the interference pattern from Maxwell's theory  and do not want to generate events according to the known intensity function
we would face a similar problem.

\section{Theoretical challenge and paradigm shift}
In general, the challenge is the following.
Given a probability distribution of observing events, construct an algorithm which runs on a digital computer and produces events with frequencies
which agree with the given distribution without the algorithm referring, in any way, to the probability distribution itself.
Traditionally, the behavior of systems is described in terms of
mathematics, making use of differential or integral equations, probability theory and so on. Although that this traditional modeling approach has been proven to be very successful
it does not seem capable of tackling this challenge. This challenge requires something as disruptive as a paradigm shift.
In scientific fields different from (quantum) optics or quantum mechanics in general, a paradigm shift has been realized in terms of a discrete-event approach
to describe the often very complex collective behavior of systems with a set of very simple rules.
Examples of this approach are the lattice Boltzmann model to describe the flow of (complex) fluids and the cellular automata of Wolfram.~\cite{WOLF02}

We have developed a discrete-event simulation method to solve the above mentioned challenging problem
by modeling physical phenomena as a chronological sequence of events whereby events can be actions of the experimenter, particles emitted by a source, signals generated by a detector,
particles impinging on material, and so on.
The basic idea of the simulation method is to try to invent an algorithm which uses the same kind of events (data) as in experiment and
reproduces the statistical results of quantum or wave theory without making use of this theory.
An overview of the method and its applications can be found in Refs.~\citen{MICH11a,RAED12a,RAED12b}.
The method provides an ``explanation'' and ``understanding'' of what is going on in terms of elementary events, logic and arithmetic.
Note that a cause-and-effect simulation on a digital computer is a ``controlled  experiment'' on a macroscopic device which is logically equivalent to a mechanical device.
Hence, an event-by-event simulation that reproduces results of quantum theory shows that there exists a macroscopic, mechanical model that mimics the underlying physical phenomena.
This is completely in agreement with Bohr's answer ``There is no quantum world. There is only an abstract quantum mechanical description.
It is wrong to think that the task of physics is to find out how nature is. Physics concerns what we can say
about nature.'' to the question whether the algorithm of quantum mechanics could be considered
as somehow mirroring an underlying quantum world.~\cite{PETE63}
Although widely circulated, these sentences are reported by Petersen~\cite{PETE63} and there is doubt that Bohr
actually used this wording.~\cite{PLOT10}

\section{Event-by-event simulation method}
\subsection{Stern-Gerlach experiment}
We explain the basics of the event-by-event simulation method using the observations made in the
Stern-Gerlach experiment.~\cite{STER22}
The experiment shows that a beam of silver atoms directed through an inhomogeneous magnetic
field splits into two components. The conclusion drawn by Gerlach and Stern is
that, independent of any theory, it can be stated, as a pure result of the experiment, and as far as the exactitude of
their experiments allows them to say so, that silver atoms in a magnetic field have only two discrete values of
the component of the magnetic moment in the direction of the field strength;
both have the same absolute value with each half of the atoms having a positive and a negative sign
respectively.~\cite{GERL24}

In quantum theory, the stationary state of the two-state system, which is the representation of the statistical experiment,
is described by the density matrix $\rho=(1+{\bf S}\cdot{\bf \sigma})/2$,
%
%
where ${\bf \sigma}=(\sigma^x,\sigma^y,\sigma^z)$ denotes the Pauli vector and ${\bf S}$ denotes the average direction of magnetic moments.
The average measured magnetic moment in the direction ${\bf a}$ is given by
${\bf S}\cdot{\bf a}={\mathrm Tr}\rho{\bf \sigma}\cdot{\bf a}$.

The fundamental question is how to go from the averages to the events observed in the experiment.
Application of Born's rule gives the probability to observe an atom in the beam (anti-)parallel to
the direction ${\bf a}$
\begin{equation}
P(w|{\bf S}\cdot {\bf a})=\frac{1+w{\bf S}\cdot{\bf a}}{2},
\label{prob}
\end{equation}
where $w=+1$ ($w=-1$) refers to the beam parallel (anti-parallel) to ${\bf a}$.

Given the probability in Eq.~(\ref{prob}) the question is how to generate a sequence of ``true''
random numbers $w_1,w_2,\ldots ,w_N$, each taking values $\pm1$, such that $\sum_{n=1}^Nw_n/N\approx {\bf S}\cdot {\bf a}$.
Probability theory postulates that such a procedure exists but is silent about how the procedure
should look like. In practice one could use a probabilistic processor,
a device which responds to and processes input in a probabilistic way,
employing pseudo-random number generators
to generate a uniformly distributed pseudo-random number $0<r_n<1$ to produce $w_n=+1$ if
$r_n<(1+{\bf S}\cdot {\bf a})/2$ and $w_n=-1$ otherwise. Repeating this procedure $N$ times gives
$\sum_{n=1}^Nw_n/N\approx {\bf S}\cdot {\bf a}$.
However, the form of $P(w|{\bf S}\cdot {\bf a})=(1+w{\bf S}\cdot{\bf a})/2$ with $w=\pm 1$ is postulated
and the procedure is deterministic thereby only giving the illusion of randomness to everyone who
does not know the details of the algorithm and the initial state of the pseudo-random generator.
Hence, we accomplished nothing and the question is whether we can do better than by using this probabilistic processor.

Let us consider a deterministic processor, a deterministic learning machine (DLM),~\cite{RAED05b,RAED06a}
that receives input in the form of identical numbers
\begin{equation}
0\le u_n\equiv u=(1+{\bf S}\cdot {\bf a})/2\le 1,
\end{equation}
for
$n=1,\ldots, N$. The processor has an internal state represented by a variable $0\le v_n\le 1$ which adapts to the
received input $u$ in a manner such that the difference with the input is minimal, namely
\begin{equation}
v_n=\gamma v_{n-1}+(1-\gamma )\Delta_n,
\label{DLMrule0}
\end{equation}
where $\Delta_n = \Theta (|\gamma v_{n-1} +(1-\gamma)-u|-|\gamma v_{n-1}-u|)$ with $\Theta (.)$
denoting the unit step function taking only the value 0 or 1 and $0\le\gamma<1$ is a learning parameter
controlling both the speed and
accuracy with which the processor learns the input value $u$.
The initial value $v_0$ of the internal state is chosen at random.
The output numbers generated by the processor are
\begin{equation}
w_n=2\Delta_n-1=\pm 1.
\end{equation}
In general the behavior of the deterministic processor defined by Eq.~(\ref{DLMrule0}) is difficult
to analyze without a computer. However, the operation of the processor can be easily
translated in simple computer code
\begin{eqnarray}
&&\mathrm {u1 = gamma*y}\nonumber\\
&&\mathrm {u2 = u1+1-gamma}\nonumber\\
&&\mathrm {if (abs(v-u1) < abs(v-u2)) then}\nonumber\\
&&\mathrm {w = -1}\nonumber\\
&&\mathrm {u = u1}\nonumber\\
&&\mathrm {else}\nonumber\\
&&\mathrm {w = + 1}\nonumber\\
&&\mathrm {u = u2}\nonumber\\
&&\mathrm {end if}
\end{eqnarray}
Also without computer this code allows getting a quick notion on how the internal state of the processor adapts to the input.
Taking as an example $u=5/8$, $\gamma = 0.5$ and $v_n=4/8$ gives $v_{n+1}=6/8$, $v_{n+2}=7/8$, $v_{n+3}=7/16$,
$\ldots$ From this step-by-step analysis it can be seen how $v_n$ comes closer to $u$, goes further away from it
to come closer again in a next step and how $v_n$ keeps oscillating around $u$ in the stationary regime.
A detailed mathematical analysis of the dynamics of the processor defined by the rule Eq.~(\ref{DLMrule0})
is given in Ref.~\citen{RAED06}. For $\gamma\rightarrow 1^{-}$ we find that
$\sum_{n=1}^Nw_n/N\approx 2u-1={\bf S}\cdot {\bf a}$.

In conclusion, we designed an event-by-event process which can reproduce the results of the quantum theoretical description
of the idealized Stern-Gerlach experiment without making use of any quantum theoretical concepts.
The strategy employed by the processor is to minimize the distance between two numbers thereby
``learning'' the input number. Hence, at least one of the results of quantum theory seems to emerge from an
event-based process, a dramatic change in the paradigm of the quantum science community.

\subsection{Malus' law}
The important question is whether this event-based approach can also be applied to other experiments which up to now
are exclusively described in terms of wave or quantum theory. To scrutinize this question we consider a basic optics experiment
with a linearly birefringent crystal, such as calcite acting as a polarizer. A beam of linearly polarized monochromatic light
impinging on a calcite crystal along a direction not parallel to the optical axis of the crystal
is split into two beams travelling in different directions and having orthogonal polarizations.
The two beams are referred to as the ordinary and extraordinary beam, respectively.~\cite{BORN64}
The intensity of the beams is given by Malus' law, which has experimentally been established in 1810,
\begin{equation}
I_o=I\sin ^2 (\psi -\phi),\quad I_e=I\cos ^2 (\psi -\phi),
\label{Malus}
\end{equation}
where $I$, $I_o$ and $I_e$ are the intensities of the incident, ordinary and extraordinary beam, respectively,
$\psi$ is the polarization of the incident light and $\phi$ specifies the orientation of the crystal.~\cite{BORN64}
Observations in single-photon experiments show that Malus' law is also obeyed at the single-photon level.

In the quantum theoretical description of these single-photon experiments in which the photons are
detected one-by-one in either the ordinary beam (represented by a detection event $w=0$) or in the extraordinary beam
(represented by a detection event $w=1$) it is postulated that the polarizer sends a photon to the extraordinary
direction with probability $\cos ^2 (\psi -\phi)$ and to the ordinary direction with probability $\sin ^2 (\psi -\phi)$.
Hence, quantum theory postulates that $\lim_{N\rightarrow\infty}\sum_{n=1}^Nw_n/N\rightarrow \cos ^2 (\psi-\phi)$.

Following a procedure similar to that of the Stern-Gerlach experiment it is obvious that we can construct a
simple probabilistic processor employing pseudo-random numbers to generate a uniform random number $0<r_n<1$
and send out a $w_n=0$ ($w_n=1$) event if $\cos^2(\psi-\phi)\le r_n$ ($\cos^2(\psi-\phi)> r_n$) so that
after repeating this procedure $N$ times we indeed
have $\lim_{N\rightarrow\infty}\sum_{n=1}^Nw_n/N\rightarrow \cos ^2 (\psi-\phi)$. However, again, by doing this
we accomplished nothing because Malus' law has been postulated from the start in the form
$P(w|\psi-\phi)=w\cos ^2 (\psi-\phi)+(1-w)\sin ^2 (\psi-\phi)$ with $w=0,1$.
Moreover, this probabilistic processor has a relatively poor performance~\cite{RAED06} and therefore in what follows we
design and analyze a much more efficient DLM that generates events according to Malus' law.

The DLM mimicking the operation of a polarizer has one input channel, two output channels and one internal vector
with two real entries.
The DLM receives as input, a sequence of angles $\psi_{n}$ for $n=1,\ldots, N$ and knows about the orientation
of the polarizer through the angle $\phi$.
Using rotational invariance, we represent these input messages by unit vectors
\begin{equation}
{\bf u}_n=(u_{0,n},u_{1,n})=(\cos(\psi_n-\phi),\sin(\psi_n-\phi)).
\end{equation}
Instead of the random number generator that is part of the probabilistic processor, the DLM has an internal
degree of freedom represented by the unit vector ${\bf v}_n=(v_{0,n},v_{1,n})$.
The direction of the initial internal vector ${\bf v}_0$ is chosen at random.
As the DLM receives input data,
it updates its internal state. The update rules are defined by
\begin{eqnarray}
v_{0,n}&=&\pm\sqrt{1+\gamma ^2 (v_{0,n-1}^2-1)},\quad v_{1,n}=\gamma v_{1,n-1}, 
\label{DLMrule1}
\end{eqnarray}
corresponding to the output event $w_n=0$ and
\begin{eqnarray}
v_{0,n}&=&\gamma v_{0,n-1},\quad v_{1,n}=\pm\sqrt{1+\gamma ^2 (v_{1,n-1}^2-1)}, 
\label{DLMrule1a}
\end{eqnarray}
corresponding to the output event $w_n=1$.
The parameter $0<\gamma <1$ controls the learning process of the DLM.
The $\pm$-sign takes care of the fact that the DLM has to decide between two quadrants.
The DLM selects one of the four possible outcomes for ${\bf v}_n=(v_{0,n}, v_{1,n})$
by minimizing the cost function defined by
\begin{equation}
C=-{\bf v}_n\cdot {\bf u}_n=-(v_{0,n}u_{0,n}+v_{1,n}u_{1,n}).
\label{DLMcost}
\end{equation}
Obviously, the cost $C$ is small (close to $-1$),  if the vectors ${\bf u}_n$ and ${\bf v}_n$ are close to each other.
In conclusion, the DLM generates output events $w_n=0,1$ by minimizing the distance between the input vector and its internal vector by means of a simple,
deterministic decision process.

\begin{figure}[pt]
\begin{center}
\includegraphics[width=12cm]{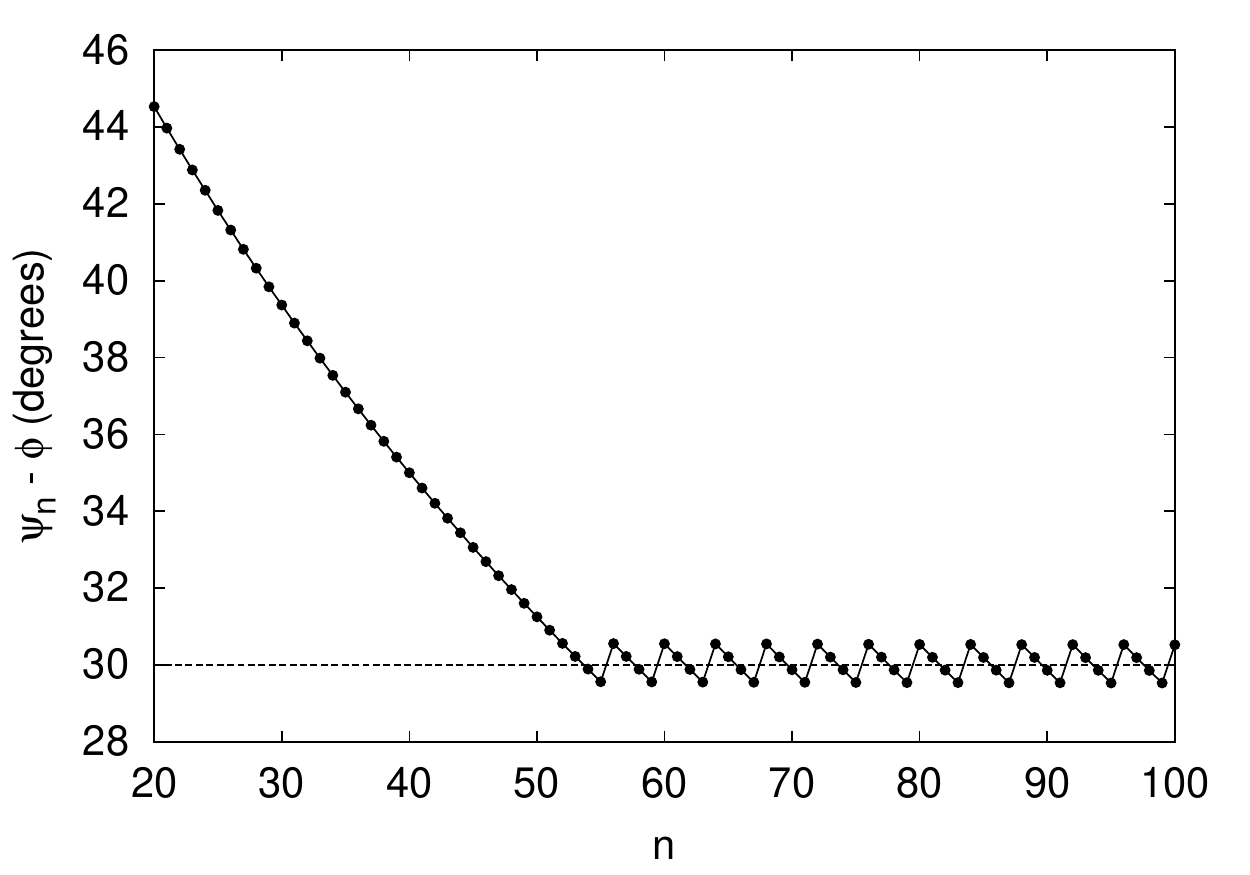}
\vspace*{8pt}
\caption{The angle $\psi_n -\phi$ representing the internal vector ${\mathbf v}_n$ of the DLM defined by Eqs.~(\ref{DLMrule1}) and (\ref{DLMcost}) as a function of the number of events $n$.
The input events are vectors ${\mathbf u}_n=(\cos 30^{\circ} , \sin 30^{\circ})$.
The direction of the initial internal vector ${\bf v}_0$ is chosen at random. In this simulation $\gamma=0.99.$
For $n>60$ the ratio of the number of 0 events to 1 events is 1/3, which is $(\sin 30^{\circ} / \cos 30^{\circ})^2$.
Data for $1\le n < 20$ lie on the decaying line but have been omitted to show the oscillating behavior more clearly. Lines are guides to the eye.
}
\label{fig1}
\end{center}
\end{figure}

In general, the behavior of the DLM defined by the rules Eqs.~(\ref{DLMrule1})--(\ref{DLMcost}) is difficult
to analyze without using a computer.
However, for a fixed input vector ${\bf u}_n=(u_0,u_1)$ for $n=1,\ldots ,N$, the DLM will minimize the cost
Eq.~(\ref{DLMcost}) by rotating its internal vector ${\bf v}_n$ towards ${\bf u}_n$
but ${\bf v}_n$ will not converge to the input vector ${\bf u}_n$ and will keep oscillating about
${\bf u}_n$. This is the stationary state of the machine. An example of a simulation is given in Fig.~\ref{fig1}.
Once the DLM has reached the stationary state the number of $w_n=0$ output events divided by the total number
of output events is $\cos^2 (\psi_n-\phi)$ and thus in agreement with Malus' law if we interpret the $w_n=0$ output events
as corresponding to the extraordinary beam.
Note that the details of the approach to the stationary
state depend on the initial value of the internal vector ${\bf v}_0$,
but the properties of the stationary state do not.
A detailed stationary-state analysis is given in Ref.~\citen{RAED06a}.

\subsection{Single particle detection}
In the event-based simulation of the Stern-Gerlach experiment and of the experiment demonstrating Malus' law
the two-valued output events $w_n$ ($n=1,\ldots ,N$) can be processed by two detectors placed behind the DLM
modeling the Stern-Gerlach magnet and the calcite crystal, respectively.
It can be easily seen that in these two experiments the only operation the detectors have to perform
is to simply count every incoming output event $w_n$.
However, real single-particle detectors
are often more complex devices with diverse properties.
In our event-based simulation approach we model
the main characteristics of these devices by rules as simple as possible
to obtain similar results as those observed in a laboratory experiment.
So far, we have designed two types of detectors,
simple particle counters and adaptive threshold devices.~\cite{MICH11a}
The adaptive threshold detector can be employed in the simulation of all single-photon experiments
we have considered so far\cite{MICH11a}  but is absolutely essential
in the simulation of for example the two-beam single
photon experiment (see Sect.~5.1).

The efficiency, defined as the ratio
of detected to emitted particles, of our model detectors
is measured in an experiment with one single-particle point
source placed far away from the detector.
If the detector is a simple particle counter
then the efficiency is 100\%, if it is an adaptive threshold detector then the efficiency is nearly
100\%.  Since no absorption effects,
dead times, dark counts, timing jitter or other effects causing
particle miscounts are simulated, these model detectors are highly idealized versions
of real single-photon detectors.

Evidently, the efficiency of a detector plays an
important role in the overall detection efficiency in an experiment,
but it is not the only determining factor.
Also the experimental configuration, as well in the laboratory experiment as in the
event-based simulation approach, in which the detector is used plays an important
role. Although the adaptive threshold detectors are ideal and have a detection
efficiency of nearly 100\%, the overall detection efficiency
can be much less than 100\% depending on the
experimental configuration. For example, using adaptive
threshold detectors in a Mach-Zehnder interferometry
experiment leads to an overall detection efficiency of
nearly 100\% (see Sect.~5.2.1), while using the same detectors in a single-photon
two-beam experiment (see Sect.~5.1.1) leads to an
overall detection efficiency of about 15\%.~\cite{MICH11a,JIN10b}
For the simple particle counters the configuration has no influence on the
overall detection efficiency.
Apart from the configuration, also the data processing procedure which is applied
after the data has been collected may have an influence
on the final detection efficiency.
An example is the postselection procedure with a time-coincidence window which is
employed to group photons, detected in two different
stations, into pairs.~\cite{WEIH98}
Even if in the event-based simulation approach simple particle counters with a 100\% detection
efficiency are used and thus all emitted photons are accounted
for during the data collection process, the final
detection efficiency is less than 100\% because some detection
events are omitted in the post-selection data procedure
using a time-coincidence window.

In conclusion, even if ideal detectors with a detection
efficiency of 100\% would be commercially available, then
the overall detection efficiency in a single-particle experiment
could still be much less than 100\% depending on
(i) the experimental configuration in which the detectors
are employed and (ii) the data analysis procedure that is
used after all data has been collected.

\section{Single particle interference}

The particle-like behavior of photons has been shown in an experiment
composed of a single 50/50 beam splitter (BS), of which only one input port is used, and a
source emitting single photons and pairs of photons.~\cite{GRAN86}
The wave mechanical character of the collection of photons has been demonstrated in single-particle interference experiments
such as the single-photon two-beam experiment~\cite{JACQ05} (see Sect.~5.1),
an experiment which shows, with minimal equipment, interference
in its purest form (without diffraction), and
the single-photon Mach-Zehnder interferometer (MZI) experiment~\cite{GRAN86} (see Sect.~5.2).

The three experiments have in common that, if one
analyzes the data after collecting $N$ detection events, long after the experiment has finished, the
averages of the detection events agree with the results obtained from wave theory,
that is with the classical theory of electrodynamics (Maxwell theory). In the first experiment
one obtains a constant intensity of 0.5 at both detectors placed at the output ports of the BS,
in the other two experiments one obtains an interference pattern. However, since the source
is not emitting waves but so-called single photons~\cite{GRAN86,JACQ05} the question arises how to interpret the output which seems to
show particle or wave character depending on the circumstances of the experiment.
This question is not limited to photons. Already in 1924, de Broglie introduced the idea that
also matter can exhibit wave-like properties.~\cite{BROG25}

To resolve the apparent behavioral contradiction, quantum theory introduces
the concept of particle-wave duality~\cite{HOME97}.
As a result, these single-particle experiments are often considered to be quantum experiments.
However, the pictorial description using concepts from quantum theory, when applied
to individual detection events (not to the averages) leads to conclusions that defy common
sense: The photon (electron, neutron, atom, molecules, $\ldots$) seems to change its representation from a particle to a wave while
traveling from the source to the detector in the single-photon interference experiments.

In 1978, Wheeler proposed a
gedanken experiment,~\cite{WHEE83} a variation on Young's double slit
experiment, in which the decision to observe wave or particle
behavior is postponed until the photon has passed the slits.
An experimental realization of Wheeler's delayed choice experiment with single-photons
traveling in an open or closed configuration of an MZI
has been reported in Refs.~\citen{JACQ07,JACQ08}.
The outcome, that is
the average result of many detection events, is in agreement with
wave theory (Maxwell or quantum theory). However, the pictorial
description using concepts of quantum theory to explain the experimental facts~\cite{JACQ07} is even more strange than in the above mentioned experiments:
The decision to observe particle or wave behavior
influences the behavior of
the photon in the past and changes the representation of the
photon from a particle to a wave.

A more sensical description of the observation of individual detection events {\bf and} of an interference pattern
after many single detection events have been collected in single-particle interference experiments,
can be given in terms of the event-based simulation approach.
This finding is not in contradiction with Feynman's statement that
electron (single particle) diffraction by a double-slit structure is ``impossible,
{\sl absolutely} impossible, to explain in any classical way, and has in it the heart of quantum mechanics''~\cite{FEYN65}.
Reading ``any classical way'' as ``any classical Hamiltonian
mechanics way'', Feynman's statement is difficult to
dispute. However, taking a broader view by allowing for
dynamical systems that are outside the realm of classical
Hamiltonian dynamics, it becomes possible to model
the gradual appearance of interference patterns through the event-by-event simulation method.

\subsection{Two-beam experiment}
\begin{figure}[pt]
\begin{center}
\includegraphics[width=12cm]{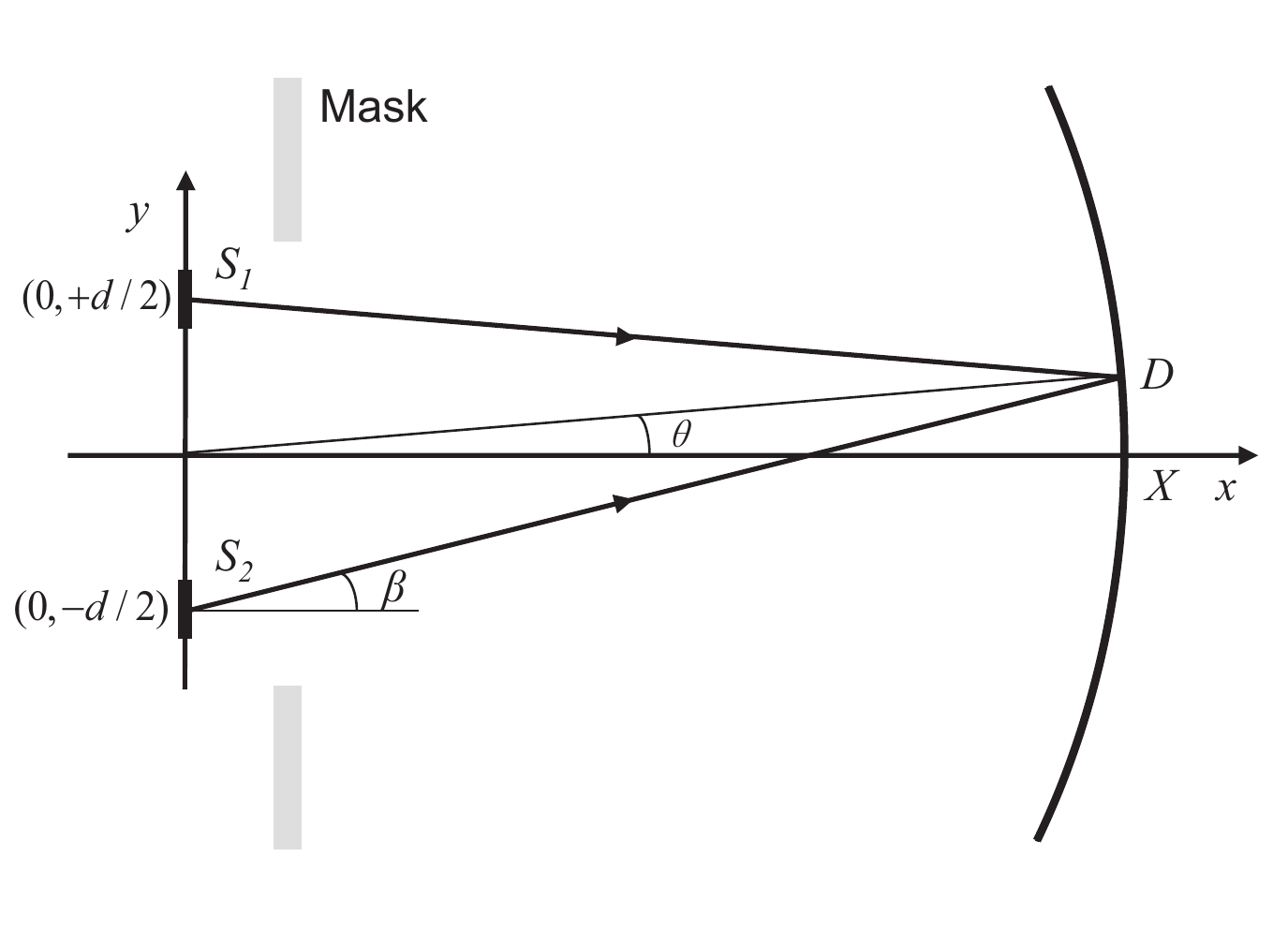}
\vspace*{8pt}
\caption{Schematic diagram of a two-beam experiment with single-particle sources $S_1$ and $S_2$ of width $a$,
separated by a center-to-center distance $d$.
In a first experiment, which can be seen as a variant of Young's double slit experiment,
$N$ single particles leave the sources $S_1$ and $S_2$ one-by-one,
at positions $y$ drawn randomly from a uniform distribution over the interval $[-d/2-a/2,-d/2+
a/2]\cup [+d/2-a/2,+d/2+a/2]$ and travel in the direction given by the angle $\beta$, a uniform pseudo-random number between
$-\pi /2$ and $\pi /2$.
In a second experiment, a movable mask is placed behind the sources which can block either $S_1$ or $S_2$.
The sources $S_1$ and $S_2$ alternately emit $M$ particles one-by-one, until a total of $N$ particles has been emitted ($M\le N/2$ and
$kM=N$ with $k$ an integer number).
In both experiments, particles are emitted one-by-one either from $S_1$ or from  $S_2$ and at any time there is only
one particle traveling from source to detector.
The particles are recorded by detectors $D$ positioned on a semi-circle with radius $X$ and center $(0,0)$. The
angular position of a detector is denoted by $\theta$.
}\label{fig2}
\end{center}
\end{figure}

We consider the experiment sketched in Fig.~\ref{fig2}.
Single particles coming from two coherent beams gradually build up an interference pattern when the particles
arrive one-by-one at a detector screen.
This two-beam
experiment can be viewed as a simplification of Young's
double-slit experiment in which the slits are regarded as
the virtual sources $S_1$ and $S_2$ (see Ref.~\citen{BORN64}) and can be used to perform Feynman's thought experiment
in which both slits are open or one is open and the other one closed.
In the event-based model of this experiment particles are created one at a time
at one of the sources and are detected by one of the detectors forming the screen.
We assume that all these detectors are identical and cannot
communicate among each other. We also do not allow for
direct communication between the particles.
This implies that this event-by-event model is locally causal by construction.
Then, if it is indeed true that individual particles
build up the interference pattern one-by-one, just
looking at Fig.~\ref{fig2} leads to the logically unescapable
conclusion that the interference pattern can only be due
to the internal operation of the detector~\cite{PFLE67}.
Detectors which simply count the incoming particles are not sufficient
to explain the appearance of an interference pattern
and apart from the detectors there is nothing else
that can cause the interference pattern to appear.
Making use of the statistical property of quantum theory one
could assume that if a detector is replaced by another one
as soon as it has detected one particle, one obtains similar
interference patterns if the detection events of all these
different detectors are combined or if only one detector detects all the particles.
However, since there is no experimental evidence confirming this assumption
and since our event-based approach is based on laboratory
experimental setups and observations we do not
consider this being a realistic option. Thus, logic dictates
that a minimal event-based model for the two-beam experiment
requires an algorithm for the detector that does
a little more than just counting particles.

\subsubsection{Event-based model}
In what follows we specify the event-by-event model for the single-photon two-beam experiment (see Fig.~\ref{fig2})
in sufficient detail such that the reader who is interested can reproduce
the simulation results (a Mathematica implementation of a slightly more sophisticated algorithm~\cite{JIN10b}
can be downloaded from the Wolfram Demonstration Project web site~\cite{DS08}).

\begin{itemize}[leftmargin=*]
\renewcommand\labelitemi{-}
\item
{\sl Source and particles:}
In the first experiment described in Fig.~\ref{fig2},
$N$ photons leave the sources one-by-one, at positions $y$ drawn randomly from
a uniform distribution over the interval $[-d/2-a/2,-d/2+a/2]\cup[+d/2-a/2,+d/2+a/2]$.
In the second experiment the sources alternately emit $M$ photons one-by-one until a total of $N$ photons has been emitted.
Here, $M\le N/2$ and $kM=N$, where $k$ denotes an integer number.
The photons are regarded as messengers, traveling in the direction specified by the angle $\beta$,
being a uniform pseudo-random number between $-\pi/2$ and $\pi/2$.
Each messenger carries a message
\begin{equation}
\mathbf{u}(t)=(\cos (2\pi f t), \sin (2\pi f t)),
\end{equation}
represented by a harmonic oscillator
which vibrates with frequency $f$ (representing the ``color'' of the light).
The internal oscillator operates as a clock to encode the time of flight $t$,
which is set to zero when a messenger is created, thereby modeling
the coherence of the two single-particle beams.

This pictorial model of a ``photon'' was used by Feynman to explain
quantum electrodynamics.~\cite{FEYN85}
The event-based approach goes one step further in that it specifies
in detail, in terms of a mechanical procedure, how the ``amplitudes''
which appear in the quantum formalism get added together.
In Feynman's path integral formulation of light propagation, which is essentially quantum
mechanical, the amplitude was obtained by summing over all possible paths.~\cite{FEYN85}

The time of flight of the particles depends on the source-detector distance.
Here, we discuss as an example, the experimental setup with a semi-circular detection screen (see Fig.~\ref{fig2})
but in principle any other geometry for the detection screen can be considered.
The messenger leaving the source at $(0, y)$ under an angle $\beta$
will hit the detector screen of radius $X$ at a position determined by the angle $\theta$ given by
$\sin\theta=(y\cos^2\beta+\sin\beta\sqrt{X^2-y^2\cos^2\beta})/{X}$,
where $|y/X|< 1$. The time of flight is then given by
$t=\sqrt{X^2-2yX\sin\theta+y^2}/c$,
where $c$ is the velocity of the messenger.
The messages $\mathbf{u}(t)$ together with the explicit expression for the time of flight are the only input to the event-based algorithm.

\item
{\sl Detector:}
Here we describe the model for one of the many identical detectors building up the detection screen.
Microscopically, the detection of a particle involves very intricate dynamical processes~\cite{NIEU13}.
In its simplest form, a light detector consists of a material that can be ionized by light.
This signal is then amplified, usually electronically, or in the case of a photographic plate
by chemical processes.
In Maxwell's theory, the interaction between the incident electric field ${\mathbf E}$ and
the material takes the form ${\mathbf P}\cdot{\mathbf E}$, where ${\mathbf P}$ is the polarization vector of the material~\cite{BORN64}.
Assuming a linear response, ${\mathbf P}(\omega)={\mathbf \chi}(\omega){\mathbf E}(\omega)$ for a monochromatic wave with
frequency $\omega$, it is clear that in the time domain, this relation expresses the fact that the material
retains some memory about the incident field, ${\mathbf \chi}(\omega)$ representing the memory kernel
that is characteristic for the material used.

In line with the idea that an event-based approach should use the simplest rules possible,
we reason as follows.
In the event-based model, the $n$th message ${\mathbf u}_n=(\cos 2\pi f t_n, \sin 2\pi f t_n)$ is taken
to represent the elementary unit of electric field ${\mathbf E}(t)$.
Likewise, the electric polarization ${\mathbf P}(t)$ of the material is represented by the
vector ${\mathbf v}_n = (v_{0,n},v_{1,n})$.
Upon receipt of the $n$th message this vector is updated according to the rule
\begin{equation}
	{\mathbf v}_n = \gamma {\mathbf v}_{n-1} + (1-\gamma) {\mathbf u}_n,
	\label{ruleofLM}
\end{equation}
where $0<\gamma<1$ and $n>0$.
Obviously, if $\gamma>0$, a message processor that operates according to the update rule Eq.~(\ref{ruleofLM}) has memory,
as required by Maxwell's theory.
It is not difficult to prove that as $\gamma\rightarrow1^-$, the internal vector ${\mathbf v}_{n}$
converges to the average of the time-series $\{{\mathbf u}_{1},{\mathbf u}_{2},\ldots\}$~\cite{JIN10b,MICH11a}.
By reducing $\gamma$, the number of messages needed to adapt decreases but also the accuracy of the DLM decreases.
In the limit that $\gamma =0$, the DLM learns nothing, it simply echoes the last message that it received.~\cite{RAED05b,RAED05d}
The parameter $\gamma$ controls the precision
with which the DLM defined by Eq.~(\ref{ruleofLM}) learns the average of the sequence of messages
${\mathbf u}_{1},{\mathbf u}_{2}, \ldots$
and also controls the pace at which new messages affect the internal state $\mathbf{v}$ of the machine~\cite{RAED05d}.
Moreover, in the continuum limit (meaning many events per unit of time),
the rule given in Eq.~(\ref{ruleofLM}) translates into the constitutive equation of the Debye model of a dielectric~\cite{JIN10b,RAED12},
a model used in many applications of Maxwell's theory~\cite{TAFL05}.

After updating the vector ${\mathbf v}_n$, the DLM uses the information stored in ${\mathbf v}_n$
to decide whether or not to generate a click.
As a highly simplified model for the bistable character of the real photodetector or
photographic plate, we let the machine generate a binary output signal $w_n$ according to
\begin{equation}
w_n = \Theta({\mathbf v}^2_k-r_n),
\label{thresholdofLM}
\end{equation}
where $\Theta(.)$ is the unit step function and $0\leq r_n <1$ is a uniform pseudo-random number.
Note that the use of pseudo-random numbers is convenient but not essential~\cite{MICH11a}.
Since in experiment it cannot be known whether a photon has gone undetected, we discard the information about
the $w_n=0$ detection events and define the total detector count as
$N^{\prime}=\sum^{n^{\prime}}_{j=1}w_j$,
%
where $n^{\prime}$ is the number of messages received. $N^{\prime}$ is the number of clicks (one's) generated by the processor.

The efficiency of the detector model is determined by simulating an experiment
that measures the detector efficiency, which for a single-photon detector is defined
as the overall probability of registering a count if a photon arrives at the detector~\cite{HADF09}.
In such an experiment a point source emitting single particles is placed far away from a single detector.
As all particles that reach the detector have the same time of flight (to a good approximation), all the
particles that arrive at the detector will carry the same message which is encoding the time of flight.
As a result ${\mathbf v}_n$ (see Eq.~(\ref{ruleofLM})) rapidly converges to the vector corresponding to
this message, so that the detector clicks every time a photon arrives.
Thus, the detection efficiency, as defined for real detectors~\cite{HADF09}, for our detector model is very close
to 100\%.
Hence, the model is a highly simplified and idealized version of a single-photon detector.
However, although the detection efficiency of the detector itself may be very close to 100\%,
the overall detection efficiency, which is
the ratio of detected to emitted photons in the simulation of an experiment, can be much less than one. This ratio depends on the
experimental setup.

\item
{\sl Simulation procedure:}
Each of the detectors of the circular screen has a predefined spatial window within which it accepts messages.
As a messenger hits a detector, this detector updates its internal state ${\mathbf v}$,
(the internal states of all other detectors do not change)
using the message ${\mathbf u}_{n}$ and then generates the event $w_n$.
In the case $w_n=1$ ($w_n=0$), the total count of the particular detector that was hit by the $n$th
messenger is (not) incremented by one and the messenger itself is destroyed.
Only after the messenger has been destroyed, the source is allowed to send a new messenger.
This rule ensures that the whole simulation complies with Einstein's criterion of local causality.
This process of creating and destroying messengers is repeated many times, building up the
interference pattern event by event.
Note that the number of emitted photons $N$ is larger than the sum of the number of clicks generated
by all the detectors forming the detection screen although no photons are lost during their travel from source to detector.
\end{itemize}

\subsubsection{Simulation results}
\begin{figure}[pt]
\begin{center}
\includegraphics[width=6cm]{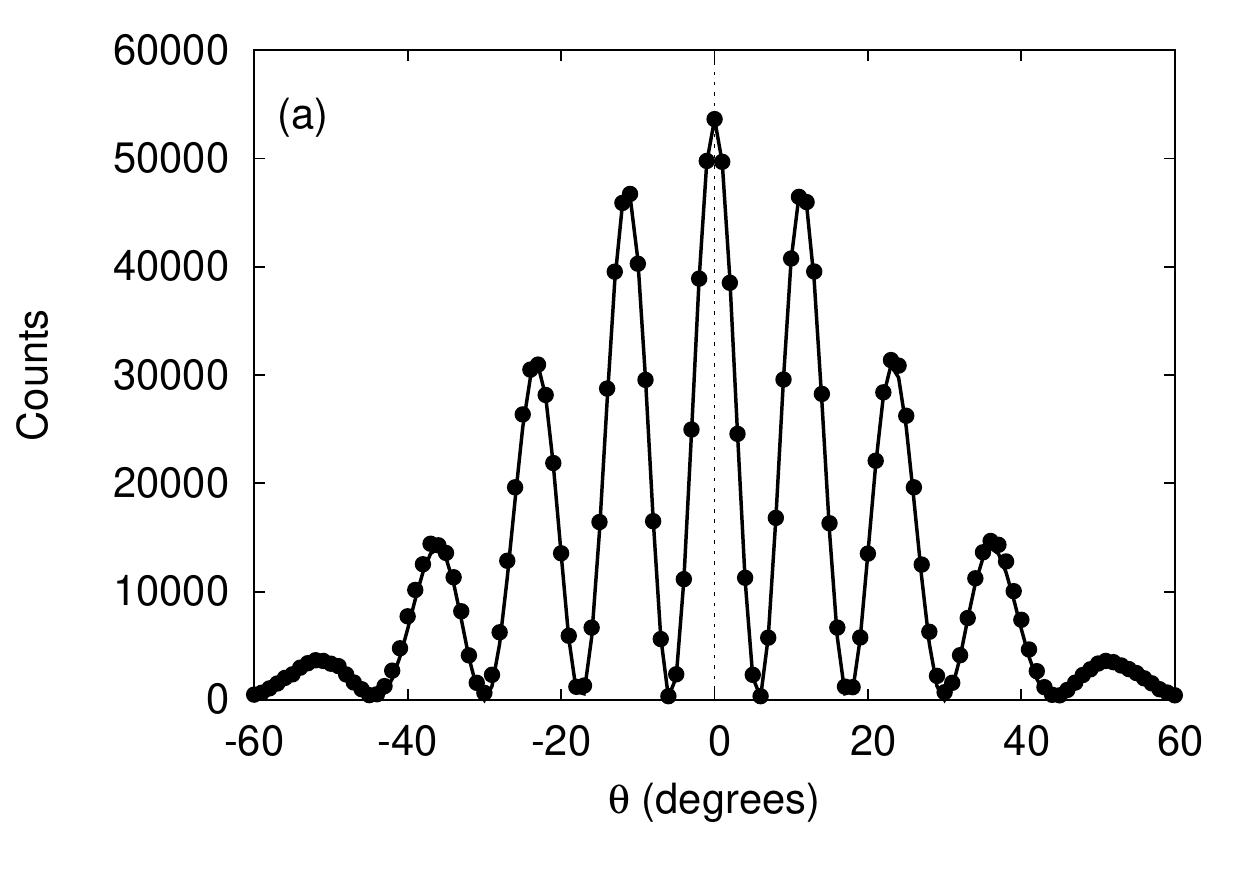}
\includegraphics[width=6cm]{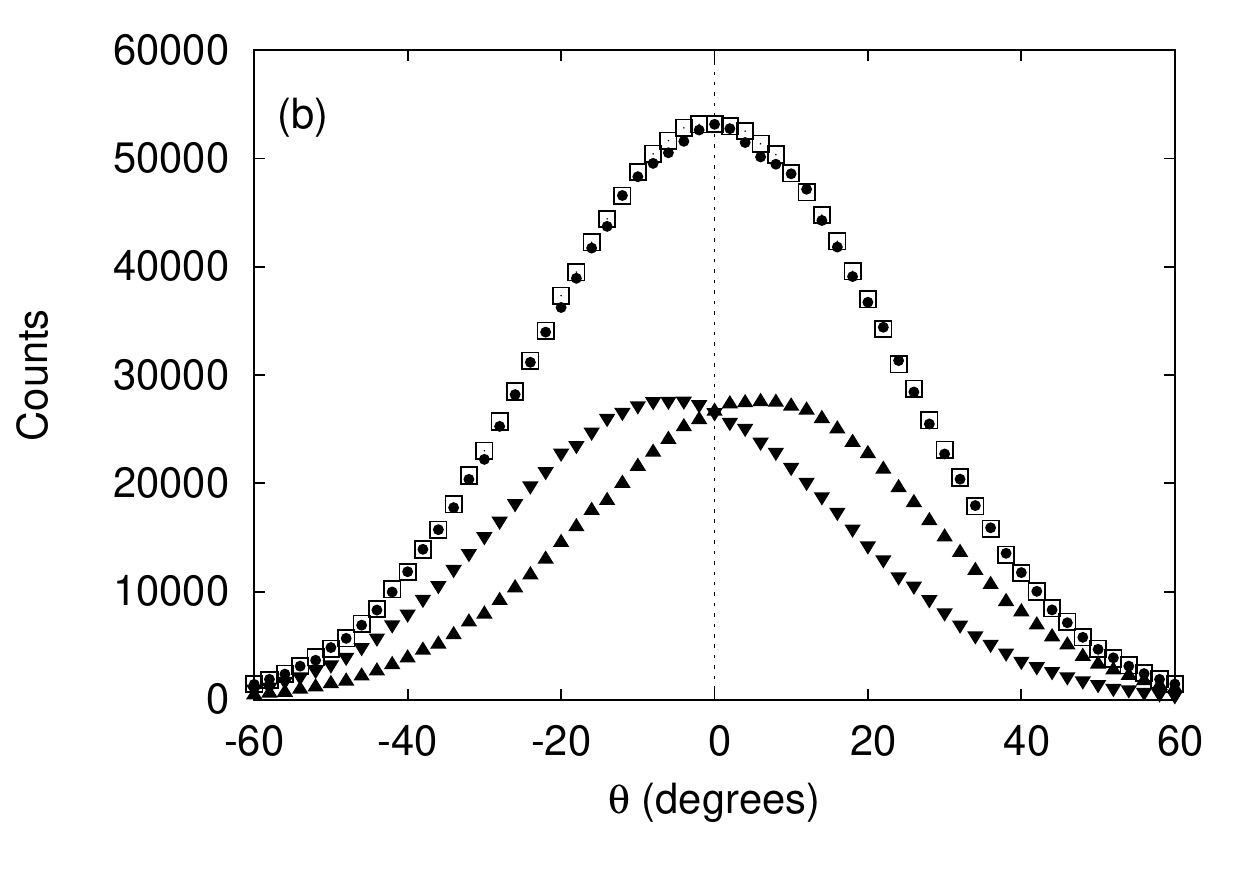}
\includegraphics[width=6cm]{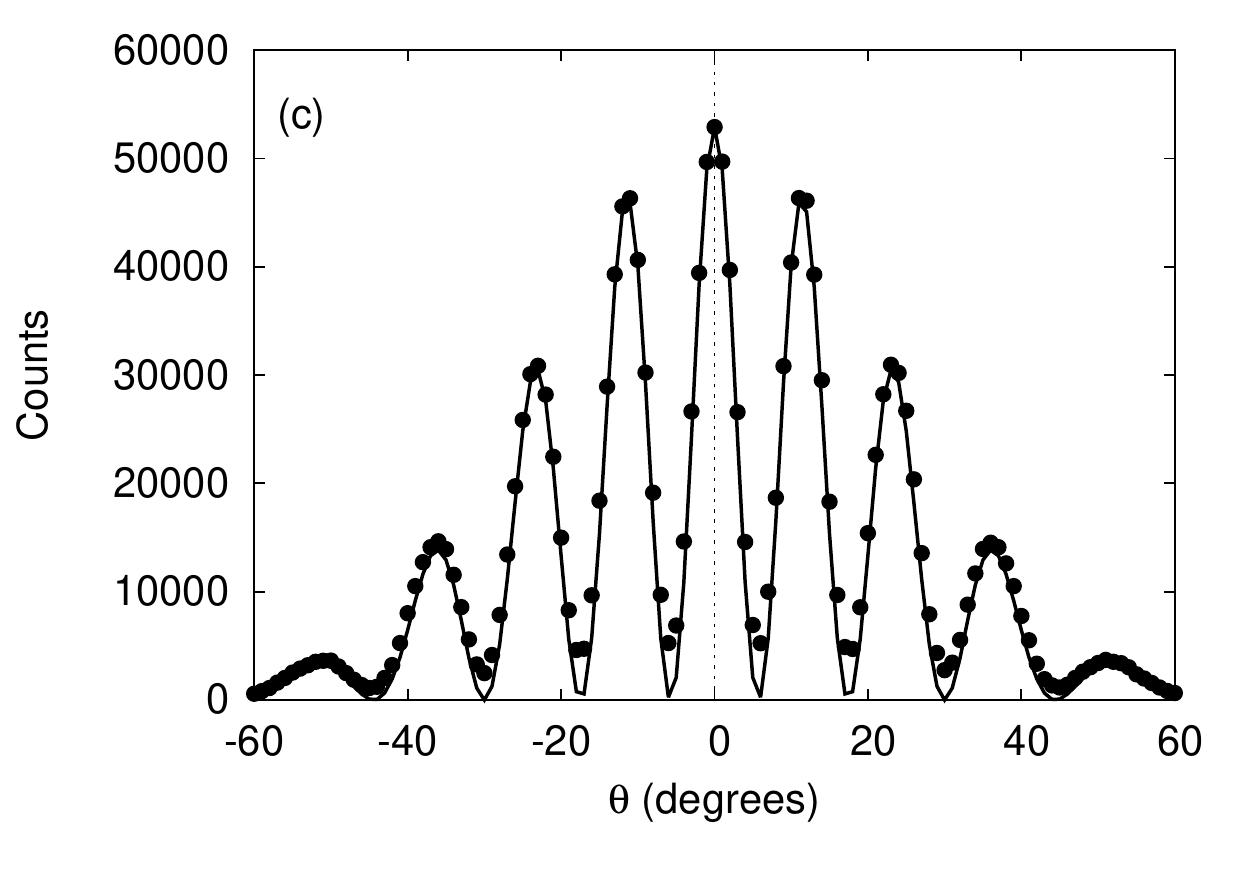}
\includegraphics[width=6cm]{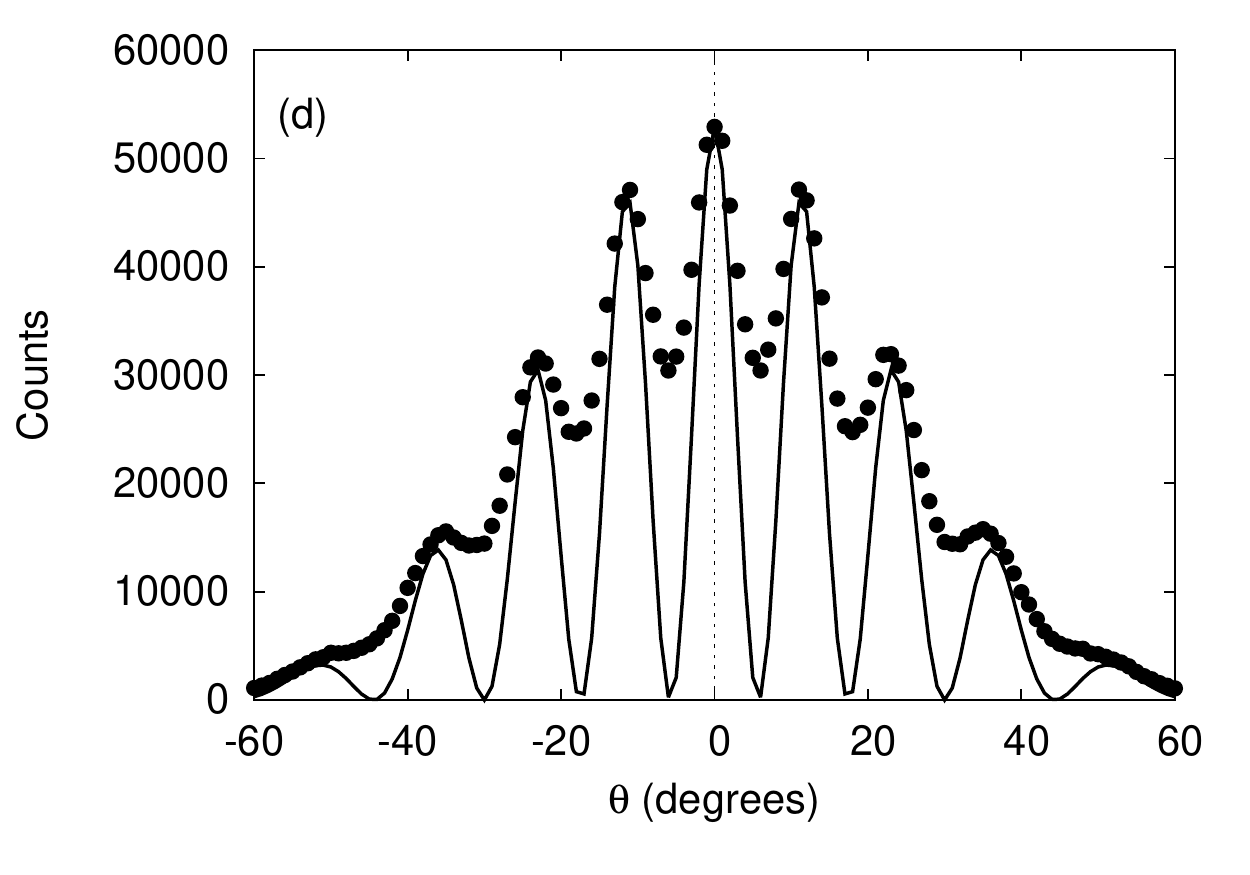}
\vspace*{8pt}
\caption{%
Detector counts (markers) as a function of $\theta$
as obtained from the event-based simulation of the two-beam interference experiments described in Fig.~\ref{fig2}.
Simulation parameters: $N=10^7$ so that on average, each of the 181 detectors,
positioned on the semi-circular screen with an angular spacing of $1^{\circ}$ in the interval $\left[-90^{\circ},90^{\circ}\right]$,
receives about $55\times 10^3$
particles, $\gamma=0.999$, $a=c/f$, $d=5c/f$, $X=75c/f$, where $c$ denotes
the velocity and $f$ the frequency of the particles ($c/f=670\;\hbox{nm}$ in our simulations).
(a): first experiment in which sources $S_1$ and $S_2$ in random order emit in total $N$ particles one-by-one.
This experiment resembles Young's (and Feynman's) two-slit experiment.
(b): first experiment in which only source $S_1$ or $S_2$ emits $N=5\times 10^6$ particles one-by-one (downward and upward triangles, respectively).
The open squares are the sum of the detector counts of the two experiments with one source emitting and the other one blocked.
This experiment resembles Feynman's two-slit experiment with first slit $S_2$ blocked and then slit $S_1$ blocked.
The closed circles are the result of the second experiment in which first $S_1$ and then $S_2$ emit a group of $M=5\times 10^6$ particles
one-by-one.
(c): second experiment with $M=10^6$.
(d): second experiment with $M=25\times 10^5$.
The solid line in (a), (c) and (d) is a least-square fit of the simulation data of (a) to the prediction of wave theory,
Eq.~(\ref{tbi2}), with only one fitting parameter.
}\label{fig3}
\end{center}
\end{figure}

In Fig.~\ref{fig3}(a), we present simulation results for the first experiment for a representative case for which the analytical
solution from wave theory is known.
Namely, in the Fraunhofer regime ($d\ll X$), the analytical expression for the light intensity at the detector on a circular screen with radius $X$ is given by~\cite{BORN64}
\begin{equation}
I(\theta) = A\sin^2\left(\frac{qa\sin\theta}{2}\right)\cos^2\left(\frac{qd\sin\theta}{2}\right)/\left(\frac{qa\sin\theta}{2}\right)^2
,
\label{tbi2}
\end{equation}
where $A$ is a constant, $q=2\pi f/c$ denotes the wavenumber with $f$ and $c$ being the frequency and velocity of the light, respectively,
and $\theta$ denotes the angular position of the detector $D$ on the circular screen, see Fig.~\ref{fig2}.
Note that Eq.~(\ref{tbi2}) is only used for comparison with the simulation data and is by no means input to the model.
From Fig.~\ref{fig3}(a) it is clear that the event-based model reproduces the results of wave theory
and this without taking recourse of the solution of a wave equation.

As the detection efficiency of the event-based detector model is very close to 100\%,
the interference patterns generated by the event-based model cannot be attributed
to inefficient detectors.
It is therefore of interest to take a look at the ratio of detected to emitted photons, the overall detection efficiency, and
compare the detection counts, observed in the event-by-event simulation
of the two-beam interference experiment, with those observed in a real experiment with single photons~\cite{JACQ05}.
In the simulation that yields the results of Fig.~\ref{fig3}(a), each of the 181 detectors making up the
detection area is hit on average by $55\times 10^3$ photons and the total number of clicks generated
by the detectors is $0.16 \times 10^7$. Hence, the ratio of the total number of detected to emitted photons is
of the order of 0.16, two orders of magnitude larger than
the ratio $0.5\times 10^{-3}$ observed in single-photon interference experiments~\cite{JACQ05}.

In Fig.~\ref{fig3}(b), we show simulation results for the experiment in which first only source $S_1$ emits $N=5\times 10^6$ photons
(downward triangles) while $S_2$ is blocked by the mask. Then in a new experiment (all detectors are reset)
$S_2$ emits $N=5\times 10^6$ photons while $S_1$ is blocked (upward triangles).
The sum of the two resulting detection curves is given by the curve with open squares. It is clear that this curve is completely
different from the curve depicted in Fig.~\ref{fig3}(a), as is also described in Feynman's thought experiment (see Sect.~2.1).
Also in Fig.~\ref{fig3}(b) we present the simulation results for the experiment in which first the source $S_1$
emits a group of $M=5\times 10^6$ particles one-by-one and then the source $S_2$ emits $M=5\times 10^6$ particles one-by-one
(no resetting of the detectors).
The resulting detection curve is drawn with closed circles. For small values of $\theta$ there is a difference
between the curves with open squares and closed circles. This difference is due to the memory effect which is
present in the detector model. Obviously this difference depends on $\gamma$ and the detector model that is used.
For more complicated detector models than the one given by Eq.~(\ref{ruleofLM}) this small difference disappears
(results not shown).

Figs.~\ref{fig3}(c),(d) depict simulation results of the experiment in which sources $S_1$ and $S_2$ alternately emit
$M$ particles one-by-one with $M=10^6$ and $M=25\times 10^5$, respectively.
It is seen that except for very large values of $M$ ($M\gtrsim 10^6$), the interference pattern is the same as the one
shown in Fig.~\ref{fig3}(a). Nevertheless, for these large values of $M$ interference can still be observed.
This is a result of the memory effects built in the detector model.
However, for any value of $M$, a simple quantum theoretical calculation would predict no interference pattern but an intensity pattern
which is the sum of two single slit patterns, as the particles pass
through one or the other slit, and never through both.
Hence, for this type of experiment the predictions of quantum theory and of the event-based model differ.

Although we are not aware of any experiment that precisely tests the above described scenario,
one experimental study in which only one slit was available to each photon~\cite{ALKO01} produced
intriguing results. In that study, an opaque barrier, all the way from the laser source to the
obstacle between the two slits, was used to make sure that photons had one or the other
slit available to them. The interference pattern observed was nevertheless essentially
unchanged despite the presence of the barrier. We are, however, not aware of any follow-up work on
that study.

\subsubsection{Why is interference produced without solving a wave problem?}
As mentioned earlier, using simple particle counters as detectors would
not result in an interference pattern.
Essential to produce an interference pattern is to account for the information about the differences
in the times of flight (or phase differences) of the particles
which encode the distance the particles travelled from one of the
two sources to one of the detectors constituting the circular
detection screen.
Simple particle counters do nothing with the information which is encoded in the messages
carried by the particles and produce a click for each incoming particle.
Since,
in the single-photon two-beam experiment the detectors
are the only apparatuses available that
can process these phase differences (there are no other apparatuses
present except for the source) we necessarily need
to employ an algorithm for the detector that exploits this
information in order to produce the clicks that gradually
build up the interference pattern. A collection of about
two hundred independent adaptive threshold detectors
defined by Eq.~(\ref{ruleofLM}) and Eq.~(\ref{thresholdofLM}) and each with a detection
efficiency of nearly 100\% is capable of doing this. As
pointed out earlier, the reason why, in this particular experiment,
this is possible is that not every particle that
impinges on the detector yields a click.

\subsection{Mach Zehnder interferometer experiment}
\subsubsection{Event-based model}
The DLM network that simulates a single-photon MZI experiment (see Fig.~\ref{fig4} (left)) consists of a source, two identical
BSs two phase shifters and two detectors.
The network of processing units
is a one-to-one image of the experimental setup.~\cite{GRAN86}
Note that the two mirrors in the MZI simply bend the paths of the photons by $\pi/2$ without introducing a phase change
or loss of particles and therefore they do not need to be considered in the event-based simulation network.
In what follows we specify the processing units in sufficient detail such that the reader who is interested can reproduce
the simulation results.
We require that the processing units for
identical optical components should be reusable within the same
and within different experiments.
Demonstration programs, including source
codes, are available for download~\cite{MZI08,MZIdemo}.

\begin{itemize}[leftmargin=*]
\renewcommand\labelitemi{-}
\item{{\sl Source and particles:}
In a pictorial description of the experiment depicted in Fig.~\ref{fig4} (left) the photons, leaving the source $S$ one-by-one, can be
regarded as particles playing the role of messengers. Each
messenger carries a message
\begin{equation}
\mathbf{u}_{k,n}=(\cos (2\pi f t_{k,n}), \sin (2\pi f t_{k,n})),
\end{equation}
where $f$ denotes the frequency of the light
source and $t_{k,n}$ the time that particles need to travel a given path.
The subscript $n>0$ numbers the
consecutive messengers and $k$ labels the channel of the BS at
which the messenger arrives (see below).
Note that in this experiment no explicit
information about distances and frequencies is required since we
can always work with relative phases.

When a messenger is created its internal clock time is set to zero ($t_{k,n}=0$) and since the source
is connected to the $k=0$ input channel of the first BS the messenger gets the label $k=0$ (see Fig.~\ref{fig4} (left)).
}%

\begin{figure}[pt]
\begin{center}
\includegraphics[width=6cm]{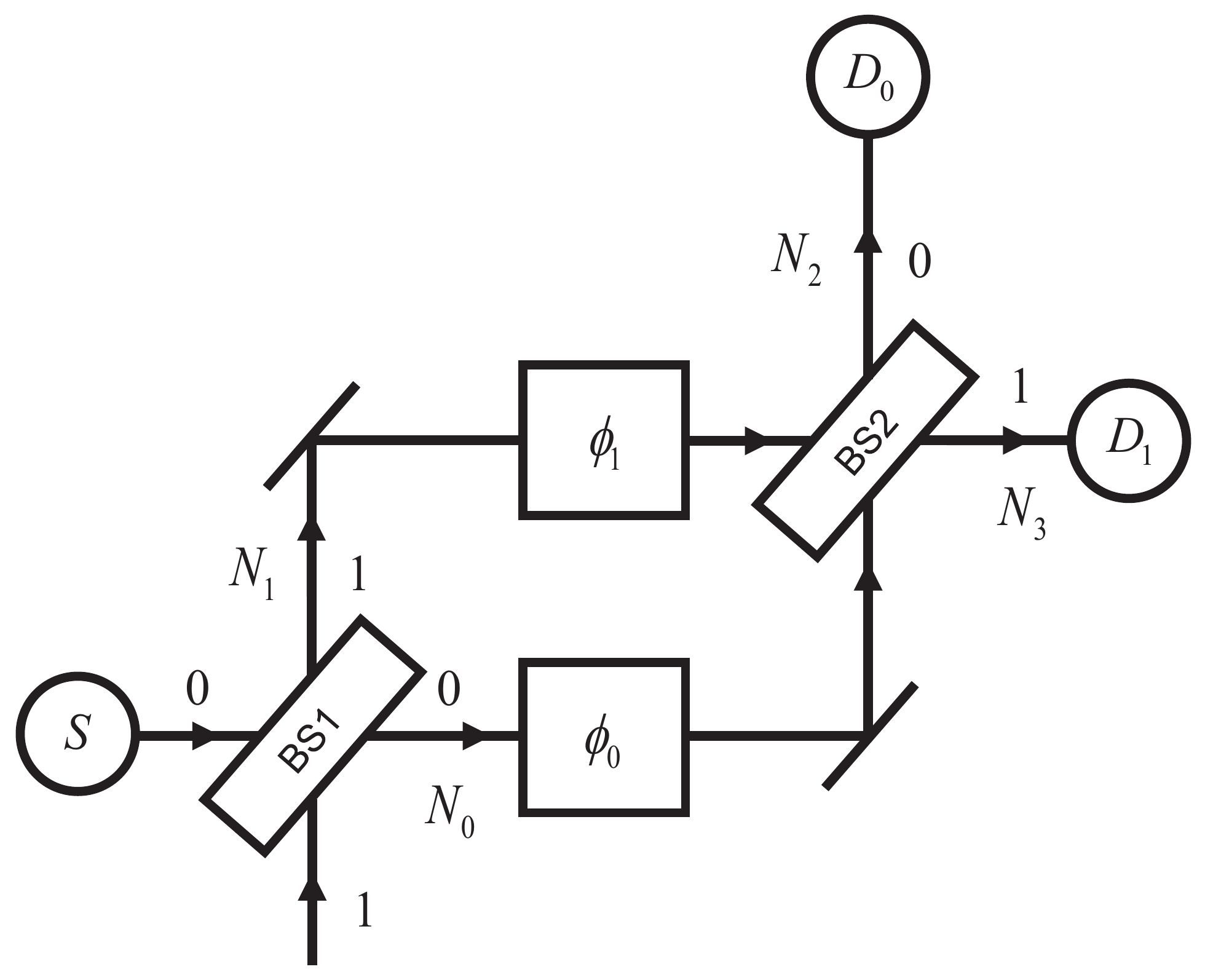}
\includegraphics[width=6.5cm]{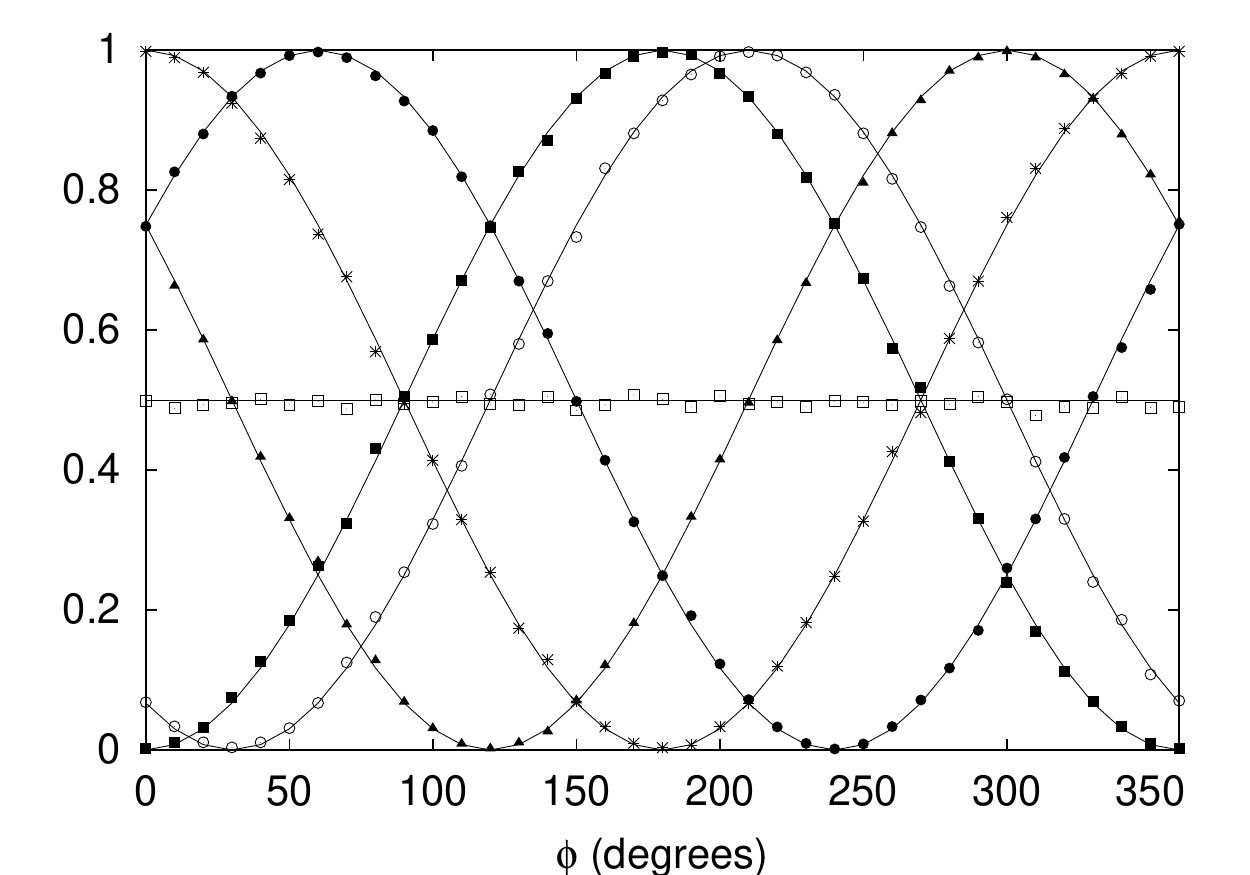}
\vspace*{8pt}
\caption{Left: Schematic diagram of a Mach-Zehnder interferometer (MZI) with a single-photon source $S$.
The MZI consists of two beam splitters, BS1 and BS2, two phase shifters $\phi_0$ and $\phi_1$ and two mirrors.
$N_0$ ($N_2$) and $N_1$ ($N_3$) count the number of events in the output channel 0 of BS1 (BS2)
and in the output channel 1 of BS1 (BS2), respectively.
Dividing $N_i$ for $i=0,\ldots,3$ by the total count $N$ yields the relative frequency
of finding a photon in the corresponding arm of the interferometer.
Since photon detectors operate by absorbing photons, in a real laboratory
experiment only $N_2$ and $N_3$ can be measured by detectors $D_0$ and $D_1$, respectively.
Right: Simulation results for the normalized detector counts (markers) as a function of $\phi = \phi_0-\phi_1$.
Input channel 0 receives $(\cos\psi_0,\sin\psi_0)$ with probability one.
One uniform random number in the range $[0,360]$ is used to choose the angle $\psi_0$.
Input channel 1 receives no events. 
The parameter $\gamma =0.98$.
Each data point represents 10000 events ($N=N_0+N_1=N_2+N_3=10000$).
Initially the rotation angle $\phi_0=0$ and after each set of 10000 events, $\phi_0$
is increased by $10^\circ$.
Open squares: $N_0/N$;
solid squares: $N_2/N$ for $\phi_1=0$;
open circles: $N_2/N$ for $\phi_1=30^\circ$;
solid circles: $N_2/N$ for $\phi_1=240^\circ$;
asterisks: $N_3/N$ for  $\phi_1=0$;
solid triangles: $N_3/N$ for $\phi_1=300^\circ$.
Lines represent the results of quantum theory~\cite{QuantumTheory}.
}\label{fig4}
\end{center}
\end{figure}

\item{{\sl Beam splitter} (BS){\sl :}
A BS is an optical component that partially transmits and partially reflects an incident light beam.
Dielectric plate BSs are often used as 50/50 BSs.
From classical electrodynamics we know that if an electric field is applied to a dielectric material the
material becomes polarized.~\cite{BORN64}
Assuming a linear response, the polarization vector of the material is given by
${\mathbf P}(\omega)={\mathbf \chi}(\omega){\mathbf E}(\omega)$ for a monochromatic wave with
frequency $\omega$. In the time domain, this relation expresses the fact that the material
retains some memory about the incident field, ${\mathbf \chi}(\omega)$ representing the memory kernel
that is characteristic for the material used.
We use this kind of memory effect in our algorithm to model the BS.

A BS has two input and two output channels labeled by 0 and 1 (see Fig~\ref{fig4} (left)).
Note that in case of the MZI experiment, for beam splitter BS1 only entrance port $k=0$ is used.
In the event-based model, the BS has two internal registers ${\mathbf R}_{k,n}=(R_{0,k,n},R_{1,k,n})$ (one for each input channel) and an
internal vector ${\mathbf v}_n=(v_{0,n},v_{1,n})$ with the additional constraints that $v_{i,n}\ge0$ for $i=0,1$ and that $v_{0,n}+v_{1,n}=1$.
As we only have two input channels, the latter constraint can be used
to recover $v_{1,n}$ from the value of $v_{0,n}$. We prefer to work with internal vectors that have as many elements as there are input
channels.
These three two-dimensional vectors ${\mathbf v}_n$, ${\mathbf R}_{0,n}$ and ${\mathbf R}_{1,n}$ are labeled by the message number
$n$ because their content is updated every time the BS receives a message.
Before the simulation starts we set ${\mathbf v}_0 = (v_{0,0}, v_{1,0}) = (r, 1 - r)$, where $r$ is a uniform pseudo-random
number. In a similar way we use pseudo-random numbers to set ${\mathbf R}_{0,0}$ and ${\mathbf R}_{1,0}$.

When the $n$th messenger carrying the message $\mathbf{u}_{k,n}$ arrives at entrance port $k=0$ or $k=1$ of the BS,
the BS first stores the message in the corresponding register ${\mathbf R}_{k,n}$ and updates its internal vector according to the rule
\begin{equation}
{\mathbf v}_n=\gamma {\mathbf v}_{n-1}+(1-\gamma){\mathbf q}_n,
\label{internalBS}
\end{equation}
where $0<\gamma<1$ is a parameter that controls the learning process
and ${\mathbf q}_n=(1,0)$ (${\mathbf q}_n=(0,1)$) if the $n$th event occurred on channel $k=0$ ($k=1$).
By construction $v_{i,n}\ge 0$ for $i=0,1$ and $v_{0,n}+v_{1,n} =1$. Hence the update rule Eq.~(\ref{internalBS})
preserves the constraints on the internal vector. Obviously,
these constraints are necessary if we want to interpret the $v_{k,n}$
as (an estimate of) the frequency for the occurrence of an event of type $k$.
Note that the BS stores information about the last message only. The information
carried by earlier messages is overwritten by updating the
internal registers.
From Eq.~(\ref{internalBS}), one could
say that the internal vector ${\mathbf v}$ (corresponding to the material polarization ${\mathbf P}$) is the response of the BS to the
incoming messages (photons) represented by the vectors ${\mathbf q}$ (corresponding to the elementary unit of electric field ${\mathbf E}$).
Therefore, the BS ``learns'' so to speak from the information
carried by the photons. The characteristics of the learning
process depend on the parameter $\gamma$ (corresponding to the
response function $\chi$).

Next, in case of a 50/50 BS,
the BS uses the six numbers stored in ${\mathbf R}_{0,n}$, ${\mathbf R}_{1,n}$ and ${\mathbf v}_n$ to calculate four numbers
$g_{0,n}=(R_{0,0,n}\sqrt{v_{0,n}}-R_{1,1,n}\sqrt{v_{1,n}})/\sqrt{2}$,
$g_{1,n}=(R_{0,1,n}\sqrt{v_{1,n}}+R_{1,0,n}\sqrt{v_{0,n}})/\sqrt{2}$,
$g_{2,n}=(R_{0,1,n}\sqrt{v_{1,n}}-R_{1,0,n}\sqrt{v_{0,n}})/\sqrt{2}$,
and $g_{3,n}=(R_{0,0,n}\sqrt{v_{0,n}}+R_{1,1,n}\sqrt{v_{1,n}})/\sqrt{2}$.
These four real-valued numbers can be considered to represent the real and imaginary part of two complex numbers
$g_{0,n}+ig_{1,n}$ and $g_{2,n}+ig_{3,n}$ which are obtained by the following matrix-vector multiplication
\begin{eqnarray}
\left(
\begin{array}{c}
g_{0,n}+ig_{1,n}\\
g_{2,n}+ig_{3,n}
\end{array}
\right)
&=&
\frac{1}{\sqrt{2}}
\left(
\begin{array}{c}
\sqrt{v_{0,n}}(R_{0,0,n}+iR_{1,0,n})+i\sqrt{v_{1,n}}(R_{0,1,n}+iR_{1,1,n})\\
i\sqrt{v_{0,n}}(R_{0,0,n}+iR_{1,0,n})+\sqrt{v_{1,n}}(R_{0,1,n}+iR_{1,1,n})
\end{array}
\right) \nonumber\\
&=&
\frac{1}{\sqrt{2}}
\left(
\begin{array}{cc}
1&i\\
i&1
\end{array}
\right)
\left(
\begin{array}{cc}
\sqrt{v_{0,n}}&0\\
0&\sqrt{v_{1,n}}
\end{array}
\right)
\left(
\begin{array}{c}
R_{0,0,n}+iR_{1,0,n}\\
R_{0,1,n}+iR_{1,1,n}
\end{array}
\right)
,
\label{BS1random}
\end{eqnarray}

Identifying $a_0$ with $\sqrt{v_{0,n}}(R_{0,0,n}+iR_{1,0,n})$ and
$a_1$ with $\sqrt{v_{1,n}}(R_{0,1,n}+iR_{1,1,n})$
it is clear that the computation of the four numbers $g_{i,n}$ for $i=0,\ldots, 3$ plays the role of the
matrix-vector multiplication in the quantum theoretical description of a beam-splitter
\begin{eqnarray}
\left(
\begin{array}{c}
b_0\\
b_1
\end{array}
\right)
=
\frac{1}{\sqrt{2}}
\left(
\begin{array}{c}
a_0+ia_1\\
a_1+ia_0
\end{array}
\right)
=
\frac{1}{\sqrt{2}}
\left(
\begin{array}{cc}
1&i\\
i&1
\end{array}
\right)
\left(
\begin{array}{c}
a_0\\
a_1
\end{array}
\right)
,
\label{BS1}
\end{eqnarray}
where $(a_0,a_1)$ and $(b_0,b_1)$ denote the input and output amplitudes, respectively.
Note however that the DLM for the BS computes the four numbers $g_{i,n}$ for $i=0,\ldots, 3$ for each incoming event
thereby always updating ${\mathbf v}_n$ and ${\mathbf R}_{0,n}$ or ${\mathbf R}_{1,n}$.
Hence, $a_0$ and $a_1$, and thus also $b_0$ and $b_1$, are constructed event-by-event and only under certain conditions
($\gamma\rightarrow 1^-$, sufficiently large number of input events $N$, stationary sequence of input events) they correspond
to their quantum theoretical counterparts $a_0=\sqrt{p_0}e^{i\psi_0}$, $a_1=\sqrt{p_1}e^{i\psi_1}$ with $p_1=1-p_0$
($0\le p_0,p_1\le 1$) and $b_0=a_0+ia_1$, $b_1=a_1+ia_0$ (see Eq.~(\ref{BS1})).

In a final step the BS uses $g_{i,n}$ for $i=0,\ldots, 3$ to create an output event.
Therefore it generates a uniform random number $r_n$ between
zero and one.
If $g_{0,n}^2+g_{1,n}^2 > r_n$, the BS sends a message
\begin{equation}
{\mathbf w}_{0,n}=(g_{0,n},g_{1,n})/\sqrt{g_{0,n}^2+g_{1,n}^2},
\end{equation}
through output channel 0. Otherwise
it sends a message
\begin{equation}
{\mathbf w}_{1,n}=(g_{2,n},g_{3,n})/\sqrt{g_{2,n}^2+g_{3,n}^2},
\end{equation}
through output channel 1.
}
\item{{\sl Phase shifters:}
These devices perform a plane rotation on the vectors (messages) carried by the
particles. As a result the phase of the particles is changed by $\phi_0$ or $\phi_1$ depending on the route followed.
}
\item{{\sl Detector:}
Detector $D_0$($D_1$) registers the output events at channel 0 (1).
The detectors are ideal particle counters,
meaning that they produce a click for each incoming
particle. Hence, we assume that the detectors have
100\% detection efficiency. Note that also adaptive
threshold detectors can be used (see Sect.~5.1.1) equally
well.~\cite{MICH11a}
}
\item{{\sl Simulation procedure:}
When a messenger is created we wait until its message has been processed by one of the detectors before creating the next messenger.
This ensures that there can be no direct communication between the messengers
and that our simulation model (trivially) satisfies Einstein's criterion of local causality.
We assume that no messengers are lost. Since the detectors are ideal particle counters the number of clicks
generated by the detectors is equal to the number of messengers created by the source.
For fixed $\phi=\phi_0-\phi_1$, a simulation run of $N$ events generates the data set $\Gamma(\phi)=\{w_n|n = 1,\ldots, N\}$.
Here $w_n=0,1$ indicates which detector fired ($D_0$ or $D_1$). Given the data set $\Gamma(\phi)$,
we can easily compute the number of 0 (1) output events $N_2$ ($N_3$).
}
\end{itemize}

\subsubsection{Simulation results}
In Fig.~\ref{fig4} (right), we present a few simulation results for the MZI and compare them to the quantum theoretical result.
According to quantum theory, the amplitudes $(b_0, b_1)$ in the output
modes 0 and 1 of the MZI are given by~\cite{BAYM74}
\begin{eqnarray}
\left(
\begin{array}{c}
b_0\\
b_1
\end{array}
\right)
=
\frac{1}{2}
\left(
\begin{array}{cc}
1&i\\
i&1
\end{array}
\right)
\left(
\begin{array}{cc}
e^{i\phi_0}&0\\
0&e^{i\phi_1}
\end{array}
\right)
\left(
\begin{array}{cc}
1&i\\
i&1
\end{array}
\right)
\left(
\begin{array}{c}
a_0\\
a_1
\end{array}
\right)
,
\label{MZ1}
\end{eqnarray}
where $a_0$ and $a_1$ denote the input amplitudes.
For the particular choice $a_0=1$ and $a_1=0$,
in which case there are no particles entering BS1 via channel 1,
it follows from Eq.~(\ref{MZ1}) that
\begin{eqnarray}
|b_0|^2=\sin^2(\frac{\phi_0-\phi_1}{2}),
\quad
|b_1|^2=\cos^2(\frac{\phi_0-\phi_1}{2}).
\label{MZ3}
\end{eqnarray}
For the results presented in Fig.~\ref{fig4} (right) we assume that input channel 0 receives $(\cos\psi_0,\sin\psi_0)$ with probability one
and that input channel 1 receives no events.
This corresponds to $(a_0, a_1) = (\cos\psi_0+i\sin\psi_0,0)$.
We use a uniform random number to determine $\psi_0$.
Note that this random number is used to generate all input events.
The data points are the simulation results
for the normalized intensity $N_i/N$ for $i=0,2,3$ as a function of $\phi = \phi_0 - \phi_1$.
Note that in an experimental setting it is impossible to
simultaneously measure ($N_0/N$, $N_1/N$)
and ($N_2/N$, $N_3/N$) because photon detectors operate by absorbing photons.
In the event-based simulation there is no such problem.
From Fig.~\ref{fig4} (right) it is clear that
the event-based processing by the DLM network reproduces the probability distribution of quantum theory, see Eq.~(\ref{MZ3})
with $|b_0|^2$ ($|b_1|^2$) corresponding to $N_2/N$ ($N_3/N$).

\subsubsection{Why is interference produced without solving a wave problem?}
We consider BS2 of the MZI depicted in Fig.~\ref{fig4} (left), the beam splitter at which, in
a wave picture, the two beams join to produce interference. The
DLM simulating a BS requires two
pieces of information to send out particles such that their distribution
matches the wave-mechanical description of the BS.
First, it needs an estimate of the ratio of particle currents
in the input channels 0 and 1 (paths 0 and 1 of the MZI), respectively.
Second, it needs to
have information about the time of flight (phase difference) along the two different
paths of the MZI.
The first piece of information is provided for by the internal
vector ${\mathbf v}=(v_0,v_1)$. Through the update rule Eq.~(\ref{internalBS}),
for a stationary sequence of input events, $v_0$ and $v_1$ converge
to the average of the number of events on input channels 0 and 1,
respectively. Thus, the ratio of the particles (corresponding to the intensities of the waves) in the two input
beams are encoded in the vector ${\mathbf v}$. Note that this information is
accurate only if the sequence of input events is stationary.
After one particle arrived at port 0 and another one arrived at
port 1, the second piece of information is available in the
registers ${\mathbf R}_0$ and ${\mathbf R}_1$. This information plays the role of the phase
of the waves in the two input beams.
Hence, all the information (intensity and
phase) is available to compute the probability for sending out
particles.
This is done by calculating the numbers $g_i$ for $i=0,\ldots ,3$
which, in the stationary state, are identical to
the wave amplitudes obtained from the
wave theory of a beam splitter.~\cite{BORN64}

\subsection{Wheeler's delayed choice experiment}
\begin{figure}[pt]
\begin{center}
\includegraphics[width=12cm]{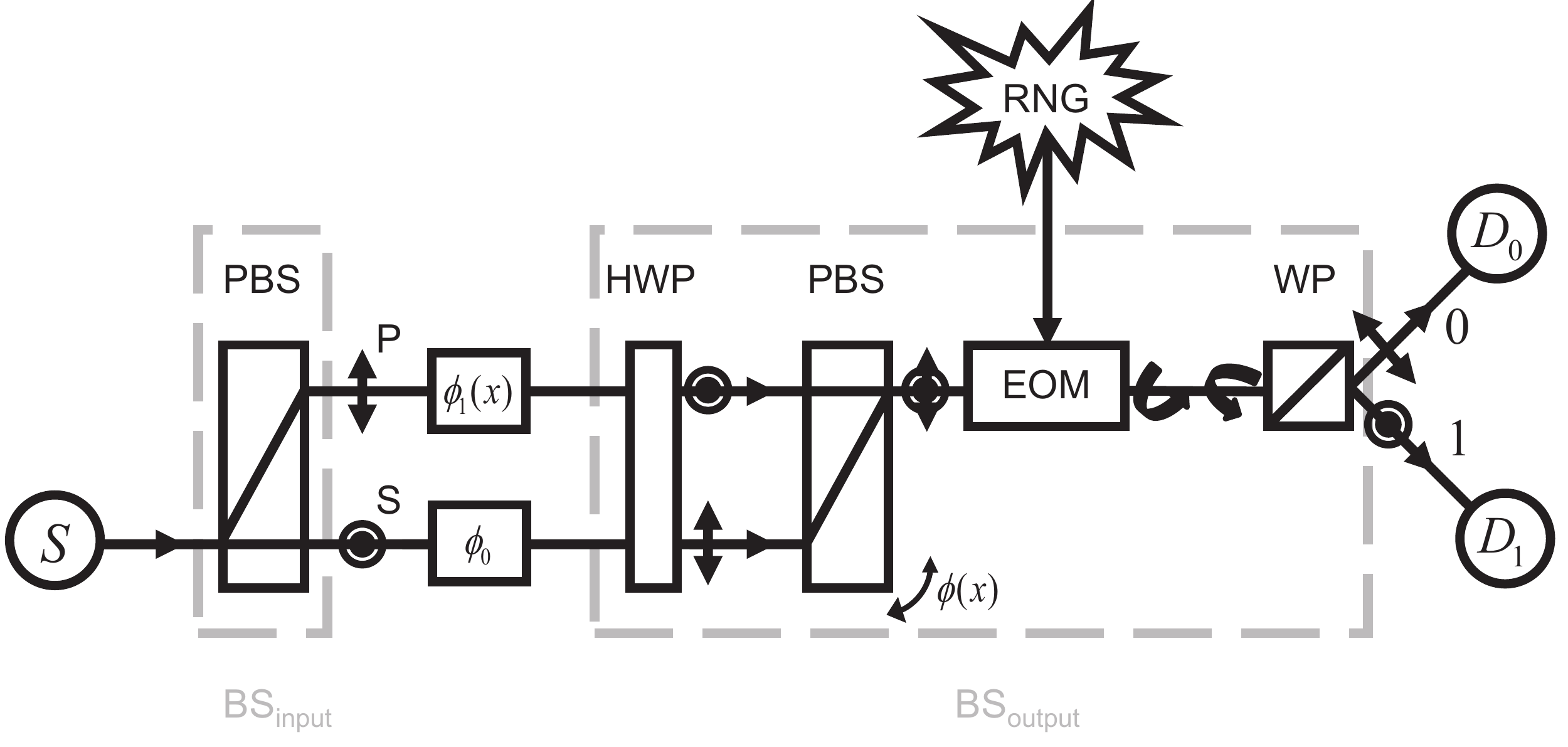}
\vspace*{8pt}
\caption{Schematic diagram of the experimental setup of Wheeler's delayed-choice experiment with single photons.~\cite{JACQ07,JACQ08}
$S$: single-photon source; PBS: polarizing beam splitter; HWP: half-wave plate; EOM: electro-optic modulator; RNG: random number generator;
WP: Wollaston prism (= PBS); $D_0$ and $D_1$: detectors; P, S: polarization state of the photons; $\phi (x)=\phi_0 -\phi_1(x)$:
phase shift between paths 0 and 1.
The diagram is that of a Mach-Zehnder interferometer composed of a 50/50 input beam splitter (BS$_{\mathrm input}$) and a variable
output beam splitter (BS$_{\mathrm output}$) with adjustable reflectivity ${\cal R}$ .
}\label{fig5}
\end{center}
\end{figure}

In a recent experimental realization of Wheeler's delayed-choice
experiment by Jacques {\sl et al.}~\cite{JACQ08} linearly polarized single photons
are sent through a polarizing beam splitter (PBS) that together with a second, movable,
variable output PBS with adjustable reflectivity ${\cal R}$ forms an
interferometer (see Fig.~\ref{fig5}). In the first realization~\cite{JACQ07} two 50/50 BSs were used.

Tilting the PBS of the variable output BS induces a time-delay in one
of the arms of the MZI, symbolically represented by the variable phase $\phi_1 (x)$ in Fig.~\ref{fig5}, and
thus varies the phase shift $\phi(x)=\phi_0-\phi_1(x)$ between the two arms of the MZI.
A voltage applied to an electro-optic modulator (EOM) controls the
reflectivity ${\cal R}$ of the variable beam
splitter BS$_{\mathrm output}$.
If no voltage is applied to the EOM then ${\cal R}=0$.
Otherwise, ${\cal R}\ne 0$ (see Eq.~(2) in Ref.~\citen{JACQ08}) and the EOM acts as a wave plate which rotates the
polarization of the incoming photon by an angle depending on the value of ${\cal R}$.
The voltage applied to the EOM is controlled by a set of pseudo-random numbers generated by the
random number generator RNG.
The key point in this experiment is that the decision to apply a voltage to the EOM
is made after the photon has passed BS$_{\mathrm input}$.

For $0\le {\cal R} \le 0.5$ measured values of the interference visibility~\cite{HELL87} $V$ and
the path distinguishability~\cite{JACQ08} $D$, a parameter that quantifies
the which-path information (WPI), were found to fulfill the
complementary relation $V^2 + D^2\le 1$.~\cite{JACQ08} For
$(V = 0,  D = 1)$ and $(V = 1, D = 0)$, obtained for ${\cal R} = 0$ and ${\cal R}= 0.5$, respectively,
full and no WPI was found, associated with particle like and wavelike
behavior, respectively. For $0\le {\cal R}\le 0.5$ partial WPI was obtained while keeping interference with
limited visibility.~\cite{JACQ08}

Although the detection events (detector ``clicks'') are the only
experimental facts and logically speaking one cannot say anything about what happens with the photons traveling
through the setup, Jacques {\sl et al.}~\cite{JACQ07,JACQ08} gave the following
pictorial description: Linearly polarized single photons are sent through a
50/50 PBS (BS$_{\mathrm input}$), spatially separating photons with S polarization
(path 0) and P polarization (path 1) with equal frequencies.
After the photon has passed BS$_{\mathrm input}$, but before the photon enters
the variable BS$_{\mathrm output}$ the decision to apply a voltage to the EOM is
made. The PBS of BS$_{\mathrm output}$ merges the paths of the orthogonally
polarized photons travelling paths 0 and 1 of the MZI, but
afterwards the photons can still be unambiguously identified by
their polarizations. If no voltage is applied to the EOM then ${\cal R} = 0$
and the EOM does nothing to the photons.
Because the polarization eigenstates of the Wollaston prism
correspond to the P and S polarization of the photons travelling
path 0 and 1 of the MZI, each detection event registered by one of
the two detectors $D_0$ or $D_1$ is associated with a specific path (path
0 or 1, respectively). Both detectors register an equal amount of
detection events, independent of the phase shift $\phi (x)$ in the MZI. This
experimental setting clearly gives full WPI about the photon within the
interferometer (particle behavior), characterized by $D = 1$. In this
case no interference effects are observed and thus $V = 0$.
When a voltage is applied to
the EOM, then ${\cal R}\ne 0$ and the EOM rotates the polarization of the incoming photon by an
angle depending on ${\cal R}$. The Wollaston prism partially recombines
the polarization of the photons that have travelled along different
optical paths with phase difference $\phi (x)$ and
interference appears ($V\ne 0$), a result expected for a wave. The
WPI is partially washed out, up to be totally erased when
${\cal R}=0.5$.
Hence, the decision to apply a voltage to the EOM after the photon
left BS$_{\mathrm input}$ but before it passes BS$_{\mathrm output}$, influences the behavior of
the photon in the past and changes the representation of the
photon from a particle to a wave~\cite{JACQ07}.

\subsubsection{Event-based model}
We construct a model for the messengers representing the linearly polarized photons and
for the processing units representing the optical components in the
experimental setup (see Fig.~\ref{fig5}) thereby fulfilling the requirements that the processing units for
identical optical components should be reusable within the same
and within different experiments and that the network of processing units
is a one-to-one image of the experimental setup.
Although, in contrast to the experiments we have considered so far, in this experiment
it is necessary to include the polarization in the model for the messengers representing the photons.
These more general messengers can also be used in a simulation of the experiments discussed previously.
In the event-based simulation of these experiments the polarization component of the message is simply not used
in the DLMs modeling the optical components of their experimental setup.
In what follows we describe the elements of the model in more detail.

\begin{itemize}[leftmargin=*]
\renewcommand\labelitemi{-}
\item{{\sl Source and particles:}
The polarization can be included in the model for the messengers representing the photons
by adding to the message a second harmonic oscillator which also vibrates with frequency $f$.
There are many different but equivalent ways to define the message.
As in Maxwell's and quantum theory, it is convenient (though) not essential to work with complex valued vectors,
that is with messages represented by two-dimensional unit vectors
\begin{equation}
{\mathbf u}=(e^{i\psi^{(1)}}\sin\xi, e^{i\psi^{(2)}}\cos\xi),
\label{polphoton}
\end{equation}
where
$\psi^{(i)} =2\pi ft +\delta_i$, for $i=1,2$.
The angle $\xi$ determines the relative magnitude of the two components and
$\delta =\delta_1-\delta_2=\psi^{(1)}-\psi^{(2)}$,
denotes the phase difference between the two components.
Both $\xi$ and $\delta$ determine the polarization of the photon.
Hence, the photon can be considered to have a polarization vector
${\mathbf P}=(\cos\delta\sin 2\xi, \sin\delta\sin 2\xi, \cos 2\xi)$.
The third degree of freedom in Eq.~(\ref{polphoton}) is used to account
for the time of flight of the photon.
Within the present model, it is thus postulated that the state of the photon is fully determined
by the angles $\psi^{(1)}$, $\psi^{(2)}$ and $\xi$ and by rules (to be specified), by which these angles change
as the photon travels through the network.

A messenger with message ${\mathbf u}$ at time $t$ and position ${\mathbf r}$
that travels with velocity $v=c/n$, where $c$ denotes the velocity of light and $n$ is the index of
refraction of the material, along the direction ${\mathbf q}$ during a time interval
$t^{\prime} - t$, changes its message according to $\psi^{(i)}\rightarrow \psi^{(i)}+\phi$ for $i=1,2$, where
$\phi =2\pi f(t^{\prime}-t)$. This suggests that we may view the two-component vectors $\mathbf {u}$ as the coordinates
of two local oscillators, carried along by the messengers and that the messenger encodes its time of flight
in these two oscillators.

It is evident that the representation used here maps
one-to-one to the plane-wave description of a classical electromagnetic
field,~\cite{BORN64} except that we assign these properties to each
individual photon, not to a wave.
As there is no communication/interaction between the messengers there can be no wave equation
(partial differential equation) that enforces a relation between the messages carried by different messages.

When the source creates a messenger, its message needs to be initialized.
This means that the three angles $\psi^{(1)}$, $\psi^{(2)}$ and $\xi$ need to be specified.
The specification depends on the type of light source that has to be simulated.
For a coherent light source, the three angles are different but the same for all the messengers being created.
Hence, three random numbers are used to specify $\psi^{(1)}$, $\psi^{(2)}$ and $\xi$ for all messengers.

In this section we will demonstrate explicitly that in the event-based
model (in general, not only for this experiment) photons always have full WPI even if interference is
observed by giving the messengers one extra label, the path label
having the value 0 or 1. The information contained in this label is
not accessible in the experiment.~\cite{JACQ08} We only use it to track the
photons in the network of processing units. The path label is set in the input BS and
remains unchanged until detection. Therefore we do not consider
this label in the description of the processing units but take it into
account when we detect the photons.
}
\item{{\sl Polarizing beam splitter} (PBS){\sl :}
A PBS is used to redirect photons depending on their polarization.
For simplicity, we assume that the coordinate system used to
define the incoming messages coincides with the coordinate system defined by two orthogonal
directions of polarization of the PBS.

In general, a PBS has two input and two output channels labeled by 0 and 1, just like an ordinary BS (see Sect.~5.2.1).
Note that in case of Wheeler's delayed choice experiment, the first PBS has only one input channel labeled by $k=0$
and therefore the second PBS has only one output channel labeled by $k=0$.
In the event-based model, the PBS has a similar structure as the BS.
Therefore, in what follows we only mention the main ingredients to construct the processing unit for the PBS. For more details we
refer to Sect.~5.2.1.

The PBS has two internal registers ${\mathbf R}_{k,n}=(R_{0,k,n},R_{1,k,n})$ with $R_{i,k,n}$ for $i=0,1$ representing a complex number, and an
internal vector ${\mathbf v}_n=(v_{0,n},v_{1,n})$, where $v_{i,n}\ge0$ for $i=0,1$, $v_{0,n}+v_{1,n}=1$
and $n$ denotes the message number.
Before the simulation starts uniform pseudo-random numbers are used to set
${\mathbf v}_0$, ${\mathbf R}_{0,0}$ and ${\mathbf R}_{1,0}$.

When the $n$th messenger carrying the message $\mathbf{u}_{k,n}$ arrives at entrance port $k=0$ or $k=1$ of the PBS,
the PBS first copies the message in the corresponding register ${\mathbf R}_{k,n}$ and updates its internal vector according to
\begin{equation}
{\mathbf v}_n=\gamma {\mathbf v}_{n-1}+(1-\gamma){\mathbf q}_n,
\label{internalPBS}
\end{equation}
where $0<\gamma<1$
and ${\mathbf q}_n=(1,0)$ (${\mathbf q}_n=(0,1)$) represents the arrival of the $n$th messenger on channel $k=0$ ($k=1$).
Note that the DLM has storage for exactly ten real-valued numbers.

Next the PBS uses the information stored in ${\mathbf R}_{0,n}$, ${\mathbf R}_{1,n}$ and ${\mathbf v}_n$ to
calculate four complex numbers
\begin{eqnarray}
\left(
\begin{array}{c}
h_{0,n}\\
h_{1,n}\\
h_{2,n}\\
h_{3,n}
\end{array}
\right)
&=&
\left(
\begin{array}{cccc}
1&0&0&0\\
0&1&0&0\\
0&0&0&i\\
0&0&i&0
\end{array}
\right)
\left(
\begin{array}{cccc}
\sqrt{v_{0,n}}&0&0&0\\
0&\sqrt{v_{1,n}}&0&0\\
0&0&\sqrt{v_{0,n}}&0\\
0&0&0&\sqrt{v_{1,n}}
\end{array}
\right)
\left(
\begin{array}{c}
R_{0,0,n}\\
R_{0,1,n}\\
R_{1,0,n}\\
R_{1,1,n}
\end{array}
\right)\nonumber \\
&=&
\left(
\begin{array}{c}
\sqrt{v_{0,n}}R_{0,0,n}\\
\sqrt{v_{1,n}}R_{0,1,n}\\
i\sqrt{v_{1,n}}R_{1,1,n}\\
i\sqrt{v_{0,n}}R_{1,0,n}
\end{array}
\right)
,
\label{PBS1}
\end{eqnarray}
and generates a uniform random number $r_n$ between
zero and one.
If $|h_{0,n}|^2+|h_{2,n}|^2 > r_n$, the PBS sends a message
\begin{equation}
{\mathbf w}_{0,n}=(h_{0,n},h_{2,n})/\sqrt{|h_{0,n}|^2+|h_{2,n}|^2},
\end{equation}
through output channel 1. Otherwise
it sends a message
\begin{equation}
{\mathbf w}_{1,n}=(h_{1,n},h_{3,n})/\sqrt{|h_{1,n}|^2+|h_{3,n}|^2},
\end{equation}
through output channel 0.
}
\item{{\sl Half-wave plate} (HWP){\sl :}
A HWP not only changes the polarization of
the light but also its phase.
In optics, a HWP is often used as a retarder. In
the event-based model, the retardation of the wave corresponds to a
change in the time of flight (and thus the phase) of the messenger.
In contrast to the BS and PBS, a HWP may be simulated without DLM.
The device has only one input and one output port (see Fig.~\ref{fig5}).
A HWP transforms the $n$th input message ${\mathbf u}_n$ into an output message
%
%
\begin {equation}
{\mathbf w}_n=-i(u_{0,n}\cos 2\theta +u_{1,n}\sin 2\theta , u_{0,n}\sin2\theta -u_{1,n}\cos 2\theta ),
\end{equation}
where $\theta$ denotes the angle of the optical axis with respect to the laboratory frame.
Hence, in order to change S polarization into P polarization, or vice versa, a HWP is used with its optical axis oriented at $\pi /4$.
This changes the phase of the photon by $-\pi /2$.
}
\item{{\sl Electro-optic modulator} (EOM){\sl :}
An EOM rotates the polarization of the photon by an angle depending on
the voltage applied to the modulator.
In the laboratory experiment, the EOM is operated such that when a
voltage is applied it acts as a HWP that rotates
the input polarizations by $\pi/4$.
We use a pseudo-random number to
mimic the experimental procedure to control the EOM, but any
other (systematic) sequence to control the EOM can
be used as well.
}
\item{{\sl Wollaston prism} (WP){\sl :}
The WP is a PBS with one input channel and two
output channels and is simulated as the PBS described earlier.
}
\item{{\sl Detector:}
Detector $D_0$($D_1$) counts the output events at channel 0 (1) of the Wollaston prism.
The detectors are ideal particle counters,
meaning that they produce a click for each incoming
particle. Hence, we assume that the detectors have
100\% detection efficiency.
Note that in this experimental configuration adaptive
threshold detectors (see Sect.~5.1.1) can be used equally
well because their detection efficiency is 100\%.~\cite{MICH11a}
}
\item{{\sl Simulation procedure:}
When a messenger is created we wait until its message has been processed by one of the detectors before creating the next messenger
(Einstein's criterion of local causality).
During a simulation run of $N$ events the data set $\Gamma(\phi (x))=\{w_n,d_n,r_n|n = 1,\ldots, N; \phi (x)=\phi_0-\phi_1(x)\}$
is generated, where $w_n=0,1$ indicates which detector fired ($D_0$ or $D_1$), $d_n=0,1$
indicates through which arm of the MZI the messenger (photon)
came that generated the detection event (note that $d_n$ is only
measured in the simulation, not in the experiment),
and $r_n$ is a pseudo-random number
that is chosen after the $n$th message has passed the first PBS,
determining which voltage is applied to the EOM. Note that in one run of $N$ events
a choice is made between no voltage (open MZI configuration) or a particular voltage (closed MZI configuration)
corresponding to a certain reflectivity ${\cal R}$ of the output BS
(see Eq.~(2) in Ref.~\citen{JACQ08}).
These choices are made such that on average the MZI configuration is as many times open as it is closed.
The angle $\phi (x)$ denotes the phase shift
between the two interferometer arms. This phase shift is varied by
applying a plane rotation on the phase of the particles entering
channel 0 of the second PBS. This corresponds to tilting the second
PBS in the laboratory experiment.~\cite{JACQ08} For each $\phi (x)$ and MZI
configuration the number of 0 (1) output events $N_0$ ($N_1$) is
calculated.
}
\end{itemize}
\subsubsection{Simulation results}
\begin{figure}[pt]
\begin{center}
\includegraphics[width=6cm]{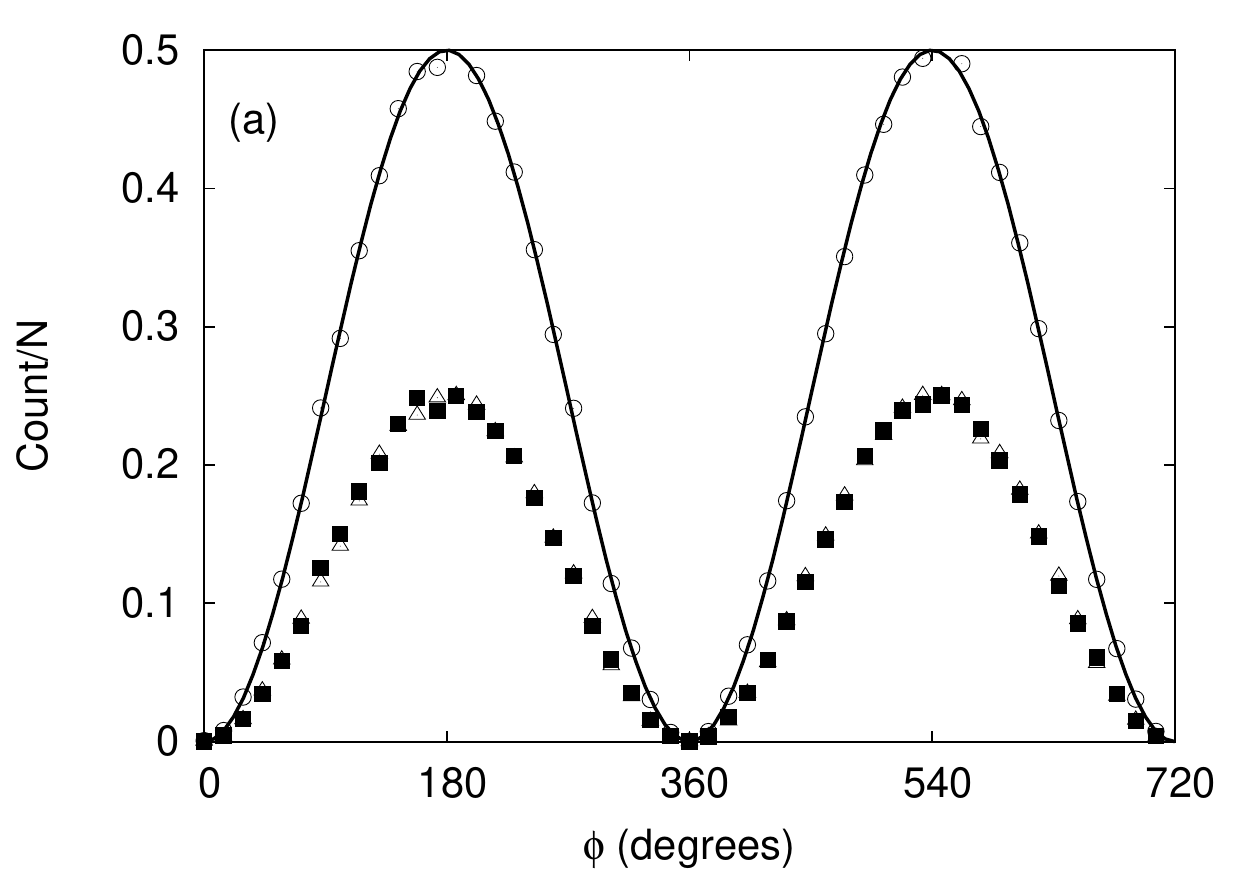}
\includegraphics[width=6cm]{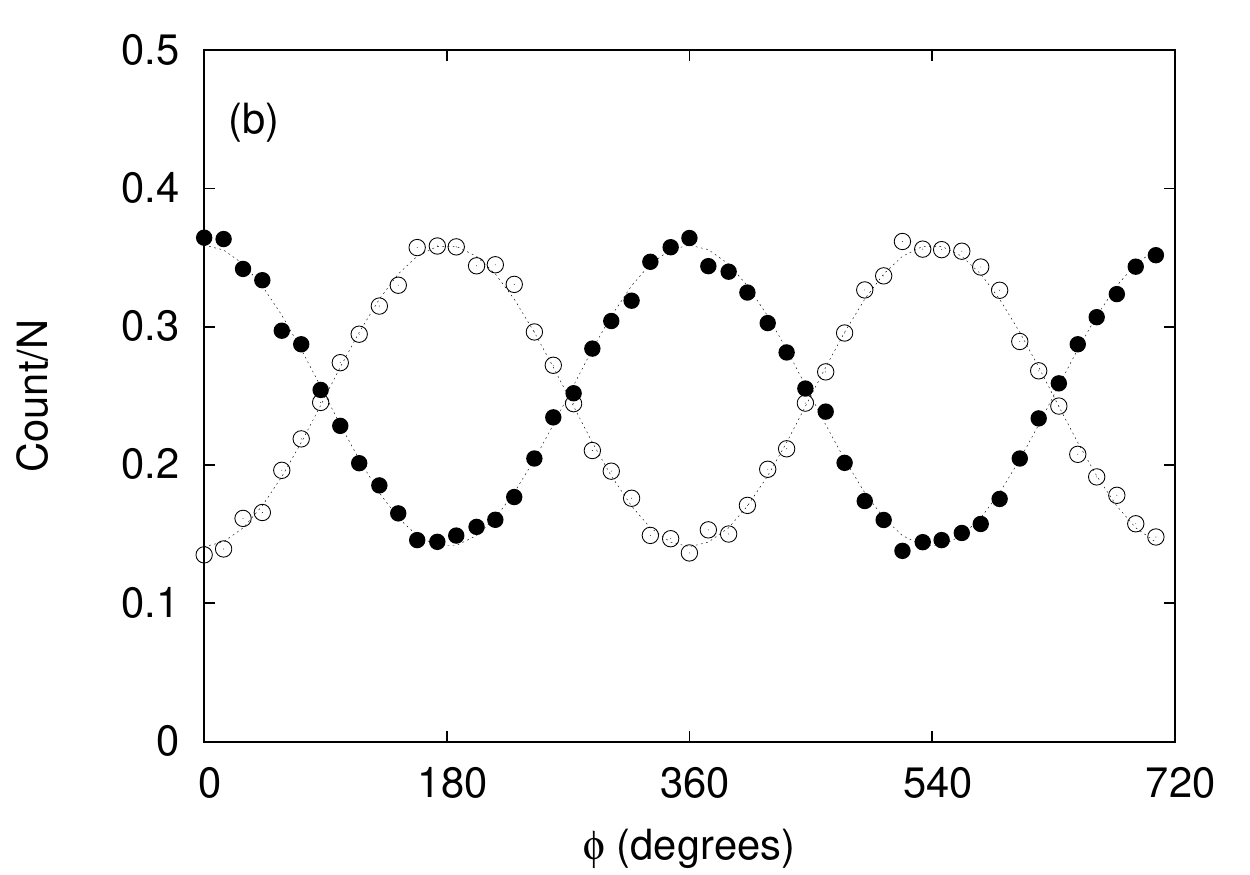}
\includegraphics[width=6cm]{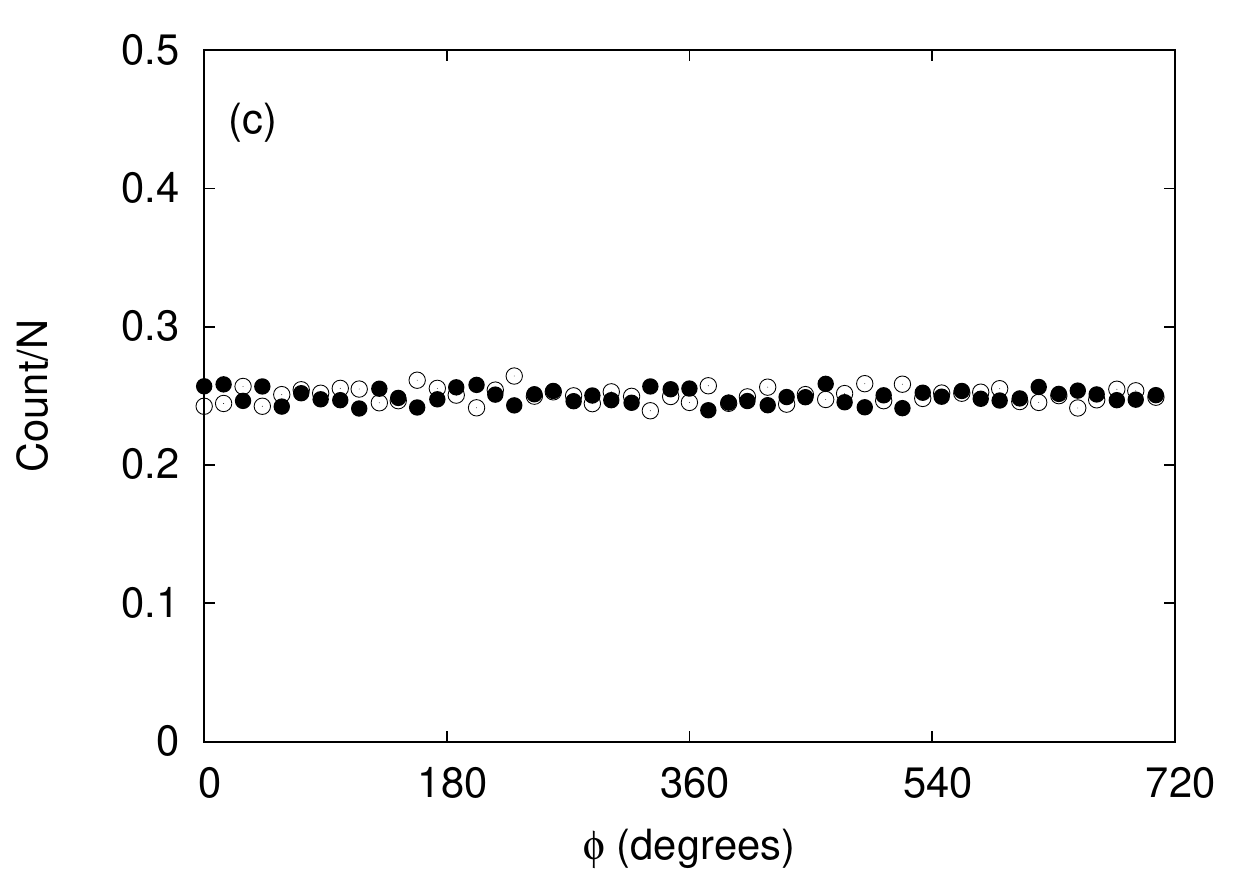}
\includegraphics[width=6cm]{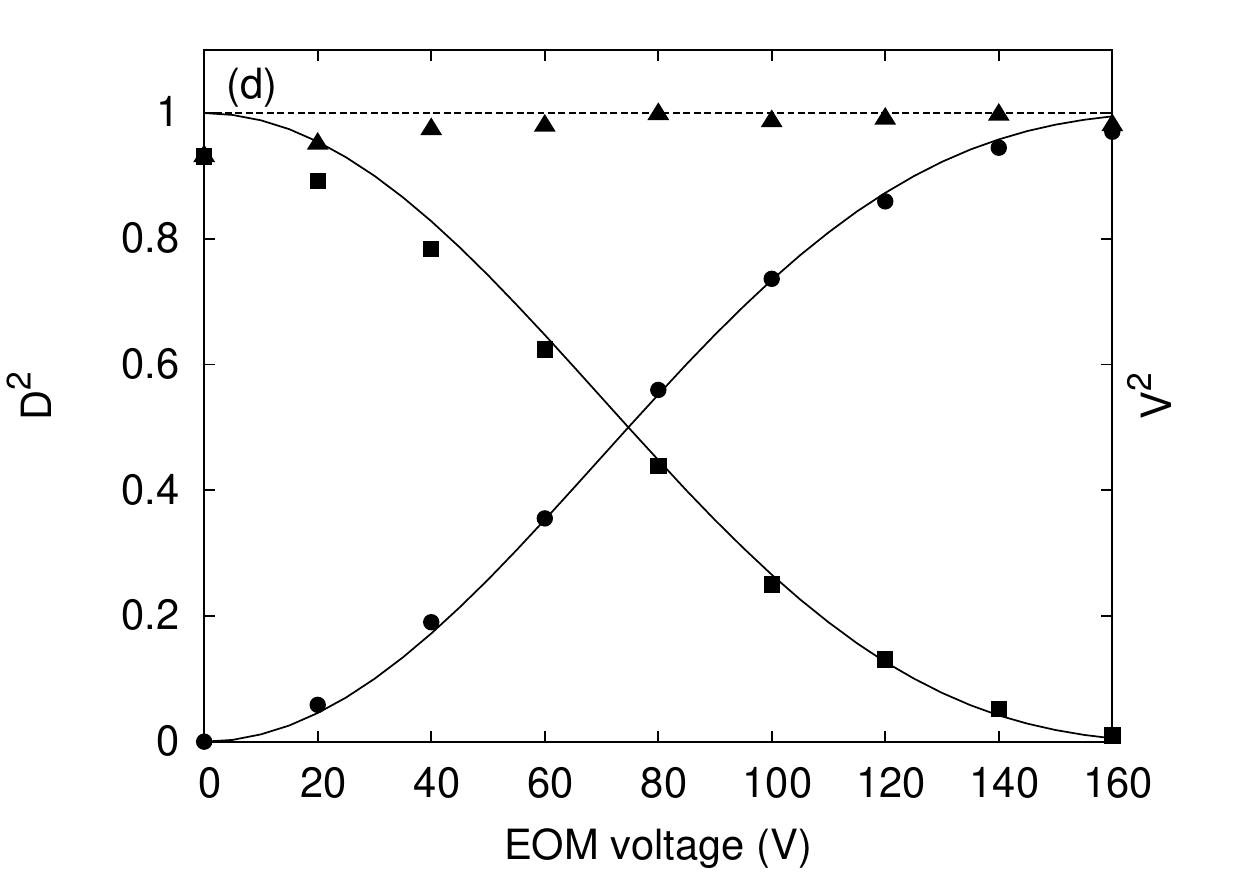}
\vspace*{8pt}
\caption{%
Event-by-event simulation results of the normalized detector counts for different values of
${\cal R}$ ((a)-(c)) and of $V^2$, $D^2$ and $V^2+D^2$ as a function of the EOM
voltage (d).
(a) Markers give the results for the normalized intensity $N_0/N$
as a function of the phase shift $\phi$, $N_0$ denoting the number of events registered at detector $D_0$.
Squares (triangles, hardly visible because they overlap with the squares) represent the detection events generated by photons which followed path 0 (1).
Open circles represent the total number of detection events.
(b)-(c) Open (closed) circles give the results for the normalized intensities $N_0/N$ ($N_1/N$) as a function of the
phase shift $\phi$, $N_0$ ($N_1$) denoting the number of events registered
at detector $D_0$ ($D_1$), for (b) ${\cal R}=0.05$ ($V\approx 0.45$) and (c) ${\cal R} =0$ ($V=0$). For each value of $\phi$,
the number of input events $N=10 000$. The number of
detection events per data point is approximately the same as in experiment.
Dashed lines represent the results of quantum theory.
(d) Squares, circles and triangles present the simulation results for $V^2$, $D^2$ and $V^2+D^2$, respectively.
Lines represent the theoretical expectations obtained from Eqs.~(2), (3) and (7) in Ref.~\citen{JACQ08} with
$\beta =24^{\circ}$ and $V_{\pi}=217$V.}\label{fig6}
\end{center}
\end{figure}
We first demonstrate that our model yields full WPI of the
photons. Fig.~\ref{fig6}(a) shows the number of detection events at $D_0$ as a
function of $\phi$ (($\phi \equiv \phi(x)$ for a given fixed position of the PBS in BS$_{\rm output}$)
for ${\cal R} = 0.5$. The events generated by photons
following paths 0 and 1 of the MZI are counted separately. It is
clear that the number of photons that followed paths 0 (squares) and 1 (triangles) is
equal and that the total intensity in output channel 0 (open circles)
shows a sinusoidal function of $\phi$. Hence, although the photons have full WPI for all $\phi$ they
can build an interference pattern by arriving one-by-one at a
detector.
Next, we calculate for ${\cal R}=0.05$ and ${\cal R}=0$ and for each
phase shift $\phi$ and configuration (open or closed) of the MZI
the number of events registered by the two detectors behind
the output BS, just like in the experiment. Figures 6(b),(c)
depict the normalized detection counts at $D_0$ (open circles) and $D_1$ (closed circles).
The simulation data
quantitatively agree with the averages calculated from
quantum theory and qualitatively agree with experiment
(see Fig.~3 in Ref.~\citen{JACQ08}). Calculation of $D$ as described in
Ref.~\citen{JACQ08} gives the results for $D^2$ and $V^2$ shown in Fig.~\ref{fig6}(d).
Comparison with Fig.~4 in Ref.~\citen{JACQ08} shows excellent qualitative
agreement.

\subsection{Single neutron interferometry}
\begin{figure}[pt]
\begin{center}
\includegraphics[width=7cm]{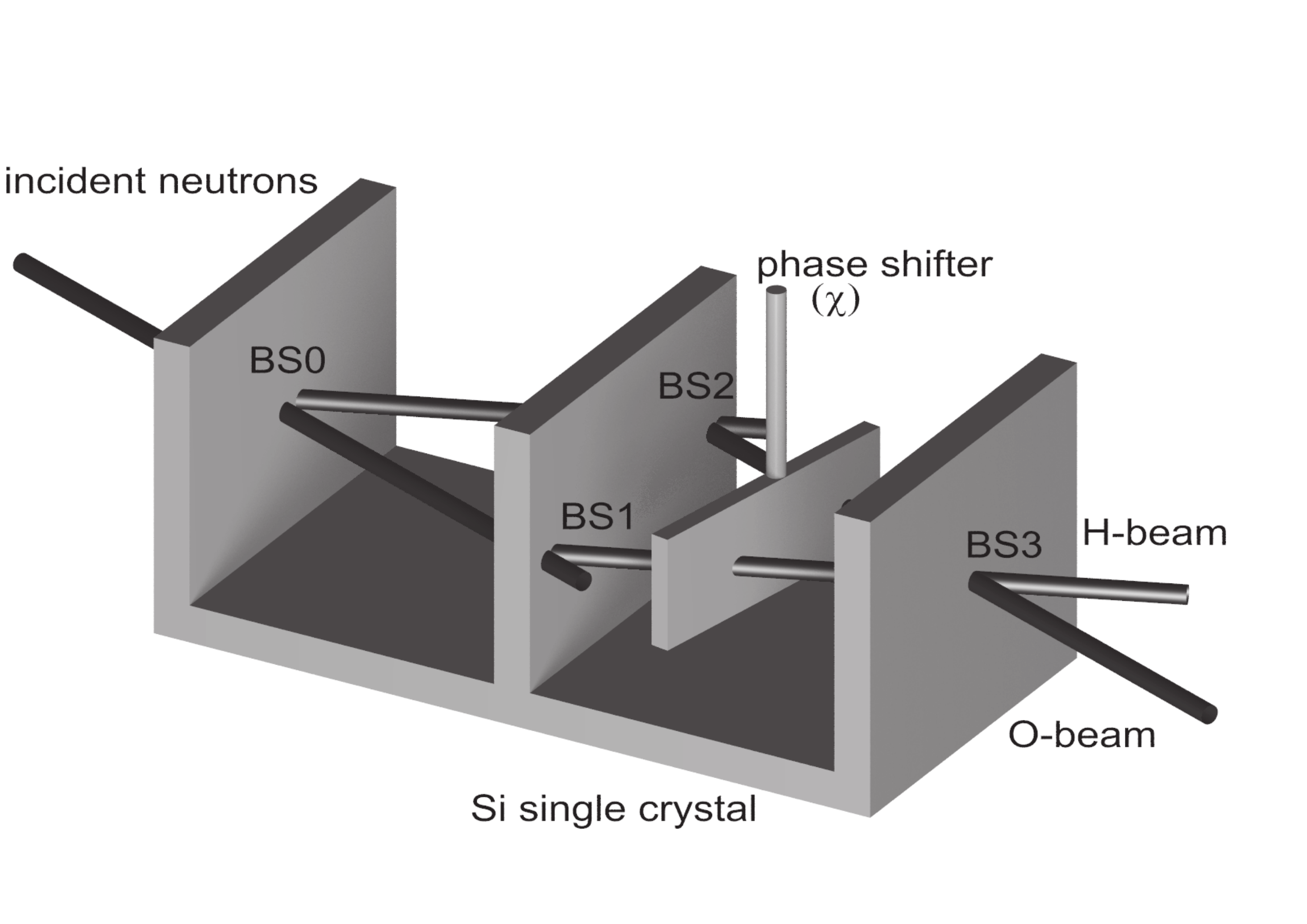}
\includegraphics[width=5.5cm]{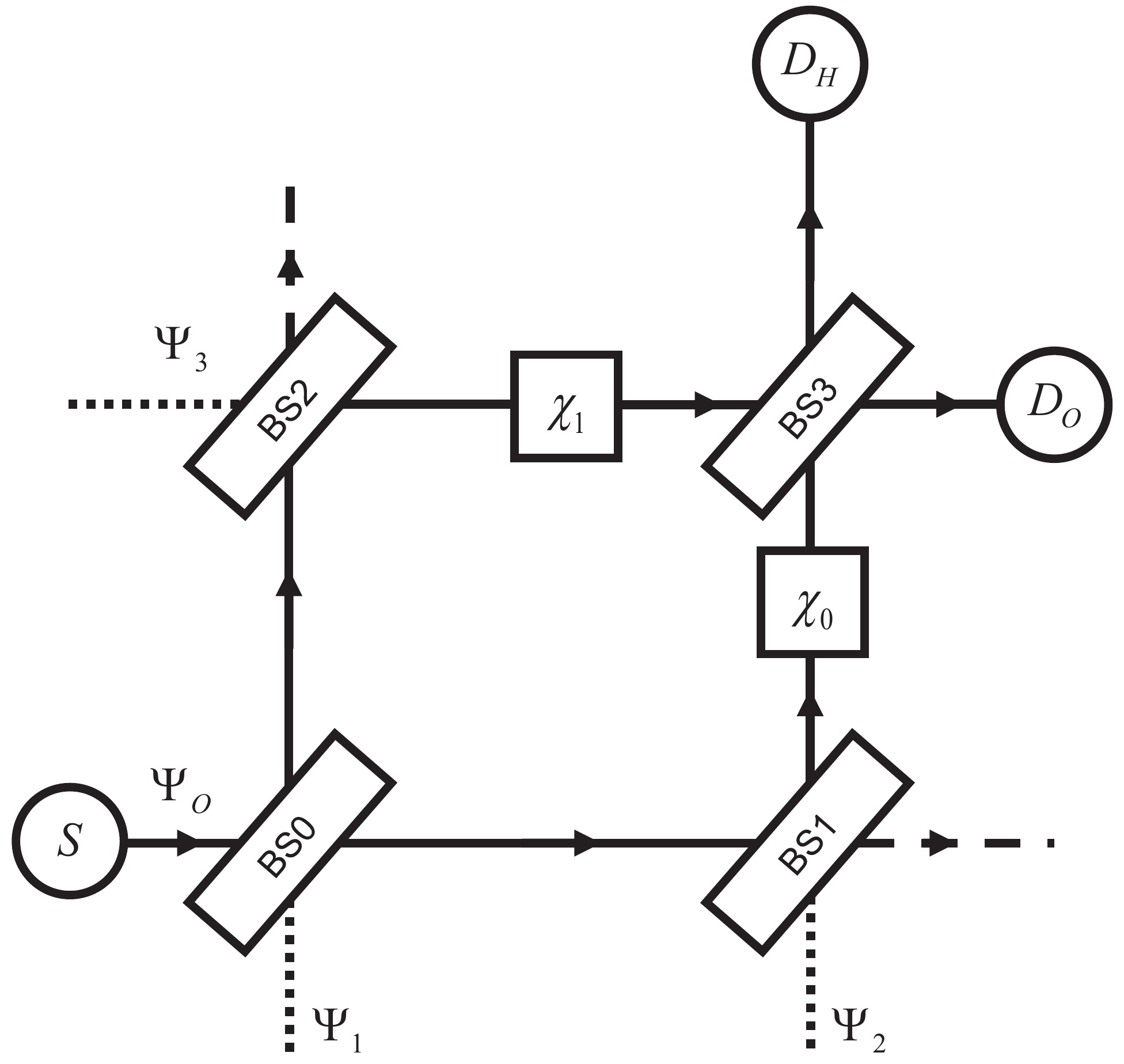}
\vspace*{8pt}
\caption{Left: Schematic picture of the silicon-perfect-crystal neutron interferometer.~\cite{RAUC74a}.
BS0, $\ldots$, BS3: beam splitters; phase shifter $\chi$: aluminum foil; neutrons that are transmitted by
BS1 or BS2 leave the interferometer and do not contribute to the interference
signal. Detectors count the number of neutrons in the O- and H-beam.
Right: Event-based network of the interferometer shown on the left.
S: single neutron source; BS0, $\ldots$ , BS3: beam splitters; $\chi_0$, $\chi_1$: phase shifters;
$D_O$, $D_H$: detectors counting all neutrons
that leave the interferometer via the O- and H-beam, respectively. In the experiment and in
the event-based simulation, neutrons enter the interferometer via the path
labeled by $\Psi_0$ only. The wave amplitudes labeled by $\Psi_1$, $\Psi_2$, and $\Psi_3$ (dotted lines) are used in the quantum
theoretical treatment only (see text). Particles leaving the interferometer via
the dashed lines are not counted.
}\label{fig7}
\end{center}
\end{figure}
Now that we have demonstrated the event-based simulation approach for the event-by-event realization
of an interference pattern in various single-photon interference
experiments, we consider
in this section one of the basic experiments in neutron interferometry, namely a Mach-Zehnder type of interferometer.
In neutron optics there exist various realizations of the Mach-Zehnder type of interferometer, but we only consider
a triple Laue diffraction type silicon perfect single crystal interferometer.~\cite{RAUC00,RAUC74a,HASE11}

Figure~\ref{fig7} (left) shows the experimental configuration. The three crystal plates, named the splitter,
mirror and analyzer plate, are assumed to be identical, which means that they have the same transmission and
reflection properties.~\cite{RAUC00}. The three crystal plates have to be parallel to high
accuracy~\cite{RAUC74a} and the whole device needs to be protected from vibrations in order to observe
interference.~\cite{KROU00}
A monoenergetic neutron beam is split by the splitter plate (BS0).
Neutrons refracted by beam splitters BS1 and BS2 (mirror plate) are directed to the analyzer plate (BS3),
also acting as a BS, thereby first passing through a rotatable-plate
phase shifter (e.~g. aluminum foil~\cite{RAUC00}).
Absorption of neutrons by the aluminum foil is assumed to be negligible.~\cite{RAUC00}
Minute rotations of the foil about an axis perpendicular to the base plane of the interferometer
induce large variations in the phase difference $\chi=\chi_0-\chi_1$.~\cite{RAUC00,LEMM10}
Finally, the neutrons are detected by one of the two detectors placed in the
so-called H-beam or O-beam.
In contrast to single-photon detectors, neutron detectors can have a very high, almost 100\%,
efficiency.~\cite{RAUC00}
Neutrons which are not refracted by BS1 and BS2 leave the interferometer and are not counted.
The intensities
in the O- and H-beam, obtained by counting individual neutrons
for a certain amount of time, exhibit sinusoidal variations as
a function of the phase shift $\chi$, a characteristic of interference.~\cite{RAUC00}

The experiment could be interpreted in different ways.
In the quantum-corpuscular view a wave packet is associated with each individual neutron.
At BS0 the wave packet splits in two components, one directed towards BS1 and one towards BS2.
At BS1 and BS2 these two components each split
in two. Two of the in total four components leave the interferometer
and the other two components are redirected towards each other at BS3 where they recombine. At BS3 the recombined wave
packet splits again in two components. Only one of these two components triggers
a detector. It is a mystery how four components of a wave packet can
conspire to do such things. Assuming that only a neutron, not
merely a part of it can trigger the nuclear reaction that causes
the detector to ``click'', on elementary logical grounds, the argument
that was just given rules out a wave-packet picture for
the individual neutron (invoking the wave function collapse only
adds to the mystery).
In the statistical interpretation of quantum mechanics there is no such conflict of interpretation.~\cite{BALL03,NIEU13}
As long as we consider
descriptions of the statistics of the experiment with many
neutrons, we may think of one single ``probability'' wave
propagating through the interferometer and as the statistical interpretation
of quantum theory is silent about single events, there
is no conflict with logic either.~\cite{RAUC00}

In what follows we demonstrate that as in the case of the single-photon interference experiments,
it is possible to construct a logically consistent,
cause-and-effect description in terms of discrete-event, particle like
processes which produce results that agree with those
of neutron interferometry experiments (individual detection events {\bf and} an interference pattern
after many single detection events have been collected) and the quantum theory
thereof (interference pattern only).

\subsubsection{Event-based model}
We construct a model for the messengers representing the neutrons and
for the processing units representing the various components in the
experimental setup (see Fig.~\ref{fig7} (right)).
\begin{itemize}[leftmargin=*]
\renewcommand\labelitemi{-}
\item{{\sl Source and particles:}
In analogy to the event-based model of a polarized photon (see Sect.~5.3.1), a neutron is regarded as a messenger carrying a
message represented by the two-dimensional unit vector
\begin{equation}
{\mathbf u}=(e^{i\psi^{(1)}}\cos (\theta/2), e^{i\psi^{(2)}}\sin (\theta/2)),
\label{neutron}
\end{equation}
where
$\psi^{(i)} =\nu t +\delta_i$, for $i=1,2$.
Here, $t$ specifies the time of flight of the neutron
and $\nu$ is an angular frequency which is characteristic for a neutron
that moves with a fixed velocity $v$.
A monochromatic
beam of incident neutrons is assumed to consist of neutrons that
all have the same value of $\nu$, that is: they have the same velocity.~\cite{RAUC00}
Both $\theta$ and $\delta=\delta_1-\delta_2=\psi^{(1)}-\psi^{(2)}$ determine the magnetic moment of the neutron, if
the neutron is viewed as a tiny classical magnet spinning around the direction
${\mathbf m}=(\cos\delta\sin \theta, \sin\delta\sin \theta, \cos \theta)$,
relative to a fixed frame of reference defined by a magnetic field.
The third degree of freedom in Eq.~(\ref{neutron}) is used to account
for the time of flight of the neutron.
Within the present model, the state of the neutron is fully determined
by the angles $\psi^{(1)}$, $\psi^{(2)}$ and $\theta$ and by rules (to be specified), by which these angles change
as the neutron travels through the network.

A messenger with message ${\mathbf u}$ at time $t$ and position ${\mathbf r}$
that travels with velocity $v$, along the direction ${\mathbf q}$ during a time interval
$t^{\prime} - t$, changes its message according to $\psi^{(i)}\rightarrow \psi^{(i)}+\phi$ for $i=1,2$, where
$\phi =\nu(t^{\prime}-t)$.

In the presence
of a magnetic field ${\mathbf B}=(B_x,B_y,B_z)$, the magnetic moment
rotates about the direction of ${\mathbf B}$ according to the
classical equation of motion. Hence, in a magnetic field the message ${\mathbf u}$ is changed into the message
${\mathbf w}=e^{ig\mu_NT \mathbf {\sigma}\cdot {\mathbf B}} {\mathbf u}$, where $g$ denotes the neutron $g$-factor, $\mu_N$
the nuclear magneton, $T$ the time during which the neutron experiences the magnetic field, and $\mathbf {\sigma}$ denotes the Pauli vector
(here we use the isomorphism between the algebra of Pauli matrices and rotations in three-dimensional space).

When the source creates a messenger, its message needs to be initialized.
This means that the three angles $\psi^{(1)}$, $\psi^{(2)}$ and $\theta$ need to be specified.
The specification depends on the type of source that has to be simulated.
For a fully coherent spin-polarized beam of neutrons, the three angles are different but the same for all the messengers being created.
Hence, three random numbers are used to specify $\psi^{(1)}$, $\psi^{(2)}$ and $\theta$ for all messengers.
}
\item{{\sl Beam splitters} BS0, $\ldots $ , BS3{\sl :}
Exploiting the similarity between the magnetic
moment of the neutron and the polarization of a photon,
we use a similar model for the BS as the
one used in Sect.~5.3.1 for polarized photons. The only difference
is that we assume that neutrons with spin up and
spin down have the same reflection and transmission
properties, while photons with horizontal and vertical
polarization have different reflection and transmission
properties.~\cite{BORN64}
Hence, what needs to be changed with respect to Sect.~5.3.1 are the complex numbers $h_{0,n},\ldots, h_{3,n}$.
For the neutrons we have
\begin{eqnarray}
\left(
\begin{array}{c}
h_{0,n}\\
h_{1,n}\\
h_{2,n}\\
h_{3,n}
\end{array}
\right)
&=&
\left(
\begin{array}{cccc}
\sqrt{{\cal T}}&i\sqrt{{\cal R}}&0&0\\
i\sqrt{{\cal R}}&\sqrt{{\cal T}}&0&0\\
0&0&\sqrt{{\cal T}}&i\sqrt{{\cal R}}\\
0&0&i\sqrt{{\cal R}}&\sqrt{{\cal T}}
\end{array}
\right)\nonumber \\
&&\times
\left(
\begin{array}{cccc}
\sqrt{v_{0,n}}&0&0&0\\
0&\sqrt{v_{1,n}}&0&0\\
0&0&\sqrt{v_{0,n}}&0\\
0&0&0&\sqrt{v_{1,n}}
\end{array}
\right)
\left(
\begin{array}{c}
R_{0,0,n}\\
R_{0,1,n}\\
R_{1,0,n}\\
R_{1,1,n}
\end{array}
\right)\nonumber \\
&=&
\left(
\begin{array}{c}
\sqrt{v_{0,n}}\sqrt{{\cal T}}R_{0,0,n}+i\sqrt{v_{1,n}}\sqrt{{\cal R}}R_{0,1,n}\\
i\sqrt{v_{0,n}}\sqrt{{\cal R}}R_{0,0,n}+\sqrt{v_{1,n}}\sqrt{{\cal T}}R_{0,1,n}\\
\sqrt{v_{0,n}}\sqrt{{\cal T}}R_{1,0,n}+i\sqrt{v_{1,n}}\sqrt{{\cal R}}R_{1,1,n}\\
i\sqrt{v_{0,n}}\sqrt{{\cal R}}R_{1,0,n}+\sqrt{v_{0,n}}\sqrt{{\cal T}}R_{1,1,n}
\end{array}
\right)
,
\label{PBS1neutron}
\end{eqnarray}
where the reflectivity $\cal{R}$ and transmissivity ${\cal T}=1-\cal{R}$ are real numbers
which are considered to be parameters to be determined from experiment.
}
\item{{\sl Phase shifter $\chi_0$, $\chi_1$:}
In the event-based model, a phase shifter is simulated without DLM.
The device has only one input and one output port and
transforms the $n$th input message ${\mathbf u}_n$ into an output message
\begin {equation}
{\mathbf w}_n=e^{i\chi_j}{\mathbf u}_n\quad j=0,1.
\end{equation}
}
\item{{\sl Detector:}
Detectors count all incoming particles.
Hence, we assume that the neutron detectors have a detection efficiency of 100\%.
This is an idealization of real neutron detectors which can have a detection efficiency of 99\% and more.~\cite{KROU00}
}
\end{itemize}

\subsubsection{Simulation results}
\begin{figure}[pt]
\begin{center}
\includegraphics[width=6cm]{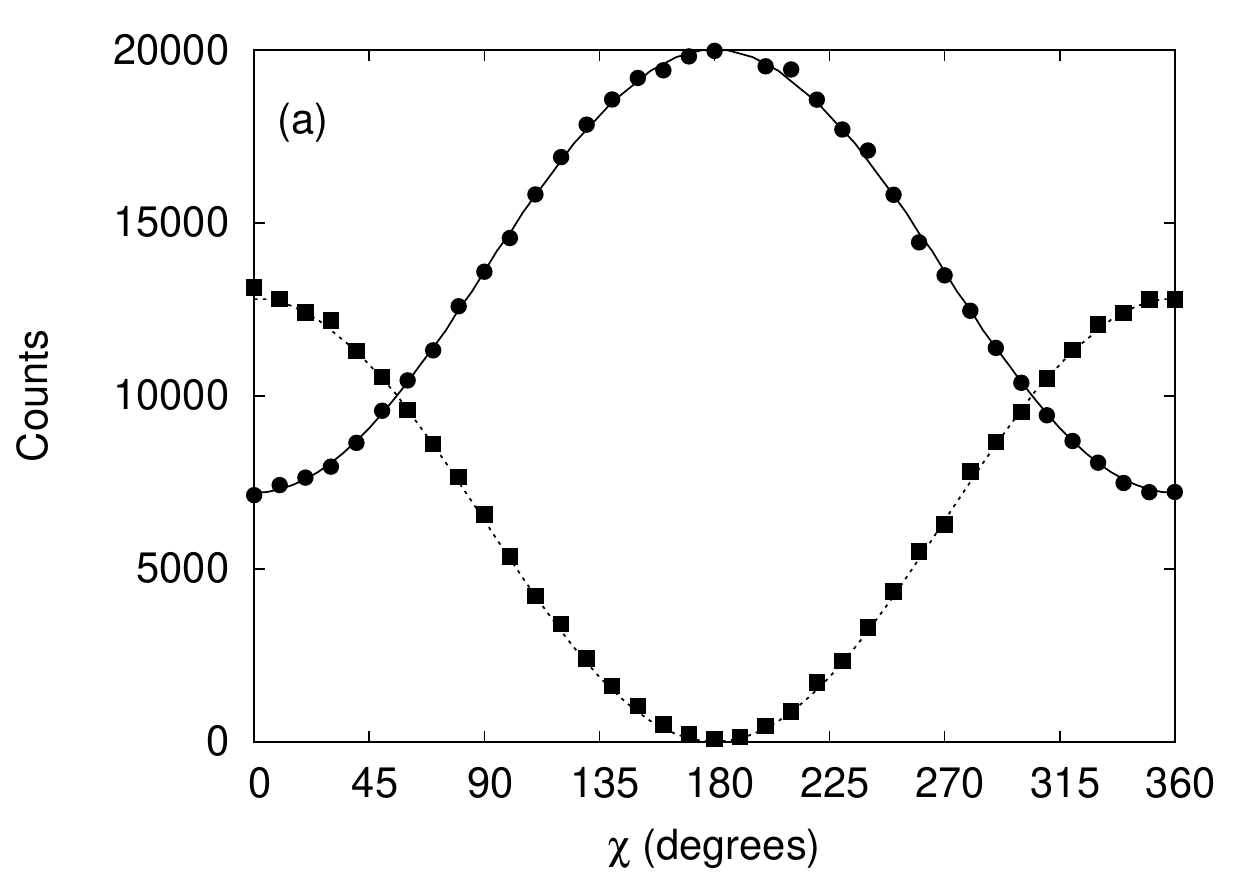}
\includegraphics[width=6cm]{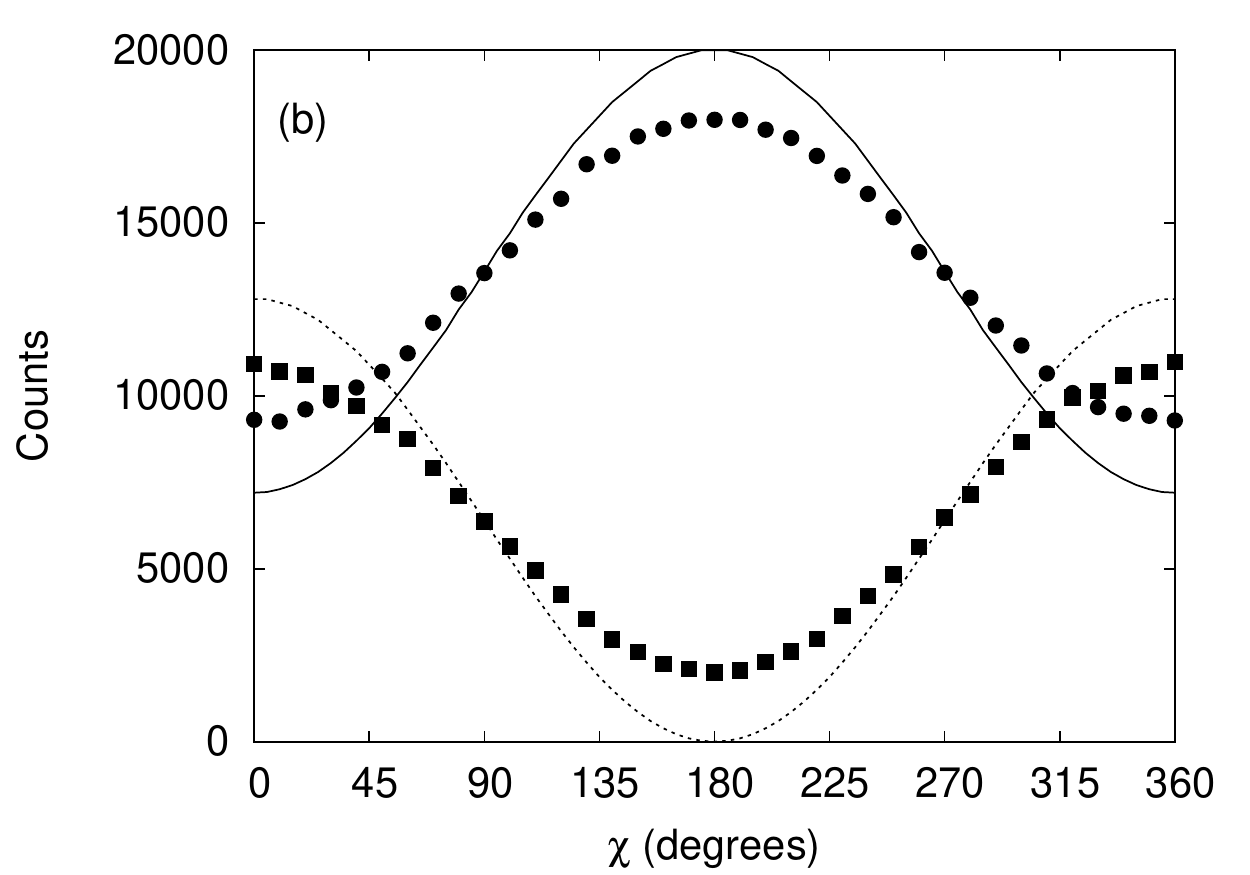}
\includegraphics[width=6cm]{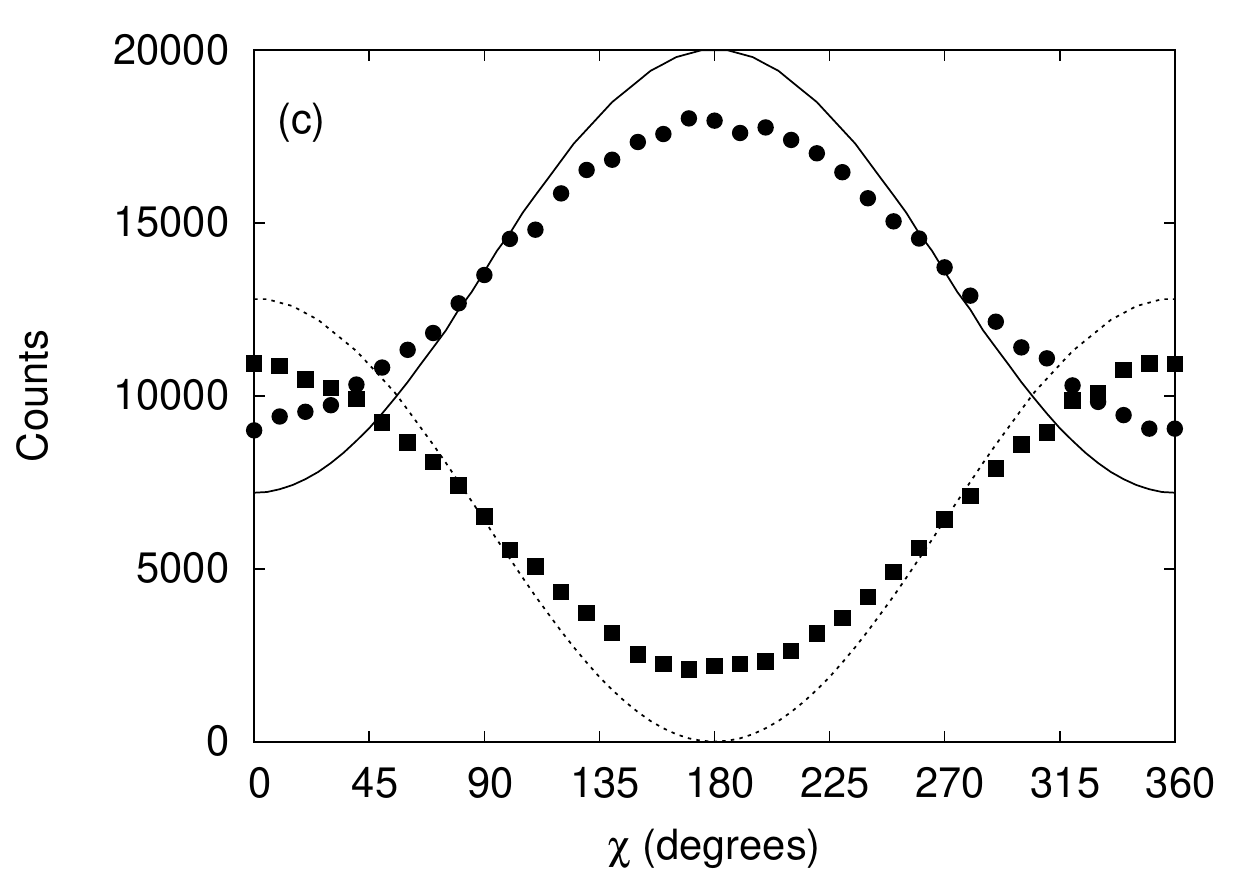}
\includegraphics[width=6cm]{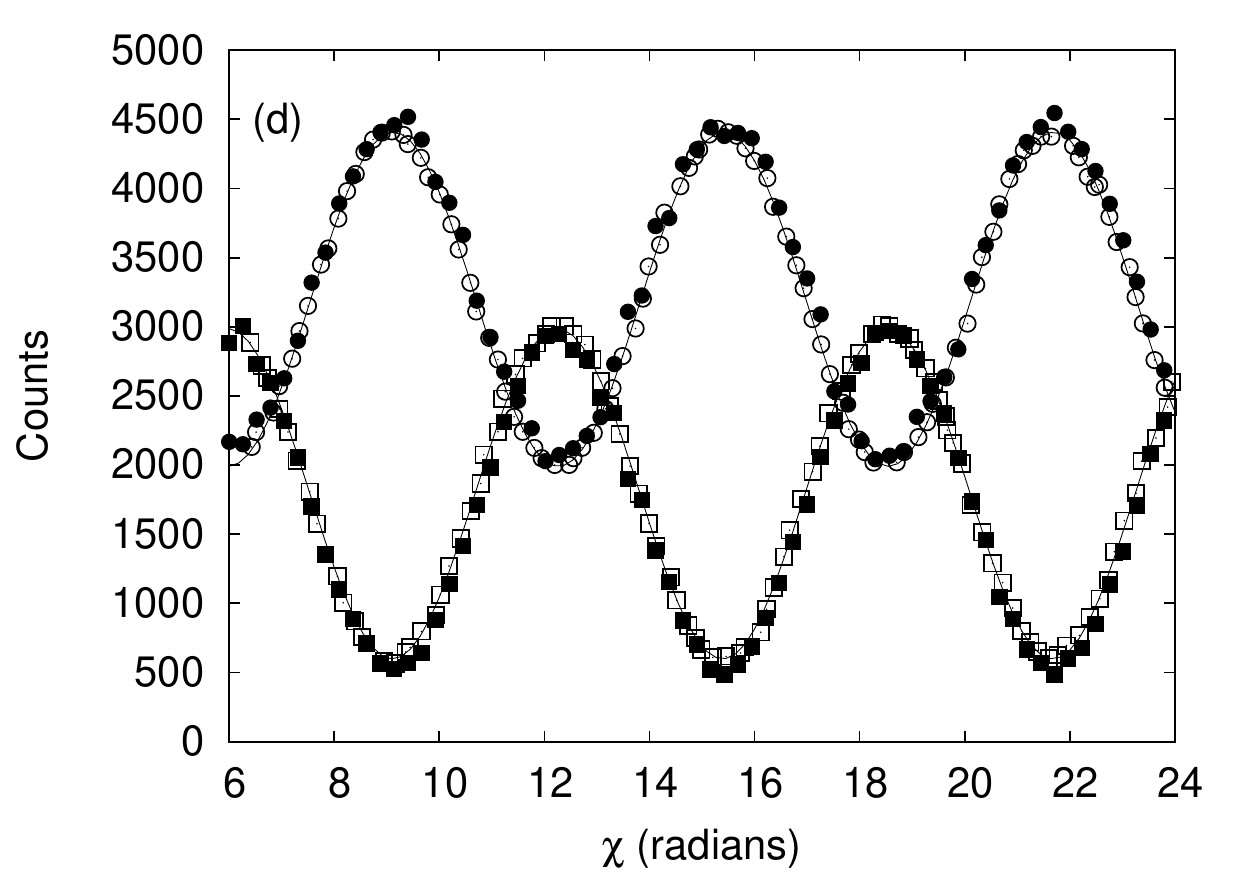}
\vspace*{8pt}
\caption{%
(a)-(c) Event-by-event simulation results of the number of neutrons leaving the interferometer via the H-beam (circles)
and O-beam (squares) as a function of the phase difference $\chi$ between the two paths inside the interferometer. For each value of
$\chi$, the number of particles generated in the simulation is
$N = 100000$.
The lines are the predictions of quantum theory. Solid line: $p_{\mathrm H}$, see Eq.~(\ref{app2z});
dotted line: $p_{\mathrm O}$, see Eq.~(\ref{app2}). (a) Model parameters:
${\cal R} = 0.2$, $\gamma = 0.99$, $\delta_1=\delta_2=0$. (b) Same as (a) except that $\gamma= 0.5$,
reducing the accuracy and increasing the response time of the DLM. (c) Same as (a) except
that to mimic the partial coherence of the incident neutron beam, the initial message carried by each particle has been modified by
choosing $\delta_1$ and $\delta_2$ uniformly random from the interval $[-\pi /3,\pi /3]$, reducing the amplitude of the interference.
(d) Comparison between the counts of neutrons per second and per square cm in the beams of a neutron interferometry experiment~\cite{KROU00}
(open symbols) and the number of neutrons per sample leaving the interferometer in an event-by-event simulation
(solid symbols). Circles: counts in the H-beam; squares: counts in the O-beam.
The experimental data has been extracted from
Figure 2 of Ref.~\citen{KROU00}. The simulation parameters ${\cal R} = 0.22$ and $\gamma= 0.5$ have
been adjusted by hand to obtain a good fit and the number of incident particles
in the simulation is $N = 22727$ per angle $\chi$. Lines through the data points are guides
to the eye.
}\label{fig8}
\end{center}
\end{figure}

In Fig.~\ref{fig8} we present a few simulation results for the neutron MZI and compare them to the quantum theoretical result
((a)-(c)) and to experiment (d).
A quantum theoretical treatment of the neutron MZI depicted in Fig.~\ref{fig7} is given in Ref.~\citen{RAUC74b}.
The quantum statistics of the neutron interferometry experiment is described in terms of the state vector
\begin{equation}
|\Psi\rangle=
\left( \Psi_{0\uparrow}, \Psi_{0\downarrow}, \Psi_{1\uparrow}, \Psi_{1\downarrow}
       \Psi_{2\uparrow}, \Psi_{2\downarrow}, \Psi_{3\uparrow}, \Psi_{3\downarrow}
\right)^T
,
\label{app0}
\end{equation}
where the components of this vector represent the complex-valued amplitudes of the wave function.
The first subscript labels the pathway and
the second subscript denotes the direction of the magnetic moment relative
to some ${\mathbf B}$-field.
The latter is not relevant for the neutron MZI experiment since the outcome of this experiment
does not depend on the magnetic moment of the neutron.
As usual, the state vector is assumed to be normalized, meaning that $\langle\Psi|\Psi\rangle=1$.
In the abstract representation of the experiment (see Fig.~\ref{fig7}(right))
we use the notation $\Psi_j=(\Psi_{j\uparrow}, \Psi_{j\downarrow})$ for $j=0,\ldots,3$.

As the state vector propagates through the interferometer, it changes according to
\begin{eqnarray}
|\Psi'\rangle
&=&
\left(\begin{array}{cc}
        \phantom{-}t^\ast &r\\
        -r^\ast & t
\end{array}\right)_{5,7}
\left(\begin{array}{cc}
        \phantom{-}t^\ast &r \\
        -r^\ast & t
\end{array}\right)_{4,6}
\left(\begin{array}{cc}
         e^{i\phi_1}&0 \\
         0 & e^{i\phi_1}
\end{array}\right)_{6,7}
\left(\begin{array}{cc}
         e^{i\phi_0}&0 \\
         0 & e^{i\phi_0}
\end{array}\right)_{4,5}
\nonumber \\ &&\times
\left(\begin{array}{cc}
        \phantom{-}t^\ast &r\\
        -r^\ast & t
\end{array}\right)_{3,7}
\left(\begin{array}{cc}
        \phantom{-}t^\ast &r \\
        -r^\ast & t
\end{array}\right)_{2,6}
\left(\begin{array}{cc}
        t & -r^\ast \\
        r & \phantom{-}t^\ast
\end{array}\right)_{1,5}
\left(\begin{array}{cc}
        t & -r^\ast \\
        r & \phantom{-}t^\ast
\end{array}\right)_{0,4}
\nonumber \\ &&\times
\left(\begin{array}{cc}
        t & -r^\ast \\
        r & \phantom{-}t^\ast
\end{array}\right)_{1,3}
\left(\begin{array}{cc}
        t & -r^\ast \\
        r & \phantom{-}t^\ast
\end{array}\right)_{0,2}
|\Psi\rangle
,
\label{app1}
\end{eqnarray}
where $t$ and $r$ denote the common transmission and reflection coefficients, respectively,
and the subscripts $i,j$ refer to the pair of elements of the
eight-dimensional vector on which the matrix acts.
Conservation of probability demands that $|t|^2 + |r|^2 =1$.

In neutron interferometry experiments, particles enter the interferometer via
the path corresponding to the amplitude $\Psi_0$ only (see Fig.~\ref{fig7} (right)),
meaning that $|\Psi\rangle=(1,0,0,0,0,0,0,0)^T$.
The probabilities to observe a particle leaving the interferometer
in the H- and O-beam are then given by
\begin{eqnarray}
p_\mathrm{H}=|\Psi'_{2}|^2&=&{\cal R}\left( {\cal T}^2+{\cal R}^2 -2 \cal{RT}\cos\chi \right)
,
\label{app2z}
\\ 
p_\mathrm{O}=|\Psi'_{3}|^2&=&2{\cal R}^2{\cal T}\left( 1+ \cos\chi \right)
,
\label{app2}
\end{eqnarray}
where $\chi=\chi_0-\chi_1$ is the relative phase shift, ${\cal R}=|r|^2$ and ${\cal T}=|t|^2=1-{\cal R}$.
Note that $p_\mathrm{H}$ and $p_\mathrm{O}$ do not depend on the imaginary part of $t$ or $r$,
leaving only one free model parameter (for instance ${\cal R}$).
In the case of a 50-50 beam splitter (${\cal T}={\cal R}=0.5$),
Eqs.~(\ref{app2z}) and (\ref{app2}) reduce to the familiar expressions
$p_\mathrm{H}=(1/2)\sin^2\chi/2$ and $p_\mathrm{O}=(1/2)\cos^2\chi/2$, respectively.
The extra factor two is due to the fact
that one half of all incoming neutrons, that is the neutrons that are
transmitted by BS1 or BS2 (see Fig.~\ref{fig7}), leave the interferometer without being counted.

The simulation results presented in Fig.~\ref{fig8}(a) demonstrate that the event-by-event simulation
reproduces the results of quantum theory if $\gamma$ approaches one~\cite{RAED05b,RAED05d,MICH11a,RAED12b}.
Indeed, there is excellent agreement with quantum theory.
In this example, the reflectivity of the beam splitters is taken to be ${\cal R}=0.2$.
The parameter $\gamma$ which controls the learning pace of the DLM-based processor
can be used to account for imperfections of the neutron interferometer.
This is illustrated in Fig.~\ref{fig8}(b) which shows simulation results for $\gamma=0.5$.

The quantum theoretical treatment assumes a fully coherent beam of neutrons.
In the event-based approach, the case of a coherent beam may be simulated by assuming
that the degree of freedom that accounts for the time of flight of the neutron
takes the same initial value each time a message is created ($\delta_1=\delta_2=0$).
In the event-based approach, we can mimic a partially coherent beam
by simply adding some noise to the message,
that is when a message is created, $\delta_i$ for $i=1,2$ is chosen random in a specified range.
In Fig.~\ref{fig8}(c), we present simulation results for the case
that $\delta_i$ is drawn randomly and uniformly from the interval $[-\pi/3,\pi/3]$,
showing that reducing the coherence of the beam reduces the visibility,
as expected on the basis of wave theory~\cite{BORN64}.
Comparing Fig.~\ref{fig8}(b) and Fig.~\ref{fig8}(c), we conclude that
the same reduced visibility can be obtained by either
reducing $\gamma$ or by adding noise to the messages.
On the basis of this interferometry experiment alone, it is difficult to exclusively
attribute the cause of a reduced visibility to one of these mechanisms.

Conclusive evidence that the event-based model reproduces the results of a
single-neutron interferometry experiment comes from comparing simulation data with experimental data.
In Fig.~\ref{fig8}(d), we present such a comparison using
experimental data extracted from Fig.~2 of Ref.~\citen{KROU00}.
It was not necessary to try to make the best fit: the parameters ${\cal R}$ and $\gamma$ and the offset
of the phase $\chi$ were varied by hand.
As shown in Fig.~\ref{fig8}(d), the event-based simulation model reproduces, quantitatively, the experimental results
reported in Fig.~2 of Ref.~\citen{KROU00}.

\section{Entanglement}
In quantum theory entanglement is the property of a state of a two or many-body quantum system in
which the states of the constituting bodies are correlated.
The most prominent example is the singlet state of two spin-$\frac{1}{2}$ particles
\begin{equation}
|\Psi\rangle = \frac {1}{\sqrt{2}}\left(\left|\uparrow\downarrow\rangle\right. - \left|\downarrow\uparrow\rangle\right.\right),
\label{singlet}
\end{equation}
which cannot be written as a product state.
According to quantum theory, if the singlet state describes the correlation between the spins of the two particles
and if we perform a measurement of both spins along the same direction,
we observe that the particles have opposite but otherwise random values of their spins.
Thus, in the quantum theoretical description,
the state of the two spin-$\frac{1}{2}$ particles may be correlated even though the particles are spatially and
temporally separated and do not necessarily interact.
Note however that this is a statistical interpretation which does not support the assumption that this singlet state is a
property of each pair of particles and does not support the idea that changing the state of one particle has a causal effect on the state
of the other.

\subsection{Einstein-Podolsky-Rosen-Bohm thought experiment}
\begin{figure}[pt]
\begin{center}
\includegraphics[width=9cm]{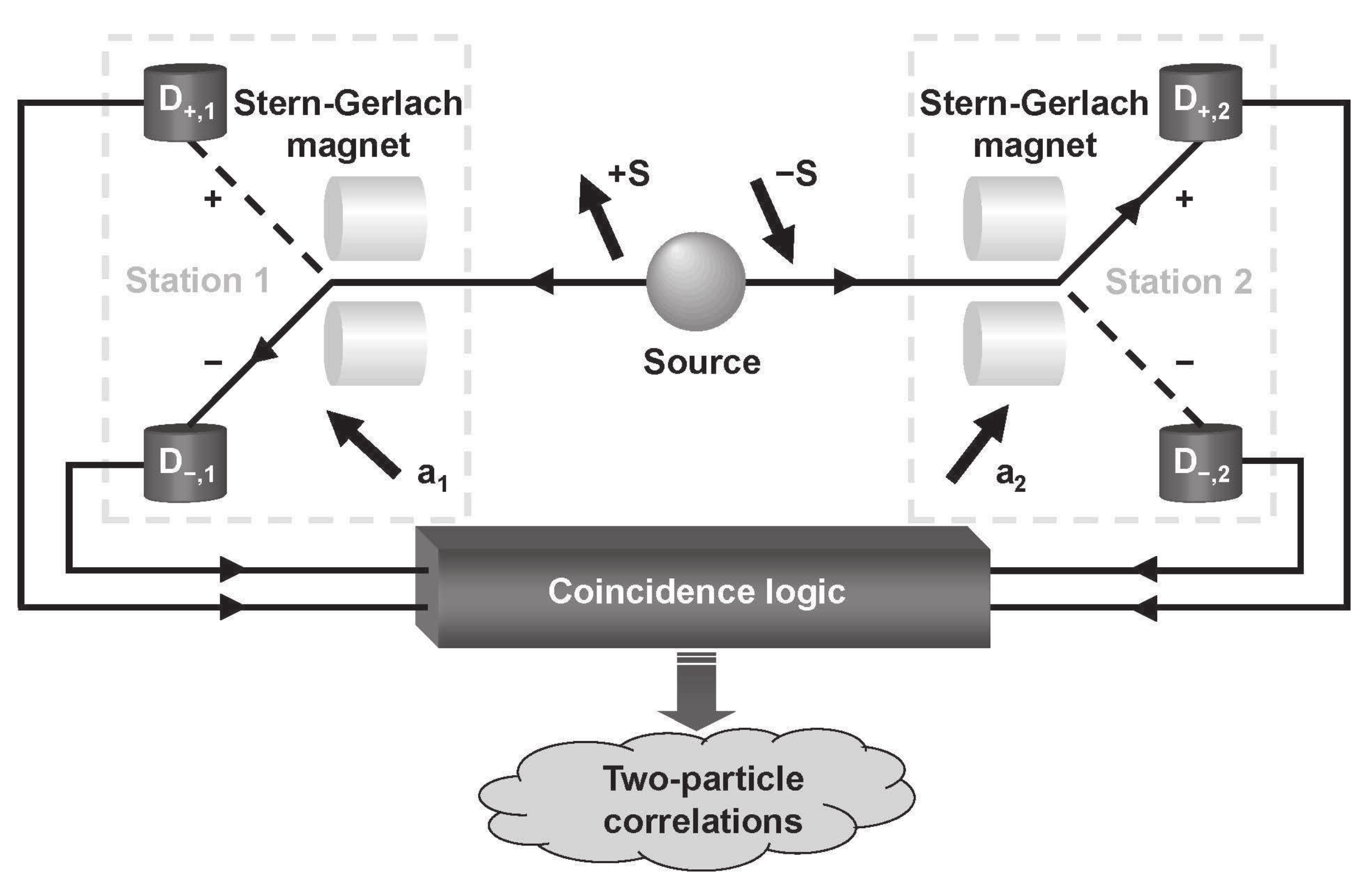}
\vspace*{8pt}
\caption{%
Schematic diagram of the Einstein-Podolsky-Rosen-Bohm (EPRB) experiment with magnetic particles~\cite{BOHM51}.
The source emits charge-neutral pairs of particles with opposite magnetic moments $+{\mathbf S}$ and $-{\mathbf S}$.
One of the particles moves to station 1 and the other one to station 2.
As the particle arrives at station $i=1,2$, it passes through a Stern-Gerlach magnet which
deflects the particle, depending on the orientation of the magnet ${\mathbf a}_i$ and the magnetic moment of the particle.
As the particle leaves the Stern-Gerlach magnet,
it generates a signal in one of the two detectors ${\mathrm D}_{\pm ,i}$.
Coincidence logic pairs the detection events of station 1 and station 2 so that they can be used to compute two-particle correlations.
}\label{fig9}
\end{center}
\end{figure}
In 1935, Einstein, Podolsky and Rosen (EPR) designed a thought experiment demonstrating the ``incompleteness'' of quantum theory.~\cite{EPR35}
The thought experiment involves the measurement of the position and momentum of two particles which interacted in the past but not at the
time of measurement. Since this experiment is not suited for designing a laboratory experiment, Bohm proposed in 1951 a more realistic
experiment which measures the intrinsic angular momentum of a correlated pair of atoms one-by-one~\cite{BOHM51}.
A schematic diagram of the experiment is shown in Fig.~\ref{fig9}.
A source emits charge-neutral pairs of particles with opposite magnetic moments $+{\mathbf S}$ and $-{\mathbf S}$.
The two particles separate spatially and propagate in free space to an observation station in which they are detected.
As the particle arrives at station $i=1,2$, it passes through a Stern-Gerlach magnet. The magnetic moment of a particle interacts
with the inhomogeneous magnetic field of the Stern-Gerlach magnet.
The Stern-Gerlach magnet deflects the particle, depending on the orientation of the magnet ${\mathbf a}_i$ and the magnetic moment of the particle.
The Stern-Gerlach magnet divides the beam of particles in two, spatially well-separated parts. As the particle leaves the Stern-Gerlach magnet,
it generates a signal in one of the two detectors ${\mathrm D}_{\pm ,i}$. The firing of a detector corresponds to a detection event.

According to quantum theory of the Einstein-Podolsky-Rosen-Bohm (EPRB) thought experiment, the results of repeated
measurements of the system of two spin-$\frac{1}{2}$ particles in the spin state
$|\Psi\rangle =\alpha_0 \left|\uparrow\uparrow\rangle\right. +\alpha_1 \left|\downarrow\uparrow\rangle\right. +\alpha_2 \left|\uparrow\downarrow\rangle\right. +\alpha_3 \left|\downarrow\downarrow\rangle\right.$
with $\sum_{j=0}^3|\alpha_j|^2=1$ are given by the single-spin expectation values
\begin{eqnarray}
\widehat E_1({\mathbf a}_1) &=& \langle\Psi |{\mathbf \sigma}_1 \cdot {\mathbf a}_1 |\Psi\rangle = \langle\Psi |{\mathbf \sigma}_1  |\Psi\rangle \cdot {\mathbf a}_1,\nonumber \\
\widehat E_2({\mathbf a}_2) &=& \langle\Psi |{\mathbf \sigma}_2 \cdot {\mathbf a}_2 |\Psi\rangle = \langle\Psi |{\mathbf \sigma}_2  |\Psi\rangle \cdot {\mathbf a}_2,
\label{twospins}
\end{eqnarray}
and the two-particle correlations $\widehat E({\mathbf a}_1,{\mathbf a}_2)=\langle\Psi |{\mathbf \sigma}_1 \cdot {\mathbf a}_1 {\mathbf \sigma}_2 \cdot {\mathbf a}_2 |\Psi\rangle
={\mathbf a}_1\cdot\langle\Psi |{\mathbf \sigma}_1\cdot {\mathbf \sigma}_2 |\Psi\rangle \cdot {\mathbf a}_2$, where
${\mathbf a}_1$ and ${\mathbf a}_2$ are unit vectors specifying the directions of the analyzers,
${\mathbf \sigma}_i$ denote the Pauli vectors describing the spin of the particles $i=1,2$, and
$\langle X\rangle ={\mathrm Tr}\rho X$ with $\rho$ being the 4x4 density matrix describing the two spin-$\frac{1}{2}$ particle system.
We have introduced the notation $\widehat {\phantom{E}}$ to make a distinction between the quantum theoretical results and the results obtained by
analysis of data sets from a laboratory experiment and from an event-based simulation (see Sect.~6.3).
Quantum theory of the EPRB thought experiment assumes that $|\Psi\rangle$ does not depend
on ${\mathbf a}_1$ or ${\mathbf a}_2$. Therefore, from Eq.~(\ref{twospins}) it follows immediately that $\widehat E_1({\mathbf a}_1)$ does not depend
on ${\mathbf a}_2$ and that $\widehat E_2({\mathbf a}_2)$ does not depend on ${\mathbf a}_1$. Note that this holds for any state $|\Psi\rangle$.
For later use, it is expedient to introduce the function
\begin{equation}
S\equiv S({\mathbf a}_1,{\mathbf a}_2,{\mathbf a}_1^{\prime},{\mathbf a}_2^{\prime})=E({\mathbf a}_1,{\mathbf a}_2)-
E({\mathbf a}_1,{\mathbf a}_2^{\prime})+E({\mathbf a}_1^{\prime},{\mathbf a}_2)+E({\mathbf a}_1^{\prime},{\mathbf a}_2^{\prime}),
\label{Bellfunction}
\end{equation}
for which it can be shown that $|S|\le2\sqrt{2}$, independent of the choice of $\rho$.~\cite{CIRE80}

The quantum theoretical description of the EPRB experiment assumes that the state of the two spin-$\frac{1}{2}$ particles is described by the singlet state Eq.~(\ref{singlet}).
For the singlet state, $\widehat E_1({\mathbf a}_1)=\widehat E_2({\mathbf a}_2)=0$, $\widehat E({\mathbf a}_1,{\mathbf a}_2)=-{\mathbf a}_1\cdot {\mathbf a}_2$
and the maximum value of $|S|$ is $2\sqrt{2}$.
Note that the singlet state is fully characterized by the three quantities $\widehat E_1({\mathbf a}_1)$, $\widehat E_2({\mathbf a}_2)=0$, and
$\widehat E({\mathbf a}_1,{\mathbf a}_2)$.
Hence, in any laboratory experiment, thought experiment or computer simulation of such an experiment, which has the goal to measure effects of the system being represented
by a singlet state, these three quantities have to be measured and computed.

\subsection{Bell and Boole inequalities}
Quantum theory yields statistical estimates for ${\widehat E}_1$, ${\widehat E}_2$ and ${\widehat E}_{12}$ and cannot say anything about individual measurements.~\cite{HOME97}
Nevertheless, for the singlet state quantum theory predicts that, if measurement of the component $\mathbf{\sigma}_1\cdot\mathbf{a}_1$ with $\mathbf{a}_1$ being a unit vector, yields the value + 1, then measurement of
${\sigma}_2\cdot\mathbf{a}_1$ must yield the value $-1$ and vice versa. The fundamental question is how to relate the statistical results of quantum theory and the individual measurements.

\subsubsection{Bell's model and inequality}
Bell made the following assumptions in constructing his model and deriving his inequality:~\cite{BELL64}
\begin{enumerate}
\item{$A(\mathbf{a_1},\lambda )=\pm 1$ and $B(\mathbf{a_2},\lambda )=\pm 1$, where $A$ ($B$) denotes the result of measuring
$\mathbf{\sigma}_1\cdot\mathbf{a_1}$ ($\mathbf{\sigma}_2\cdot\mathbf{a}_2$) and $\lambda$ denotes a variable or a set of variables which only
depend on the preparation (source) and not on the measurement (magnet settings) of the spin components.
Note that this assumption already includes the hypothesis that the orientation of one magnet does not influence the measurement
result obtained with the other magnet (often referred to as the locality condition).
In other words, $A$ ($B$) does not depend on $\mathbf{a}_2$ ($\mathbf {a}_1$).
}
\item{If $\rho (\lambda )$ is the probability distribution of $\lambda$ ($\int\rho (\lambda ) d\lambda = 1$)
then the expectation value of the product of the two components
$\mathbf{\sigma}_1\cdot\mathbf{a}_1$ and $\mathbf{\sigma}_2\cdot\mathbf{a}_2$ can be written as
$P(\mathbf{a}_1,\mathbf{a}_2)=\int d\lambda\rho (\lambda )A(\mathbf{a}_1,\lambda )B(\mathbf{a}_2,\lambda )$.
Note that one could also introduce variables  $\lambda^{\prime}$ and $\lambda^{\prime\prime}$
depending on the characteristics of the instruments on both sides. Averaging over these instrument dependent variables would result in new
variables having values between $-1$ and $+1$.
However, this is only the case if $\lambda^{\prime}$ and $\lambda^{\prime\prime}$ are completely independent.
For example, if  $\lambda^{\prime}$ and $\lambda^{\prime\prime}$ are sets of variables including the detection times,
used for coincidence measurements in a laboratory experiment, then assumption 2 does not hold~\cite{LARS04}.
}
\item{$A(\mathbf{a}_1,\lambda )=-B(\mathbf{a}_1,\lambda )$ so that $P(\mathbf{a}_1,\mathbf{a}_2)=-\int d\lambda\rho (\lambda )A(\mathbf{a}_1,\lambda )A(\mathbf{a}_2,\lambda )$.
This assumption follows from the observation that $P(\mathbf{a}_1,\mathbf{a}_2)=\int d\lambda\rho (\lambda )A(\mathbf{a}_1,\lambda )B(\mathbf{a}_2,\lambda )$ reaches $-1$
at $\mathbf{a}_1=\mathbf{a}_2$ only if $A(\mathbf{a}_1,\lambda )=-B(\mathbf{a}_1,\lambda )$.
Note that $P(\mathbf{a}_1,\mathbf{a}_1)=-1$ if and only if $A(\mathbf{a}_1,\lambda )=-B(\mathbf{a}_1,\lambda )$, making both these assumptions equivalent.
Hence, what Bell assumed is that the results of the measurements at both sides of the source can be represented by one and the same symbol ``$A$''
that depends only on the respective magnet setting and on $\lambda$. Moreover, also the measurement outcomes of an experiment with another setting of (only one of) the magnets,
can be represented by the same symbol ``$A$''.
}
\end{enumerate}

Using the above hypotheses and considering a third unit vector $\mathbf{a}_3$ Bell derived
the inequality~\cite{BELL64}
\begin{equation}
|P(\mathbf{a}_1,\mathbf{a}_2)-P(\mathbf{a}_1,\mathbf{a}_3)|\le 1+P(\mathbf{a}_2,\mathbf{a}_3),
\label{Bell}
\end{equation}
which is violated for certain magnet settings $\mathbf{a}_1,\mathbf{a}_2,\mathbf{a}_3$ if $P(\mathbf{a}_1,\mathbf{a}_2)$ is replaced by $\widehat E(\mathbf{a}_1,\mathbf{a}_2)=-\mathbf{a}_1\cdot\mathbf{a}_2$,
the quantum theoretical two-particle expectation value describing the averaged two-particle correlations obtained in EPRB experiments.
Note that 1, 2 and 3 are sufficient conditions for the Bell inequality to be obeyed.
Hence, if the Bell inequality is obeyed then one cannot say anything about the validity of the assumptions,
but if it is violated then one can say that at least one of the
assumptions must be false, thereby refuting Bell's model.
It is worth mentioning that Bell analyzed a very restricted class of classical models,
namely models which do not account for (i) the physics of the detection process and/or (ii) the use of time-coincidences
to define particle pairs (see below).
Although the above conclusion is the only logical conclusion that can be drawn, it is common but erroneous practice to take a violation
of a Bell inequality as a ``proof'' of the quantum nature of the system under study.
Far reaching conclusions drawn from Bell's results, such as violations of Bell-like inequalities having implications for action-on-a-distance, locality, realism , $\ldots$, have all been shown
to be logical fallacies.~\cite{PENA72,FINE74,FINE82,BROD93,KUPC86,JAYN89,SICA99,HESS01a,HESS05,KRAC05,SANT05,RAED07c,NIEU09,KARL09,NIEU11,RAED11a,KARL12}

\subsubsection{Boole inequality for the correlations of two-valued variables}
We consider two-valued variables $S(x,n)=\pm 1$ where $x$ can be considered to represent the orientations
$\mathbf {a}_1,\mathbf {a}_2,\mathbf {a}_3$ of the magnets in an EPRB experiment and $n=1,\ldots N$ simply numbers the measurements in an experimental run.
From
the variables
$S(x,n)$ with $x=\mathbf{a}_1,\mathbf{a}_2,\mathbf{a}_3$ we compute the averages $F_{\mathbf {a}_1,\mathbf {a}_2}=\sum_{n=1}^N S(\mathbf {a}_1,n)S(\mathbf {a}_2,n)/N$,
$F_{\mathbf {a}_1,\mathbf {a}_3}=\sum_{n=1}^N S(\mathbf {a}_1,n)S(\mathbf {a}_3,n)/N$ and $F_{\mathbf {a}_2,\mathbf {a}_3}=\sum_{n=1}^N S(\mathbf {a}_2,n)S(\mathbf {a}_3,n)/N$.
According to Boole~\cite{BO1862} it is impossible to violate
\begin{equation}
|F_{\mathbf{a}_1,\mathbf{a}_2}\pm F_{\mathbf{a}_1,\mathbf{a}_3}|\le 1\pm F_{\mathbf{a}_2,\mathbf{a}_3},
\label{Boole}
\end{equation}
if there is a one-to-one correspondence between the two-valued variables $S(\mathbf {a}_1,n)$, $S(\mathbf {a}_2,n)$, $S(\mathbf {a}_3,n)$ of the mathematical description and each
triple $\{X(\mathbf {a}_1,n), X(\mathbf {a}_2,n), X(\mathbf {a}_3,n)\}$ of binary data collected in the experimental run denoted by $n$.
This one-to-one correspondence is a necessary and sufficient condition for the inequality to be obeyed.
Note that inequalities Eq.~(\ref{Bell}) and Eq.~(\ref{Boole}) have the same structure.
We emphasize that it is essential that the correlations $F_{\mathbf{a}_1,\mathbf{a}_2}$, $F_{\mathbf{a}_1,\mathbf{a}_3}$ and $F_{\mathbf{a}_2,\mathbf{a}_3}$ have been calculated from one data set
that contains triples instead
of from three sets in which the data has been collected in pairs.~\cite{RAED11a}

\subsubsection{An inequality within quantum theory}
From the algebraic identity $(1\pm xy)^2=(x\pm y)^2 +(1-x^2)(1-y^2)$ it follows that
$|x\pm y|\le 1\pm xy$ for real numbers $x$ and $y$ with $|x|\le 1$ and $|y|\le 1$.
From this inequality it immediately follows that
\begin{equation}
|xz\pm yz|\le 1\pm xy,
\label{ineq}
\end{equation}
for real numbers $x$, $y$, $z$ such that $|x|\le1$, $|y|\le1$ and $|z|\le1$.

If we now assume that the two spin-$\frac{1}{2}$ particle system is in a product state $|\Psi\rangle=|\Psi\rangle_1|\Psi\rangle_2$ with
$|\Psi\rangle_j=\alpha_{0,j}\left|\uparrow\rangle_j\right.+\alpha_{1,j}\left|\uparrow\rangle_j\right.$ with $|\alpha_{0,j}|^2+|\alpha_{1,j}|^2=1$ for $j=1,2$, then
\begin{eqnarray}
\widehat E_1({\mathbf a}_1) &=& \langle\Psi |{\mathbf \sigma}_1  |\Psi\rangle_1 \cdot {\mathbf a}_1,\nonumber \\
\widehat E_2({\mathbf a}_2) &=& \langle\Psi |{\mathbf \sigma}_2  |\Psi\rangle_2 \cdot {\mathbf a}_2,\nonumber \\
\widehat E({\mathbf a}_1,{\mathbf a}_2) &=& \langle\Psi |{\mathbf \sigma}_1  |\Psi\rangle_1 \cdot {\mathbf a}_1 \langle\Psi |{\mathbf \sigma}_2
|\Psi\rangle_2 \cdot {\mathbf a}_2 = \widehat E_1({\mathbf a}_1) \widehat E_2({\mathbf a}_2),
\label{QTproduct}
\end{eqnarray}
and the correlation $\widehat E({\mathbf a}_1,{\mathbf a}_2)-\widehat E_1({\mathbf a}_1)\widehat E_2({\mathbf a}_2)=0$.
Using Eq.~(\ref{ineq}) and unit vectors $\mathbf{a}_1$, $\mathbf{a}_2$, $\mathbf {a}_3$ we obtain a Bell-type inequality
\begin{equation}
|\widehat E({\mathbf a}_1,{\mathbf a}_2)-\widehat E({\mathbf a}_1,{\mathbf a}_3)|\le 1+\widehat E({\mathbf a}_2,{\mathbf a}_3),
\label{Bellproduct}
\end{equation}
and similarly the Bell-CHSH inequality~\cite{CLAU69}
\begin{equation}
|S|=|\widehat E({\mathbf a}_1,{\mathbf a}_2)-\widehat E({\mathbf a}_1,{\mathbf a}_2^{\prime})+\widehat E({\mathbf a}_1^{\prime},{\mathbf a}_2)+\widehat E({\mathbf a}_1^{\prime},{\mathbf a}_2^{\prime})|\le 2,
\label{CHSH}
\end{equation}
for unit vectors ${\mathbf a}_1$, ${\mathbf a}_1^{\prime}$, ${\mathbf a}_2$, and ${\mathbf a}_2^{\prime}$.

Hence, if the state of the two spin-$\frac{1}{2}$ particle system is a product state, then the Bell and Bell-CHSH inequality hold.
On the other hand, if the Bell or Bell-CHSH inequality is violated then the two-particle quantum system is not in a product state.
Note that these logical statements are made entirely within the framework of quantum theory.

\subsubsection{Bell inequality tests}
In a typical ideal EPRB experiment three runs are performed in which $N$ detection events are collected on both sides (referred to by 1 and 2) of the source.
The outcomes of the detection events take the values $+1$ or $-1$ and are represented by the symbol $X$.
This results in the three data sets
\begin{eqnarray}
\Gamma_{\mathbf {a},\mathbf {b}}&=&\{X(\mathbf{a},n,1),X(\mathbf{b},n,2)|n=1,\ldots ,N\},
\nonumber\\
{\widetilde\Gamma}_{\mathbf {a},\mathbf {c}} &=&\{{\widetilde X}(\mathbf{a},{\widetilde n},1),{\widetilde X}(\mathbf{c},{\widetilde n},2)|{\widetilde n}=1,\ldots ,N\},
\nonumber\\
{\widehat\Gamma}_{\mathbf {b},\mathbf {c}}&=&\{{\widehat X}(\mathbf{b},{\widehat n},1),{\widehat X}(\mathbf{c},{\widehat n},2)|{\widehat n}=1,\ldots ,N\}.
\end{eqnarray}
Note that in real experiments the measurement outcomes are also labeled by the time of measurement but for simplicity we omit this label here.
Using these data sets for testing the validity of Bell's inequality Eq.~(\ref{Bell}) and of the structurally equivalent Boole inequality Eq.~(\ref{Boole}), requires making the following assumptions:
\begin{enumerate}
\item{The same symbol $X$ can be used for all the data collected in the three runs. This results in the data set
$\Upsilon=\{X(\mathbf{a},n,1),X(\mathbf{a},{\widetilde n},1),X(\mathbf{b},n,2),X(\mathbf{b},{\widetilde n},1),X(\mathbf{c},{\widetilde n},2),$
$X(\mathbf{c},{\widehat n},2)|n,{\widetilde n},{\widehat n}=1,\ldots ,N\}$.
}
\item{The data can be rearranged such that $X(\mathbf{a},n,1)=X(\mathbf{a},{\widetilde n},1)$, $X(\mathbf{b},{\widehat n},1)=X(\mathbf{b},n,1)$ and
$X(\mathbf{c},{\widetilde n},2)=X(\mathbf{c},{\widehat n},2)=X(\mathbf {c},n,2)$.
This results in the data set
$\Upsilon^{\prime}=\{X(\mathbf{a},n,1),X(\mathbf{b},n,2),X(\mathbf{b},n,1),X(\mathbf{c},n,2)|n=1,\ldots ,N\}$, a data set containing quadruples, not yet triples, as
used in the derivation of Bell's inequality and as required by Boole for his inequality to be obeyed.
Reduction to a set of triples requires the extra assumption:
}
\item{$X(\mathbf{b},n,1)=X(\mathbf{b},n,2)$
}
\end{enumerate}

Since the data in EPRB laboratory experiments are not collected as one set of triples but as three sets of pairs,
in case a violation of Boole's inequality Eq.~(\ref{Boole}) is found, at least one of the assumptions 1, 2 or 3 is false.
In other words, if the data sets collected in an EPRB experiment satisfy these three conditions,
the one-to-one correspondence between
the two-valued variables in the mathematical description and the observed two-valued experimental data is guaranteed, and hence Boole's and thus also Bell's inequality are satisfied.
If the Bell inequality is violated then at least one of the sufficient conditions 1, 2 or 3 to derive the Bell inequality is false, but then also at least one of the assumptions listed above is false.

\subsubsection{Summary}
One could ask the question how to translate the inequality Eq.~(\ref{Bellproduct}) together with its accompanying assumptions, derived within the context of quantum theory, into an experimental test.
The answer is one simply cannot.
It is not legitimate to replace the quantum theoretical expectations that appear in Eq.~(\ref{Bellproduct}) by certain empirical data,
simply because Eq.~(\ref{Bellproduct}) has been derived within the mathematical framework of quantum theory, not for sets of data collected, grouped and characterized by experimenters.
Since the collected data have values $+1$ or $-1$ they can be tested against the Boole inequalities only and the conclusions that follow from their violation
(see Sect.~6.2.4) have no bearing on the quantum theoretical model, without making additional assumptions which are not self-evident.

In conclusion, an inequality cannot be blindly applied to any set of experimental data, a model or theory.
The inequality should be derived in the proper context and conditions and conclusions belonging to the respective derivations cannot simply be mixed.

\subsection{EPRB experiment with single photons}
In this experiment, the polarization of each photon plays the role of the spin-$\frac{1}{2}$ degree-of-freedom in Bohm's version~\cite{BOHM51}
of the EPR thought experiment~\cite{EPR35}.
Using the fact that the two-dimensional vector space with basis vectors $\{|H\rangle,|V\rangle\}$,
where $H$ and $V$ denote the horizontal and vertical polarization of the photon, respectively, is isomorphic to the vector space
with basis vectors $\{\left|\uparrow\rangle\right. ,\left|\downarrow\rangle\right.\}$ of spin-$\frac{1}{2}$ particles, we may use the quantum theory of the latter to describe the EPRB
experiments with photons.
The expressions for the single-photon expectation values and the two-photon correlations are similar to those of the
genuine spin-$\frac{1}{2}$ particle problem except for the restriction of $\mathbf{a}_1$ and $\mathbf{a}_2$ to lie in planes
orthogonal to the direction of propagation of the photons and that the polarization is defined modulo $\pi$, not modulo $2\pi$ as
in the case of the spin-$\frac{1}{2}$ particles. The latter results in a multiplication of the angles by a factor of two.
For simplicity it is often assumed that ${\mathbf a}_i=(\cos a_i,\sin a_i,0)$ for $i=1,2$.
For the singlet state we then have $\widehat E_1({\mathbf a}_1)=\widehat E_2({\mathbf a}_2)=0$ and $\widehat E({\mathbf a}_1,{\mathbf a}_2)=-\cos 2(a_1-a_2)$.

We take the EPRB experiment with single photons, carried out by Weihs {\it et al.}~\cite{WEIH98,WEIH00},
as a concrete example.
We first describe the data collection and analysis procedure of the experiment and present results
demonstrating that the conclusion that the experimental results can be described by quantum theory is premature.
Next we illustrate how to construct an event-based model of an idealized version of this EPRB experiment which
reproduces the predictions of quantum theory for the single and two-particle averages
for a quantum system of two spin-$\frac{1}{2}$ particles in the singlet state and
a product state~\cite{ZHAO08,MICH11a}, without making reference to concepts of quantum theory.
\begin{figure}[pt]
\begin{center}
\includegraphics[width=9cm]{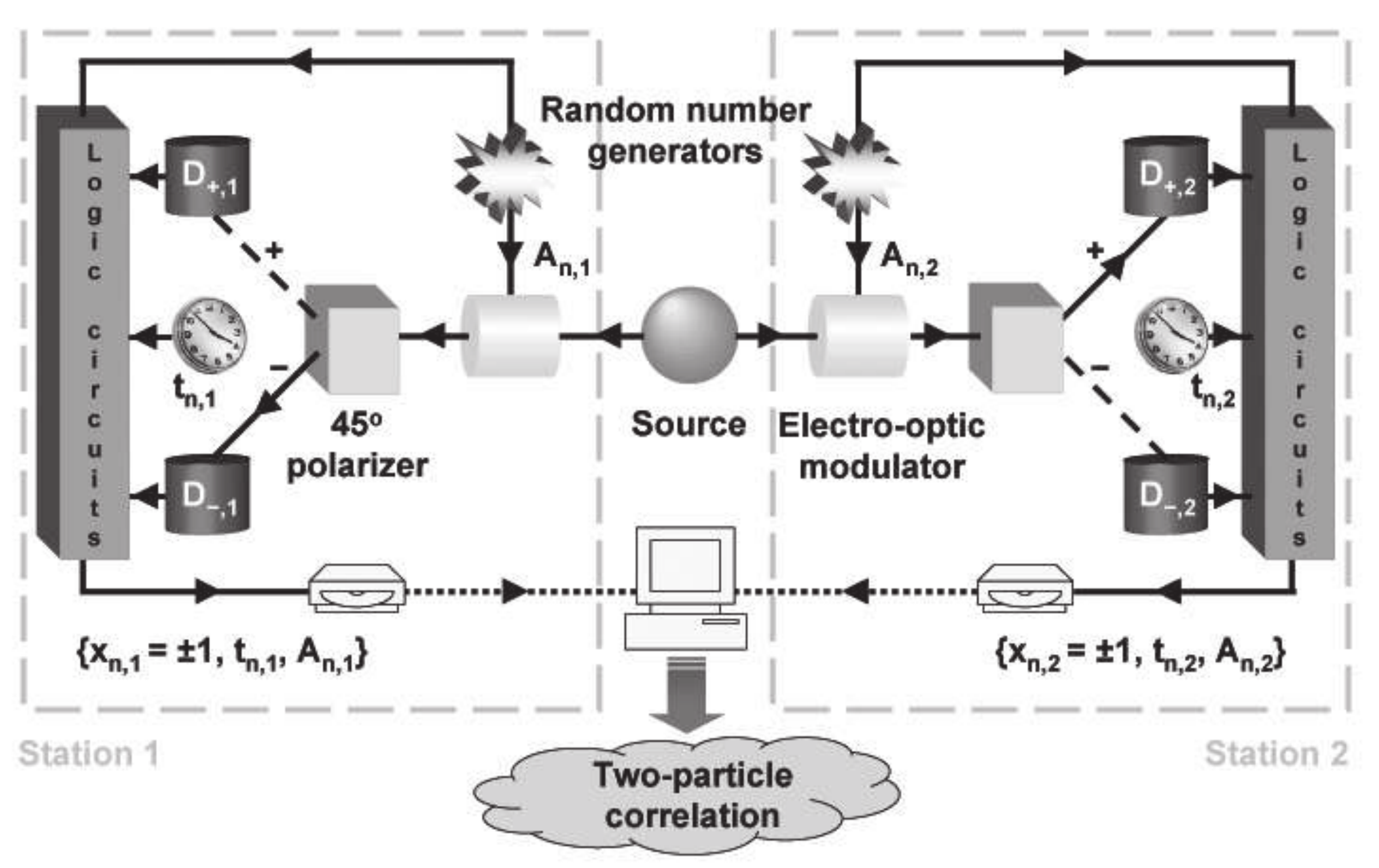}
\vspace*{8pt}
\caption{%
Schematic diagram of the EPRB experiment with single photons~\cite{WEIH98,WEIH00}.
The source emits pairs of photons. The photon pair splits and
one of the photons moves to station 1 and the other one to station 2.
As the photon arrives at station $i=1,2$ it first passes through an electro-optic modulator (EOM) which
rotates the polarization of the photon by an angle $\theta_i$ depending on the voltage applied to the EOM.
This voltage is controlled by a binary variable $A_i$, which is chosen at random.
As the photon leaves the EOM, a polarizing beam splitter directs it to one of the two detectors ${\mathrm D}_{\pm ,i}$.
The detector produces a signal $x_{n,i}=\pm 1$ where the subscript $n$ labels the $n$th detection event.
Each station has its own clock which assigns a time-tag $t_{n,i}$ to each detection signal.
A data set $\left\{ {x_{n,i},t_{n,i},A_{n,i} \vert n =1,\ldots ,N_i } \right\}$ is stored on a hard disk for each station.
Long after the experiment is finished both data sets can be analyzed and among other things, two-particle correlations can be computed.
}\label{fig10}
\end{center}
\end{figure}

\begin{itemize}[leftmargin=*]
\renewcommand\labelitemi{-}
\item{{\sl Data collection:}
Figure~\ref{fig10} shows a schematic diagram of the EPRB experiment with single photons carried out by Weihs {\sl et al.}~\cite{WEIH98,WEIH00}
The source emits pairs of photons. The photon pair splits and each photon travels in free space to
an observation station, labeled by $i=1$ or $i=2$, in which it is manipulated and
detected. The two stations are assumed to be identical and are separated spatially and temporally.
Hence, the observation at station 1 (2) cannot have a causal effect on the data registered at station 2 (1).~\cite{WEIH98}
As the photon arrives at station $i=1,2$ it first passes through an electro-optic modulator (EOM) which
rotates the polarization of the photon by an angle $\theta_i$ depending on the voltage applied to the EOM.~\cite{WEIH98,WEIH00}
This voltage is controlled by a binary variable $A_i$, which is chosen at random.~\cite{WEIH98,WEIH00}
Optionally, a bias voltage is added to the randomly varying voltage.~\cite{WEIH98,WEIH00}
The relation between the voltage applied to the EOM and the resulting rotation of the polarization
is determined experimentally, hence there is some uncertainty in relating the applied voltage
to the rotation angle.~\cite{WEIH98,WEIH00}
As the photon leaves the EOM, a polarizing beam splitter directs it to one of the two detectors.
The detector produces a signal $x_{n,i}=\pm1$ where the subscript $n$ labels the $n$th detection event.
Each station has its own clock which assigns a time-tag $t_{n,i}$ to each signal generated by one of the two detectors.~\cite{WEIH98,WEIH00}
Effectively, this procedure discretizes time in intervals, the width of which is
determined by the time-tag resolution $\tau$.
In the experiment, the time-tag generators are synchronized before each run.~\cite{WEIH98,WEIH00}

The firing of a detector is regarded as an event.
At the $n$th event at station $i$,
the dichotomic variable $A_{n,i}$,
controlling the rotation angle $\theta_{n,i}$,
the dichotomic variable $x_{n,i}$ designating which detector fires,
and the time tag $t_{n,i}$ of the detection event
are written to a file on a hard disk,
allowing the data to be analyzed long after the experiment has terminated.~\cite{WEIH98,WEIH00}
The set of data collected at station $i$ may be written as
\begin{eqnarray}
\label{Ups}
\Upsilon_i=\left\{ {x_{n,i},t_{n,i},\theta_{n,i} \vert n =1,\ldots ,N_i } \right\}
,
\label{eprb1}
\end{eqnarray}
where we allow for the possibility that the number of detected events $N_i$
at stations $i=1,2$ need not (and in practice is not) to be the same
and we have used the rotation angle $\theta_{n,i}$ instead
of the corresponding experimentally relevant dichotomic variable $A_{n,i}$ to facilitate the
comparison with the quantum theoretical description.
}%
\item{{\sl Data analysis procedure:}
A laboratory EPRB experiment requires some criterion to decide which detection
events are to be considered as stemming from a single or two-particle system.
In EPRB experiments with photons, this decision is taken on the basis of coincidence in time.~\cite{WEIH98,CLAU74}
Here we adopt the procedure employed by Weihs {\sl et al.}~\cite{WEIH98,WEIH00}
Coincidences are identified by comparing the time differences
$t_{n,1}-t_{m,2}$ with a window $W$,~\cite{WEIH98,WEIH00,CLAU74}
where $n=1,\ldots,N_1$ and $m=1,\ldots,N_2$.
By definition, for each pair of rotation angles $a_1$ and $a_2$,
the number of coincidences between detectors $D_{x,1}$ ($x =\pm $1) at station 1 and
detectors $D_{y,2}$ ($y =\pm $1) at station 2 is given by
\begin{eqnarray}
\label{Cxy}
C_{xy}&=&C_{xy}(a_1,a_2) \nonumber \\
&=&
\sum_{n=1}^{N_1}
\sum_{m=1}^{N_2}
\delta_{x,x_{n ,1}} \delta_{y,x_{m ,2}}
\delta_{a_1 ,\theta_{n,1}}\delta_{a_2,\theta_{m,2}}
\Theta(W-\vert t_{n,1} -t_{m ,2}\vert)
,
\end{eqnarray}
where $\Theta (t)$ denotes the unit step function.
In Eq.~(\ref{Cxy}) the sum over all events has to be carried out such that each event (= one detected photon) contributes only once.
Clearly, this constraint introduces some ambiguity in the counting procedure as there is a priori, no clear-cut criterion
to decide which events at stations $i=1$ and $i=2$ should be paired.
One obvious criterion might be to choose the pairs such that $C_{xy}$ is maximum, but
such a criterion renders the data analysis procedure (not the data production) acausal.
It is trivial though to analyze the data generated by the experiment of Weihs {\sl et al.}
such that conclusions do not suffer from this artifact.~\cite{RAED12}
In general, the values for the coincidences
$C_{xy}(a_1,a_2)$ depend on the time-tag resolution $\tau$
and the window $W$ used to identify the coincidences.

The single-particle averages and correlation between the coincidence counts
are defined by
\begin{eqnarray}
\label{Exy}
E_1(a_1,a_2)&=&
\frac{\sum_{x,y=\pm1} x C_{xy}}{\sum_{x,y=\pm1} C_{xy}}
=
\frac{C_{++}-C_{--}+C_{+-}-C_{-+}}{C_{++}+C_{--}+C_{+-}+C_{-+}}
\nonumber \\
E_2(a_1,a_2)&=&
\frac{\sum_{x,y=\pm1} yC_{xy}}{\sum_{x,y=\pm1} C_{xy}}
=
\frac{C_{++}-C_{--}-C_{+-}+C_{-+}}{C_{++}+C_{--}+C_{+-}+C_{-+}}
\nonumber \\
E(a_1,a_2)&=&
\frac{\sum_{x,y=\pm1} xy C_{xy}}{\sum_{x,y=\pm1} C_{xy}}
=
\frac{C_{++}+C_{--}-C_{+-}-C_{-+}}{C_{++}+C_{--}+C_{+-}+C_{-+}}
,
\end{eqnarray}
where the denominator $N_c=N_c(a_1,a_2)=C_{++}+C_{--}+C_{+-}+C_{-+}$
in Eq.~(\ref{Exy}) is the sum of all coincidences.

In practice, coincidences are determined by a four-step procedure~\cite{WEIH00}:
\begin{enumerate}
\item{Compute a histogram of time-tag differences $t_{n,1}-t_{m,2}$ of pairs of detection events.}
\item{Determine the time difference $\Delta_G$ for which this histogram shows a maximum.}
\item{Add $\Delta_G$ to the time-tag data $t_{n,1}$, thereby moving the position of the maximum of the histogram to zero.}
\item{Determine the coincidences using the new time-tag differences,
each photon contributing to the coincidence count at most once.}
\end{enumerate}
The global offset, denoted by $\Delta_G$, may be attributed
to the loss of synchronization of the clocks used in the stations 1 and 2.~\cite{WEIH00}

Local-realistic treatments of the EPRB experiment assume that the correlation,
as measured in the experiment, is given by~\cite{BELL93}
\begin{eqnarray}
\label{CxyBell}
C_{xy}^{(\infty)}(a_1,a_2)&=&\sum_{n=1}^N\delta_{x,x_{n ,1}} \delta_{y,x_{n ,2}}
\delta_{a_1,\theta_{n,1}}\delta_{a_2,\theta_{m,2}}
,
\end{eqnarray}
which is obtained from Eq.~(\ref{Cxy})
(in which each photon contributes only once)
by assuming that $N=N_1=N_2$, pairs are defined by $n=m$ and
by taking the limit $W\rightarrow\infty$.
However, the working hypothesis that the value of $W$ should not matter
because the time window only serves to identify pairs may not apply to real experiments.
The analysis of the data of the experiment of Weihs {\sl et al.} shows that
the average time between pairs of photons is of the order of $30\mu$s or more,
much larger than the typical values (of the order of a few nanoseconds)
of the time-window $W$ used in the experiments.~\cite{WEIH00}
In other words, in practice, the identification of photon pairs
does not require the use of $W$'s of the order of a few nanoseconds.
}%
\begin{figure}[pt]
\begin{center}
\includegraphics[width=6cm]{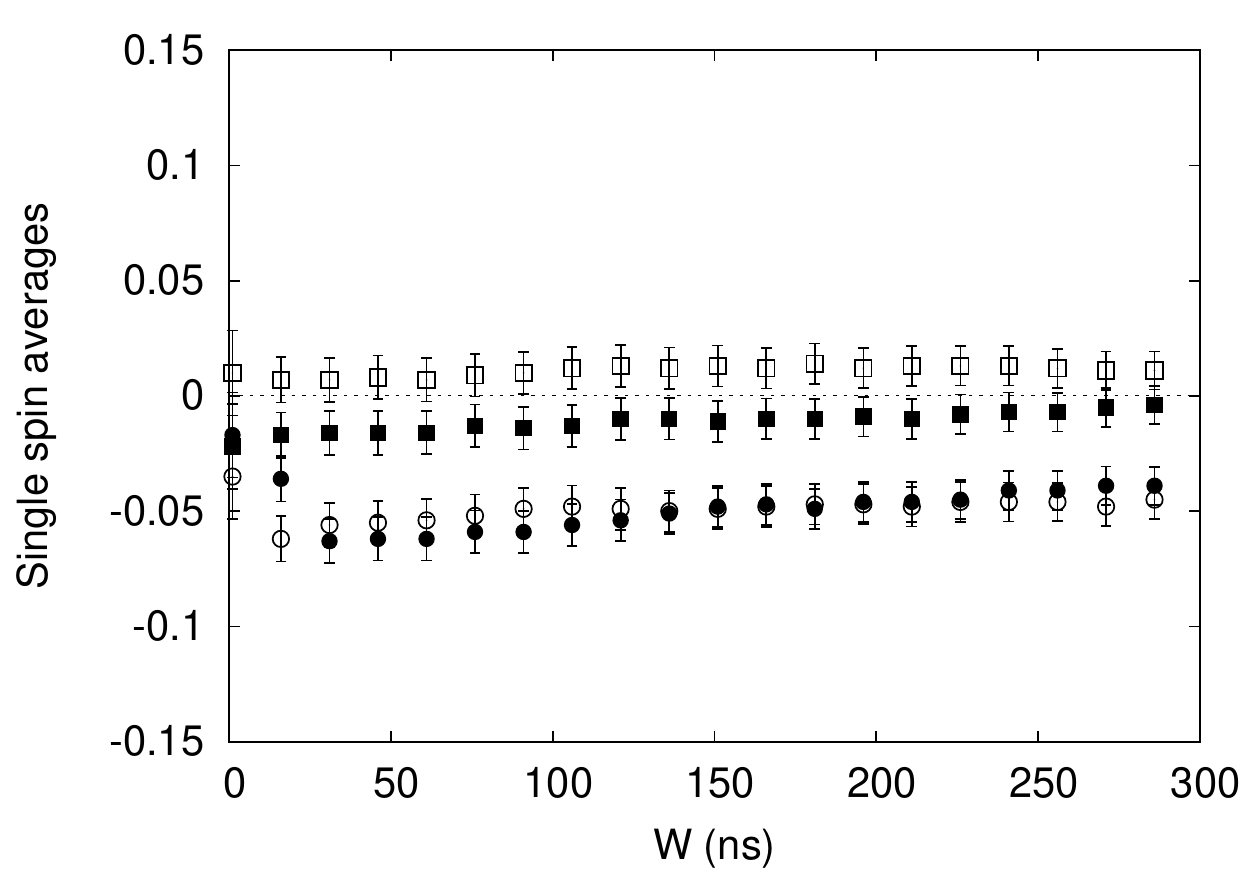}
\includegraphics[width=6cm]{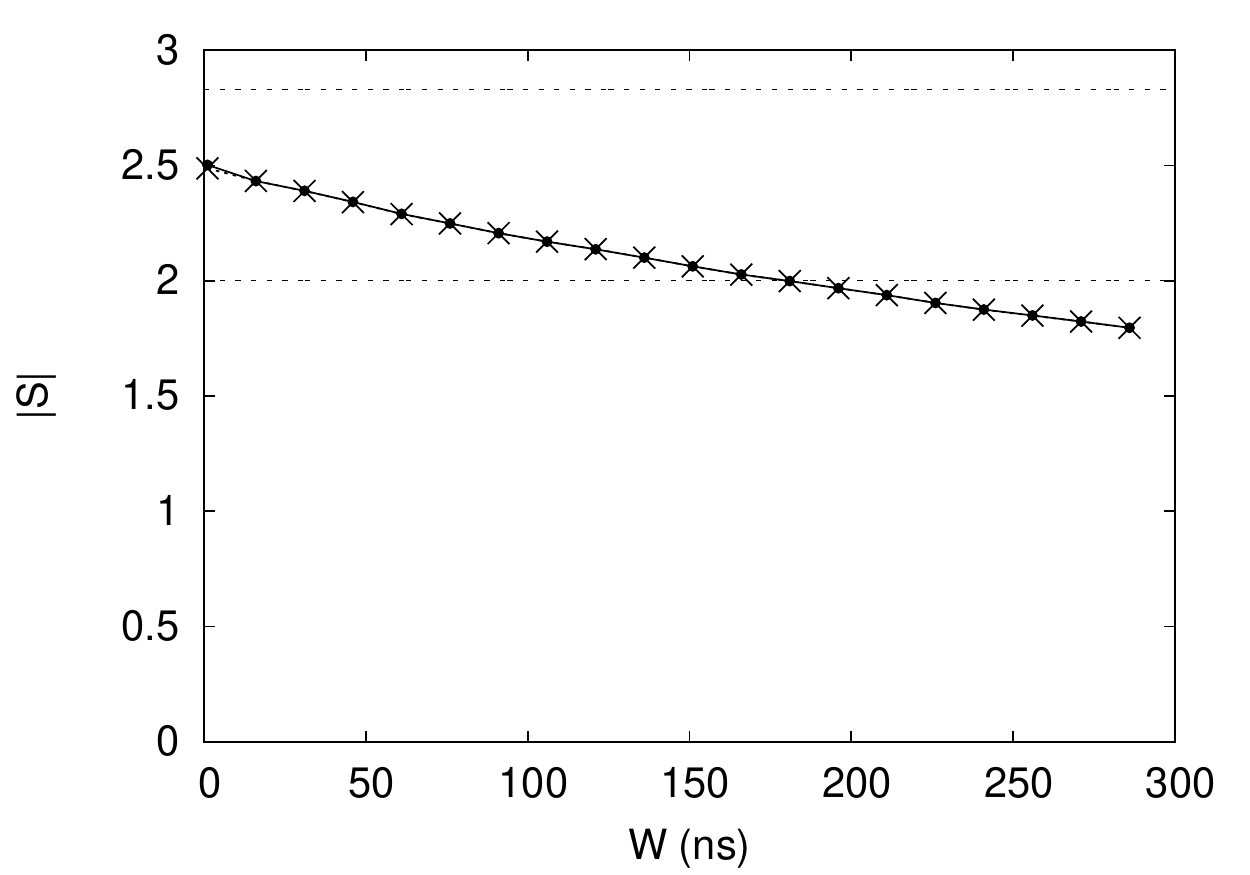}
\vspace*{8pt}
\caption{%
Analysis of the data set {\bf newlongtime2}.
Left: Selected single-particle averages as a function of $W$ for $\Delta_G = 0$ and $a_1=0$, $a_1^{\prime}= \pi /4$, $a_2= \pi /8$ and $a_2^{\prime}= 3\pi /8$.
Open squares: $E_1(a_1,a_2)$; open circles: $E_1(a_1,a_2^{\prime})$; solid squares: $E_2(a_1,a_2)$; solid circles: $E_2(a_1^{\prime},a_2)$.
The error bars correspond to 2.5 standard deviations.
Right: $|S|=|E(a_1,a_2)-E(a_1,a_2^{\prime})+E(a_1^{\prime},a_2)+E(a_1^{\prime},a_2^{\prime})|$
as a function of the time window $W$. The dashed lines
represent the maximum value for a quantum system of two S = 1/2 particles in a separable (product) state
($|S| = 2$) and in a singlet state ($|S| = \sqrt {2}/2$), respectively. Crosses: $\Delta_G = 0$; solid circles connected by
the solid line: $\Delta_G = 0.5$ns.
}\label{fig11}
\end{center}
\end{figure}
\item{{\sl Data analysis results:}
Here, we present only a very limited set of results of our analysis of the experimental data of Weihs {\sl et al.}.
This data has already been analyzed in Refs.~\citen{HNIL02,HNIL07,ADEN07,RAED07c,ZHAO08,BIGE09,AGUE09,RAED10a,BIGE11,RAED12}.

In order to test whether the experimental results are compatible with the predictions
of quantum theory for a system of two spin-$\frac{1}{2}$ particles we first check whether
$E_1(a_1,a_2)$ is independent of $a_2$ and $E_2(a_1,a_2)$ is independent of $a_1$ because quantum theory predicts
that this is the case independent of the state of the two-particle system (see Eq.~(\ref{twospins})).
Since we are dealing with real data we need a criterion to decide whether the data complies with this quantum theoretical prediction.
We consider the data  $E_1(a_1,a_2)$ ($E_2(a_1,a_2)$) to be in conflict with the quantum theoretical prediction if the data show
a dependency on $a_1$ ($a_2$) that exceeds five times the upper bound $1/\sqrt{N_C(a_1,a_2)}$ to the standard deviation $\sigma_{N_c}$.

We analyze a selection of single-particle expectations as a function
of $W$ for the dataset {\bf newlongtime2} (see Fig.~\ref{fig11}(left)).
For small $W$, the total number of coincidences is
too small to yield statistically meaningful results.
For $W>20$ns it is clear that the curves for $E_1(a_1=0,a_2=\pi /8)$ and $E_1(a_1,a_2^{\prime}=3\pi/8)$ (open symbols),
and for $E_2(a_1=0,a_2=\pi /8)$ and $E_2(a_1^{\prime}=\pi /4=0,a_2=\pi /8)$ (closed symbols) are not independent
of the settings $a_2$ and $a_1$, respectively.
The change of these single-spin
averages observed in station 1 (station 2) when the settings are changed in station 2 (station 1), systematically exceeds five
standard deviations, clearly violating our criterion for the data to be compatible with
the prediction of quantum theory of the EPRB model. According to standard practice
of hypothesis testing, the likelihood that this data set can be described by the quantum
theory of the EPRB experiment should be considered as extremely small.
An analysis of in total 23 data sets produced by the experiment of Weihs {\sl et al.}
shows that none of these data sets satisfies our hypothesis test for being compatible
with the predictions of quantum theory of the EPRB model.
Based on the observation of dependency of $E_1(a_1,a_2)$ on $a_2$ and $E_2(a_1,a_2)$ on $a_1$ one could conclude that the data exhibits a spurious kind of ``non-locality''
which cannot be described by the quantum theory of the EPRB experiment.
In trying to find an explanation for this ``non-locality'' we demonstrated elsewhere~\cite{RAED12,RAED13a} that
including a model for the detection efficiencies of the detectors
cannot resolve the conflict between the experimental data of Weihs {\sl et al.} and the quantum theoretical description of the
EPRB experiment.

Although the results for the single particle expectations demonstrate that the experimental data cannot be described
by a quantum theoretical model of two spin-$\frac{1}{2}$ particles (independent of the state which the two photons are in),
in what follows we nevertheless investigate the function $S$ (see Eq.~(\ref{Bellfunction}))
as a function of the time window $W$. Our motivation to do this is two-fold.
First, the goal of the experiment of Weihs {\sl et al.}
was to demonstrate a violation of the Bell-CHSH inequality. We show that the amount of violation depends on $W$, a parameter
absent in the data collection procedure but chosen in the data analysis procedure.
Second, in Sect.~6.3.1 we demonstrate that the Bell-CHSH inequality can also be violated in an event-based model,
a classical dynamical system outside the realm of classical Hamiltonian dynamics,
of the type of EPRB experiment performed by Weihs {\sl et al.}.

Figure~\ref{fig11}(right) shows results of the function $S$ as a function of $W$ for the dataset {\bf newlongtime2}.
For $W<150$ ns, the Bell-CHSH inequality $|S|\le 2$ is clearly violated.
For $W>200$ ns, much less than the average time ($>30\mu$s) between two coincidences,
the inequality $|S|\le 2$ is satisfied, demonstrating that the ``nature''
of the emitted pairs is not an intrinsic property of the pairs themselves but also depends on
the choice of $W$ made by the experimenter.
For $W>20$ ns, there is no significant statistical evidence that the ``noise'' on the data depends on $W$
but if the only goal is to maximize $|S|$, it is expedient to consider $W<20$ ns.

In other words, depending on the value of $W$, chosen by the experimenter when analyzing the data,
the inequality $|S|\le 2$ may or may not be violated.
Hence, also the conclusion about the state of the system depends on the value of $W$.
Analysis of the data of the experiment by Weihs {\sl et al.} shows that $W$ can be as large as 150 ns for the Bell-CHSH inequality to be violated
and in the time-stamping EPRB experiment of Ag\"uero {\sl et al.}~\cite{AGUE09} $|S|\le 2$ is clearly violated for $W<9\mu$s.
Hence, the use of a time-coincidence window does not create a ``loophole''.
Nevertheless, very often it is mentioned that these single-photon Bell test experiments suffer from the fair sampling loophole,
being the result of the usage of a time window $W$ to filter out coincident photons or being the result of the usage of inefficient detectors~\cite{ADEN07}.
The detection loophole was first closed in an experiment with two entangled trapped ions~\cite{Rowe01} and later in a single-neutron
interferometry experiment~\cite{HASE03} and in an experiment with two entangled qubits~\cite{ANSM10}.
However, the latter three experiments are not Bell test experiments performed according to the CHSH protocol~\cite{CLAU69} because the two degrees of freedom
are not manipulated and measured independently.

The narrow time window $W$ in the experiment by Weihs {\it et al.} mainly acts as a filter that selects pairs
of which the individual photons differ in their time tags by the order of nanoseconds.
The possibility that such a filtering mechanism can lead to correlations
that are often thought to be a characteristic of quantum systems only was,
to our knowledge, first pointed out by P.~ Pearle~\cite{PEAR70} and later by A. Fine~\cite{FINE82},
opening the route to a description in terms of locally causal, classical models.
A concrete model of this kind was proposed by S. Pascazio who showed
that his model approximately reproduces the correlation of the singlet state~\cite{PASC86}
with an accuracy that seems beyond what is experimentally achievable to date.
Larson and Gill showed that Bell-like inequalities need to be modified
in the case that the coincidences are determined by a time-window filter~\cite{LARS04}.
We found models that exactly reproduce the results of quantum theory for the singlet
and uncorrelated state.~\cite{RAED06c,RAED07b,ZHAO08,MICH11a}
Here, we closely follow Refs.~\citen{RAED07b,ZHAO08,RAED12a}.
}%
\end{itemize}
\subsubsection{Event-based simulation}
A minimal, discrete-event simulation model of the EPRB experiment by Weihs {\sl et al.} (see Fig.~\ref{fig10})
requires a specification of the information carried by the particles,
of the algorithm that simulates the source and
the observation stations, and of the procedure to analyze the data.
Since in the description of the experiment the orientation of the polarization vectors and the orientations
of the optical axis of the polarizers ${\mathbf a}_i=(\cos a_i, \sin a_i,0)$ for $i=1,2$ is limited to the $xy$-plane
we omit the $z$-component in the simulation.

\begin{itemize}[leftmargin=*]
\renewcommand\labelitemi{-}
\item{{\sl Source and particles:}
Each time, the source emits two particles which carry a vector
${\mathbf u}_{n,i}=(\cos(\xi_{n}+(i-1)\pi/2) ,\sin(\xi_{n}+(i-1)\pi/2))$,
representing the polarization of the photons.
This polarization is completely characterized by the angle $\xi _{n}$ and
the direction $i=1,2$ to which the particle moves.
A uniform pseudo-random number generator is used to pick the angle $0\le\xi _{n}<2\pi$.
Clearly, the source emits two particles with a mutually orthogonal, hence correlated but otherwise
random polarization.
Note that for the simulation of this experiment it is not necessary that the particles carry information about the phase $2\pi ft_{i,n}$, although it would be possible.
In this case the time of flight $t_{i,n}$ is determined by the time-tag model (see below).
}
\item{{\sl Electro-optic modulator} (EOM){\sl :}
The EOM in station $i=1,2$ rotates the polarization of the incoming particle by an angle $\theta_i$,
that is its polarization angle becomes $\xi^{\prime}_{n,i}\equiv\mathrm{EOM}_i(\xi_{n}+(i-1)\pi/2,\theta_i)=\xi_{n}+(i-1)\pi/2-\theta_i$ symbolically.
Mimicking the experiment of Weihs {\sl et al.} in which $\theta_1$ can take the values $a_1,a_1^{\prime}$ and $\theta_2$ can take the values $a_2,a_2^{\prime}$,
we generate two binary uniform pseudo-random numbers $A_i=0,1$ and use them
to choose the value of the angles $\theta_i$, that is
$\theta_1=a_1(1-A_1)+a_1^{\prime}A_1$ and $\theta_2=a_2(1-A_2)+a_2^{\prime}A_2$.
}
\item{{\sl Polarizing beam splitter:}
The simulation model for a polarizing beam splitter is defined by the rule
\begin{eqnarray}
x_{n,i}=\left\{
\begin{array}{lll}
+1 & \mbox{if} & r_n\le \cos^2\xi^{\prime}_{n,i}\\
-1 & \mbox{if} & r_n > \cos^2\xi^{\prime}_{n,i}
\end{array}
\right.
,
\label{sg1}
\end{eqnarray}
where $0< r_n<1$ are uniform pseudo-random numbers.
It is easy to see that for fixed $\xi^{\prime}_{n,i}=\xi^{\prime}_i$, this rule generates
events such that
\begin{eqnarray}
\lim_{N\rightarrow\infty}
\frac{1}{N}\sum_{n=1}^N x_{n,i} = \cos^2\theta_{n,i}
,
\label{sg2}
\end{eqnarray}
with probability one, showing that
the distribution of events complies with Malus law.
Note that this model for the PBS does not make use of a DLM and is therefore much more simple than the event-based model of the PBS described in Sect.~5.3.1.
This simplified mathematical model suffices to simulate the EPRB experiment but cannot be used to simulate other optics experiments (for instance  Wheeler's delayed choice experiment).
However, the PBS described in Sect.~5.3.1 can be used to simulate the EPRB experiment.~\cite{MICH11a}
}
\item{{\sl Time-tag model:} %
As is well-known, as light passes through an EOM (which is essentially a tuneable wave plate), it experiences a retardation
depending on its initial polarization and the rotation by the EOM.
However, to our knowledge, time delays caused by retardation properties of waveplates, being components of various
optical apparatuses, have not yet been explicitly measured for single photons.
Therefore, in the case of single-particle experiments, we hypothesize that for each particle this delay is
represented by the time tag~\cite{RAED07b,ZHAO08}
\begin{eqnarray}
t_{n,i}&=&\lambda(\xi^{\prime}_{n,i}) r^{\prime}_n
,
\label{sg3b}
\end{eqnarray}
that is, the time tag is distributed uniformly ($0<r^{\prime}_n <1$ is a uniform pseudo-random number) over the interval $[0, \lambda(\xi^{\prime}_{n,i})]$.
For $\lambda(\xi^{\prime}_{n,i})=T_0\sin^4 2\xi^{\prime}_{n,i}$ this time-tag model, in combination with the model
of the polarizing beam splitter, rigorously reproduces the results
of quantum theory of the EPRB experiments in the limit $W\rightarrow0$~\cite{RAED07b,ZHAO08}.
We therefore adopt the expression $\lambda(\xi^{\prime}_{n,i})=T_0 \sin^4 2\xi^{\prime}_{n,i}$
leaving only $T_0$ as an adjustable parameter.
}
\item{{\sl Detector:} %
The detectors are ideal particle counters, meaning that they produce a click for each incoming particle.
Hence, we assume that the detectors have 100\% detection efficiency.
Note that adaptive threshold detectors can be used (see Sect.~5.1.1) equally well.~\cite{MICH11a}
}
\item{{\sl Simulation procedure:} %
The simulation algorithm generates the data sets $\Upsilon_i$, similar to the ones obtained in the experiment (see Eq.~(\ref{eprb1})).
In the simulation, it is easy to generate the events such that $N_1=N_2$.
We analyze these data sets in exactly the same manner as the experimental data are analyzed, implying that we
include the post-selection procedure to select photon pairs by a time-coincidence window $W$.
The latter is crucial for our simulation method to give results that are very similar to those observed in a laboratory experiment.
Although in the simulation the ratio of detected to emitted photons is equal to one, the final detection efficiency is reduced
due to the time-coincidence post-selection procedure.
}
\end{itemize}
\subsubsection{Simulation results}
\begin{figure}[pt]
\begin{center}
\includegraphics[width=6cm]{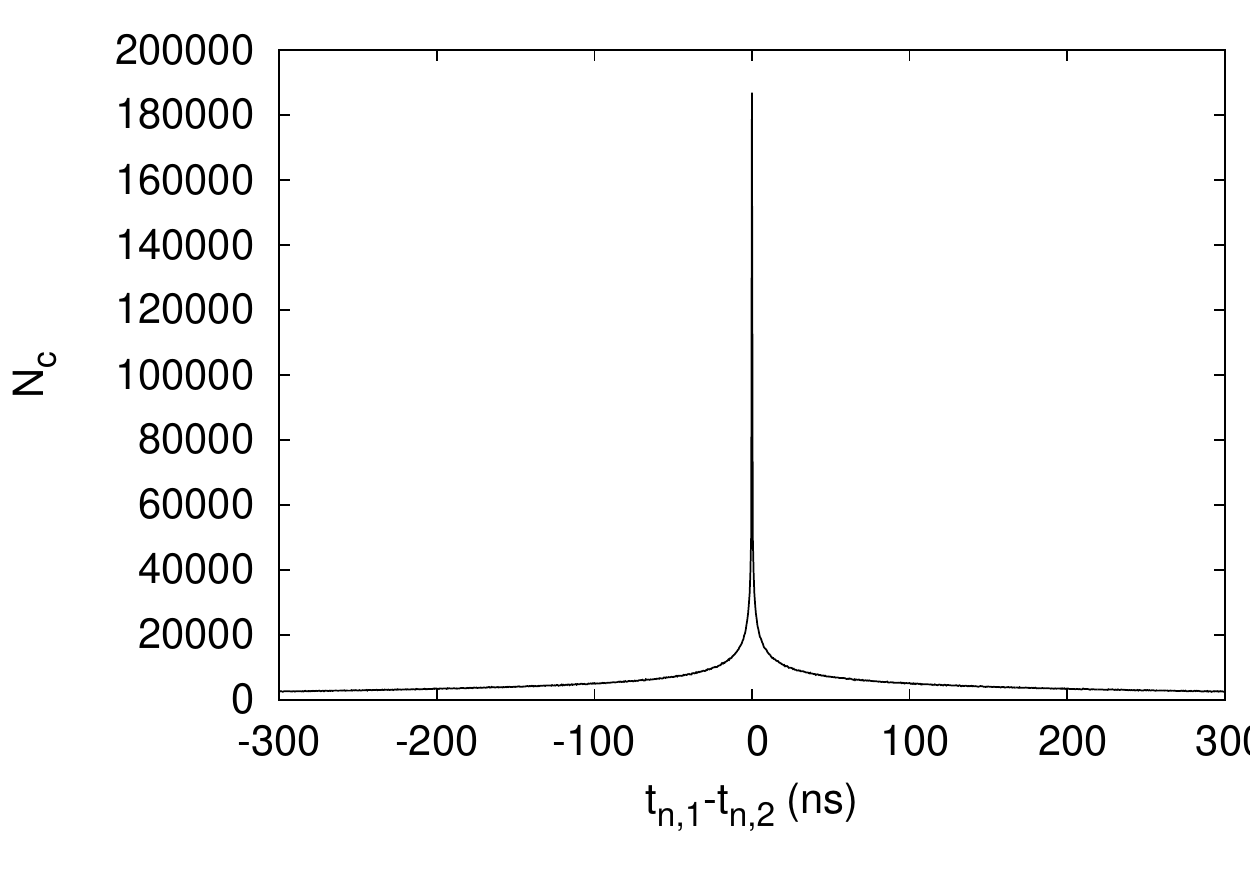}
\includegraphics[width=6cm]{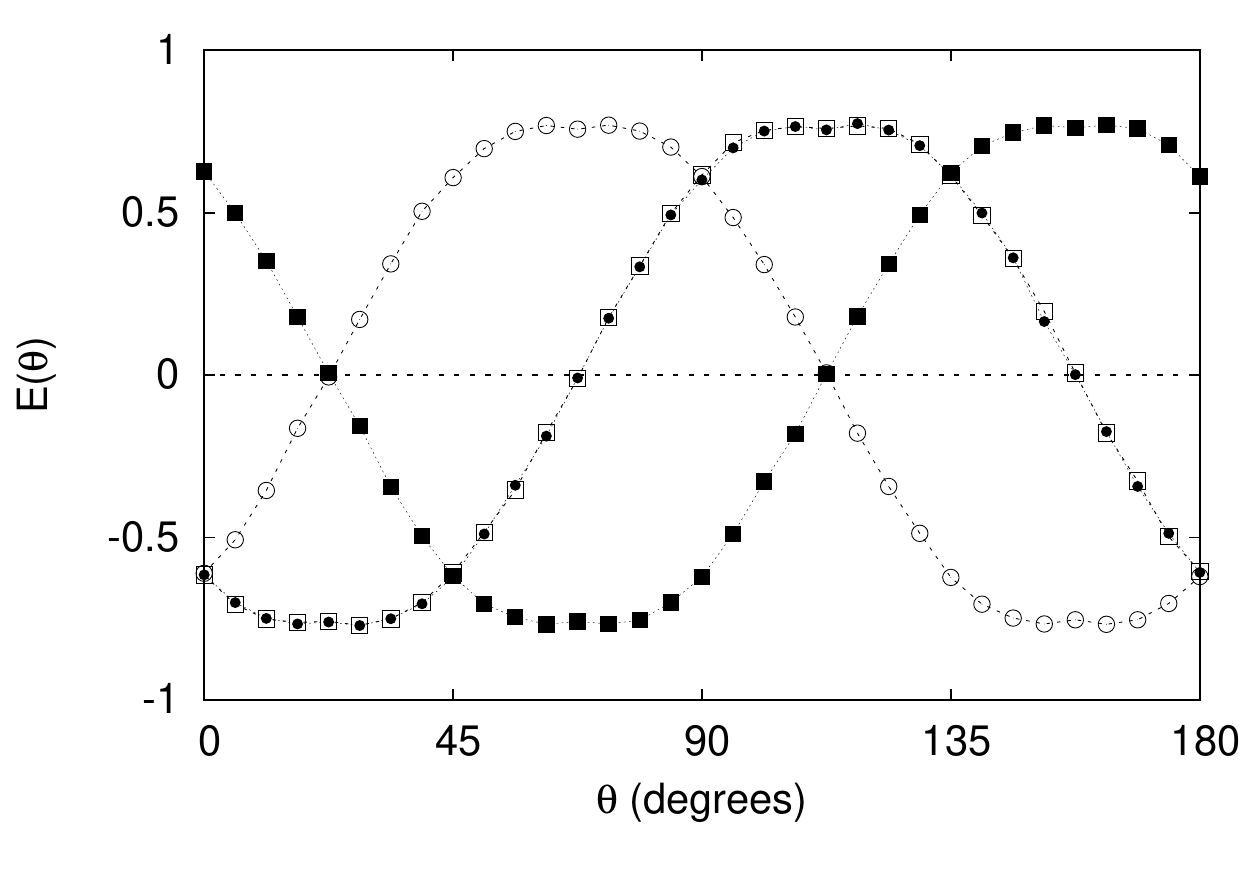}
\vspace*{8pt}
\caption{%
Simulation results using the time-tag model Eq.~(\ref{sg3b}) with $T_0=1000$ ns.
The total number of pairs generated by the source is $3\times 10^5$, roughly the same as in experiment.~\cite{WEIH98}
(a) Coincidence count $N_c$ as a function of the time-tag difference $t_{n,1}-t_{n,2}$.
(b) Two-particle correlations as a function of $\theta$ for $W=50$ ns.
Open squares: $E(\theta)=E(a_1=\theta, a_2=\pi /8)$; open circles: $E(\theta)=E(a_1^{\prime}=\theta +\pi /4, a_2=\pi /8)$;
solid squares: $E(\theta)=E(a_1=\theta, a_2^{\prime}=3\pi /8)$; solid circles: $E(\theta)=E(a_1^{\prime}=\theta +\pi /4, a_2^{\prime}=3\pi /8)$.
}\label{fig12}
\end{center}
\end{figure}
In Fig.~\ref{fig12}(a) we present simulation results for the distribution of time-tag differences, as obtained by using time-tag model Eq.~(\ref{sg3b}).
The distribution is sharply peaked and displays long tails, in qualitative agreement with experiment.~\cite{WEIH00}
The single-particle averages $E_1(a_1,a_2)$ and $E_2(a_1,a_2)$ (results not shown) are zero up to the usual statistical
fluctuations and do not show any statistically relevant dependence on $a_2$ or $a_1$, respectively, in concert with a rigorous
probabilistic treatment of this simulation model.~\cite{ZHAO08}

Some typical simulation results for the two-particle correlations are depicted in Fig.~\ref{fig12}(b) for $W=50$ ns.
For this value of the time-window $W$, the minimum and maximum value of the two-particle correlations is not $-1$ and $+1$, respectively,
as would be expected from the quantum theoretical description. Moreover, the two-particle correlations look more like flattened cosine functions.
For $W=50$ ns we find $|S|=2.62$ which compares very well with the values between 2 and 2.57 extracted from different sets of experimental data of Weihs {\sl et al.}.
However, for $W=2$ ns (results not shown), the results for the two-particle correlations fit very well to the prediction of quantum theory for the EPRB experiment.
From these data we extract $|S|=2.82$.

\begin{figure}[pt]
\begin{center}
\includegraphics[width=6cm]{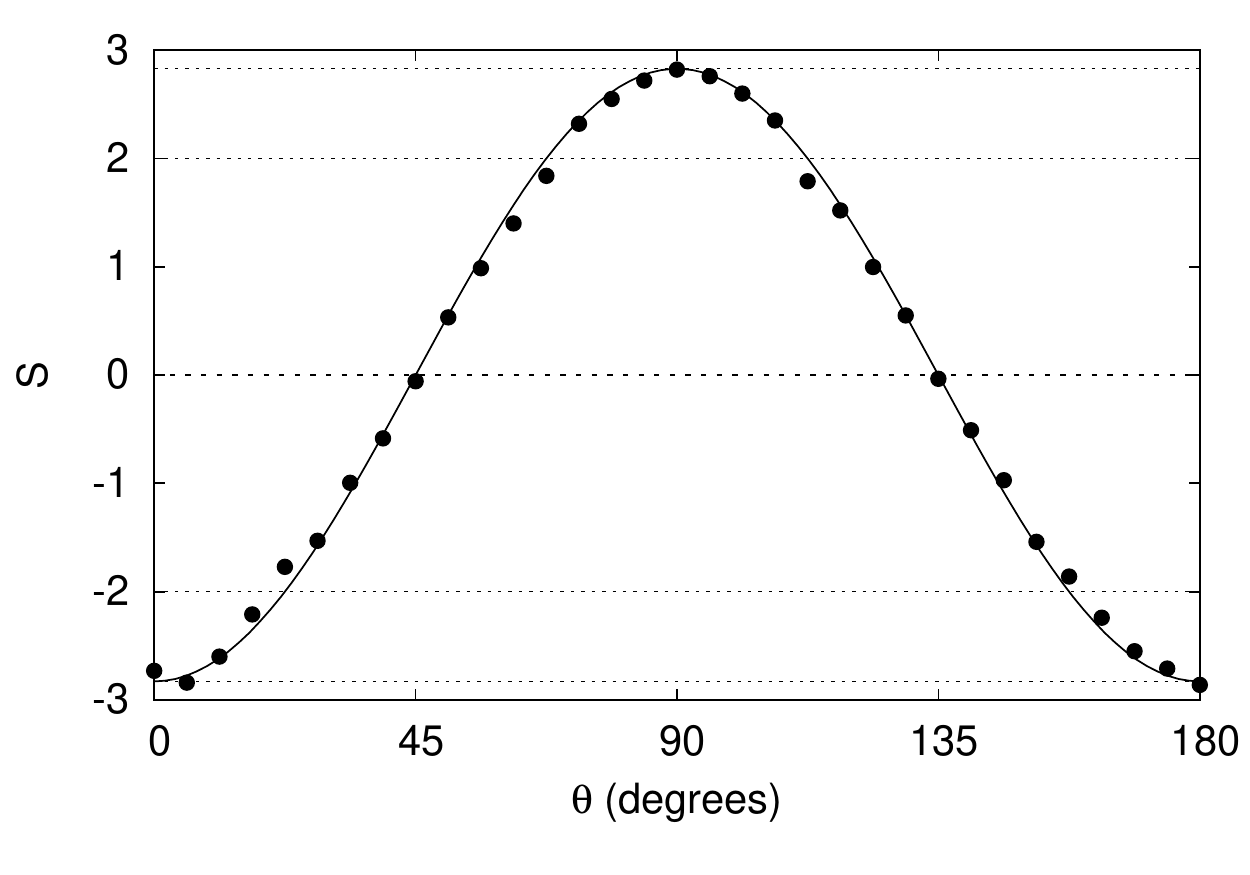}
\includegraphics[width=6cm]{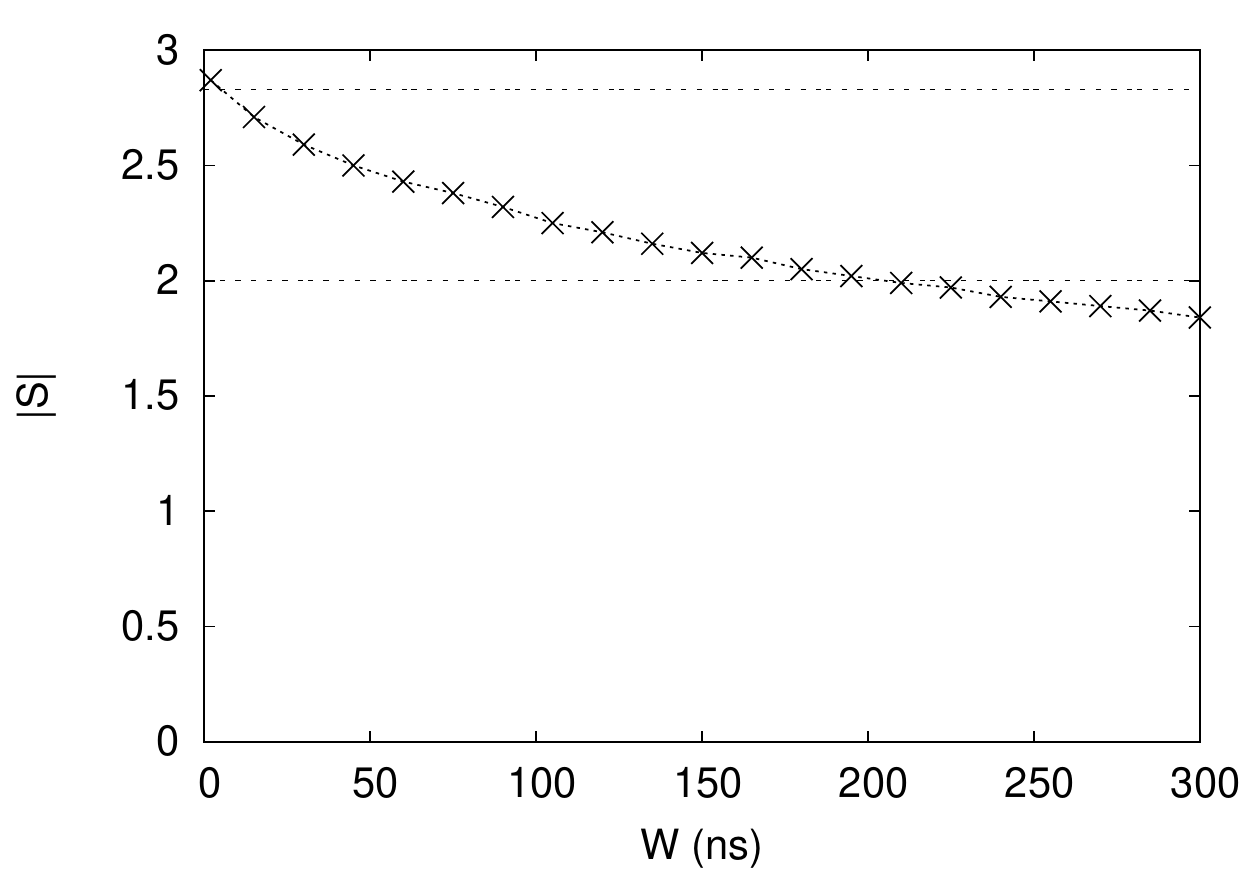}
\vspace*{8pt}
\caption{%
Simulation results for the function $S=E(a_1,a_2)-E(a_1,a_2^{\prime})+E(a_1^{\prime},a_2)+E(a_1^{\prime},a_2^{\prime})$ using the time-tag model Eq.~(\ref{sg3b}) with $T_0=1000$ ns.
The total number of pairs generated by the source is $3\times 10^5$, roughly the same as in experiment.~\cite{WEIH98}
(a) $S$ as a function of $\theta$ with $a_1=\theta$, $a_1^{\prime}= \pi /4+\theta$, $a_2= \pi /8$ and $a_2^{\prime}= 3\pi /8$
for a time window $W=2$ ns. The line connecting the solid circles is the result $-2\sqrt{2}\cos\theta$ predicted by quantum theory.
(b) $|S|$ as a function of $W$ for $a_1=0$, $a_1^{\prime}= \pi /4$, $a_2= \pi /8$ and $a_2^{\prime}= 3\pi /8$.
The line connecting the
crosses is a guide to the eye only.
The dashed horizontal lines indicate the maximum value for a quantum system of two spin-$\frac{1}{2}$
particles in a product state ($|S|=2$) and in a singlet state ($|S|=2\sqrt{2}$).
}\label{fig13}
\end{center}
\end{figure}

Figure \ref{fig13}(a) depicts $S(\theta)$ for $W=2$ ns and shows that the event-based model reproduces the result predicted by quantum theory for the singlet state (solid line),
namely $S=-2\sqrt{2}\cos\theta$. Note that the comparison between the simulation results and quantum theory becomes perfect if more pairs are generated by the source ($10^6$ pairs
is sufficient for most purposes).

From Fig.~\ref{fig13}(b), it follows that a violation of the Bell-CHSH inequality $|S|\le2$
depends on the choice of $W$, a parameter which is absent in the
quantum theoretical description of the EPRB thought experiment.
There are two limiting cases for which $S$ become independent of $W$.
If $W\rightarrow\infty$, it is impossible to let a digital computer violate the inequality $|S|\le2$
without abandoning the rules of Boolean logic or arithmetic~\cite{RAED11a}.
For relatively small $W$ ($W<150$ns), the inequality $|S|\le2$ may be violated.
When $W\rightarrow0$ the discrete-event models which generate the same type
of data as real EPRB experiments,
reproduce exactly the single- and two-spin averages of the singlet state and therefore also violate the
inequality $|S|\le2$.
Obviously, as the discrete-event model does not rely on any concept of quantum theory, a violation of the inequality $|S|\le2$ does not
say anything about the ``quantumness'' of the system under observation~\cite{KARL09,KARL10,RAED11a}.
Similarly, a violation of this inequality cannot say anything about locality and realism~\cite{KARL09,KARL10,RAED11a,NIEU11}.
Clearly, the event-based model is contextual, literally meaning ``being dependent of the (experimental) measurement arrangement''.
The fact that the event-based model reproduces, for instance, the correlations of the singlet state
without violating Einstein's local causality criterion suggests that the
data $\{x_{n,1},x_{n,2}\}$ generated by the event-based model cannot be represented by a single Kolmogorov probability space.
This complies with the idea that contextual, non-Kolmogorov models can lead to
violations of Bell's inequality without appealing to nonlocality or nonobjectivism~\cite{KHRE09,KHRE11,NIEU11}.

In conclusion, event-based simulation models provide a
cause-and-effect description of real EPRB experiments at a level of detail which is not covered
by quantum theory, such as the effect of the choice of the time-window.
Some of these simulation models exactly reproduce the results of quantum theory of the EPRB experiment,
indicating that there is no fundamental obstacle for an EPRB experiment
to produce data that can be described by quantum theory.
However, as we have shown, it is highly unlikely that quantum theory describes the
data of the EPRB experiment by Weihs {\sl et al.}
This suggests that in the real experiment, there may be processes at work which have not been identified yet.

\subsubsection{Why is Bell's inequality violated?}
In Ref.~\citen{ZHAO08}, we have presented a probabilistic description of our simulation model that (i) rigorously proves that
for up to first order in $W$ it exactly reproduces the single particle averages and the two-particle correlations of
quantum theory for the system under consideration; (ii) illustrates how the presence of the time-window $W$ introduces
correlations that cannot be described by the original Bell-like ``hidden-variable'' models~\cite{BELL93}.
Here, we repeat the discussion presented in Ref.~\citen{RAED12a}.

The time-coincidence post-selection procedure with the time-window $W$ filters out the ``coincident'' photons based on the time-tags
$t_{n,i}$ thereby reducing the final detection efficiency to less than 100\%, although in the simulation a measurement
always returns a $+1$ or $-1$ for both photons in a pair (100\% detection efficiency of the detectors).
Hence, even in case of a perfect detection process the data set that is finally retained consists only of a subset of the entire
ensemble of correlated photons that was emitted by the source, exactly as in the laboratory experiments.

We briefly elaborate on point (ii) (see Ref.~\citen{ZHAO08} for a more extensive discussion).
Let us assume that there exists a probability $P(x_1,x_2,t_1,t_2|\theta_1,\theta_2)$ to observe the data $\{x_i,t_i\}$
conditional on the settings $\theta_i$ at stations $i$ for $i=1,2$.
The probability $P(x_1,x_2,t_1,t_2|\theta_1,\theta_2)$ can always be expressed as an integral over the mutually exclusive events
$\xi_1$, $\xi_2$, representing the polarization of the photons

\begin{eqnarray}
P(x_1,x_2,t_1,t_2|\theta_1,\theta_2)&=&\frac{1}{4\pi^2}\int_0^{2\pi}\int_0^{2\pi}P(x_1,x_2,t_1,t_2|\theta_1,\theta_2,\xi_1,\xi_2)\nonumber \\
&&\times P(\xi_1,\xi_2|\theta_1,\theta_2)d\xi_1d\xi_2.
\label{propmod1}
\end{eqnarray}
We now assume that in the probabilistic version of our simulation model, for each event, (i) the values of $\{x_1,x_2,t_1,t_2\}$ are
independent of each other, (ii) the values of $\{x_1,t_1\}$ ($\{x_2,t_2\}$) are independent of $\theta_2$ and $\xi_2$ ($\theta_1$ and $\xi_1$),
(iii) $\xi_1$ and $\xi_2$ are independent of $\theta_1$ or $\theta_2$.
With these assumptions Eq.~(\ref{propmod1}) becomes

\begin{eqnarray}
P(x_1,x_2,t_1,t_2|\theta_1,\theta_2)
&\stackrel{\rm{(i)}}{=}&\frac{1}{4\pi^2}\int_0^{2\pi}\int_0^{2\pi}P(x_1,t_1|\theta_1,\theta_2,\xi_1,\xi_2)\nonumber\\
&&\times P(x_2,t_2|\theta_1,\theta_2,\xi_1,\xi_2)P(\xi_1,\xi_2|\theta_1,\theta_2)d\xi_1d\xi_2\nonumber\\
&\stackrel{\rm{(ii)}}{=}&\frac{1}{4\pi^2}\int_0^{2\pi}\int_0^{2\pi}P(x_1,t_1|\theta_1,\xi_1)P(x_2,t_2|\theta_2,\xi_2)\nonumber\\
&&\times P(\xi_1,\xi_2|\theta_1,\theta_2)d\xi_1d\xi_2\nonumber\\
&\stackrel{\rm{(i)}}{=}&\frac{1}{4\pi^2}\int_0^{2\pi}\int_0^{2\pi}P(x_1|\theta_1,\xi_1)P(t_1|\theta_1,\xi_1)P(x_2|\theta_2,\xi_2)\nonumber\\
&&\times P(t_2|\theta_2,\xi_2)P(\xi_1,\xi_2|\theta_1,\theta_2)d\xi_1d\xi_2\nonumber\\
&\stackrel{\rm{(iii)}}{=}&\frac{1}{4\pi^2}\int_0^{2\pi}\int_0^{2\pi}P(x_1|\theta_1,\xi_1)P(t_1|\theta_1,\xi_1)
P(x_2|\theta_2,\xi_2)\nonumber\\
&&\times P(t_2|\theta_2,\xi_2)P(\xi_1,\xi_2)d\xi_1d\xi_2,
\label{propmod2}
\end{eqnarray}
which is the probabilistic description of our simulation model.
According to our simulation model, the probability distributions that describe the polarizers are given by
$P(x_i|\theta_i,\xi_i)=[1+x_i\cos 2 (\theta_i -\xi_i)]/2$ and
those for the time-delays $t_i$ that are distributed randomly over the interval $[0,\lambda (\xi_i +(i-1)\pi/2 -\theta_i)]$
are given by
$P(t_i|\theta_i,\xi_i)=\Theta (t_i)\Theta (\lambda (\xi_i +(i-1)\pi/2 -\theta_i)-t_i)/\lambda (\xi_i +(i-1)\pi/2 -\theta_i)$,
where $\Theta (.)$ denotes the unit step function.
In the experiment~\cite{WEIH98} and therefore also in our simulation model, the events are selected using a time window $W$ that
the experimenters try to make as small as possible~\cite{WEIH00}.
Accounting for the time window, that is multiplying Eq.~(\ref{propmod2}) by a step function and integrating over all $t_1$ and $t_2$,
the expression for the probability for observing the event $(x_1,x_2)$ reads

\begin{equation}
P(x_1,x_2|\theta_1,\theta_2)=\int_0^{2\pi}\int_0^{2\pi}P(x_1|\theta_1,\xi_1)P(x_2|\theta_2,\xi_2)
\rho(\xi_1,\xi_2|\theta_1,\theta_2)d\xi_1d\xi_2,
\label{propmod3}
\end{equation}
where the probability density $\rho(\xi_1,\xi_2|\theta_1,\theta_2)$ is given by
\begin{eqnarray}
&&\rho(\xi_1,\xi_2|\theta_1,\theta_2)=
\nonumber\\
&&\frac{\int_{-\infty}^{+\infty}\int_{-\infty}^{+\infty}P(t_1|\theta_1,\xi_1)P(t_2|\theta_2,\xi_2)
\Theta (W-|t_1-t_2|)P(\xi_1,\xi_2)dt_1dt_2}
{\int_0^{2\pi}\int_{0}^{2\pi}\int_{-\infty}^{+\infty}\int_{-\infty}^{+\infty}P(t_1|\theta_1,\xi_1)P(t_2|\theta_2,\xi_2)
\Theta (W-|t_1-t_2|)P(\xi_1,\xi_2)d\xi_1d\xi_2dt_1dt_2}.
\nonumber\\
\end{eqnarray}
The simple fact that $\rho(\xi_1,\xi_2|\theta_1,\theta_2)\ne\rho(\xi_1,\xi_2)$ brings the derivation of the original Bell (CHSH)
inequality to a halt. Indeed, in these derivations it is assumed that the probability distribution for $\xi_1$ and $\xi_2$
does not depend on the settings $\theta_1$ or $\theta_2$~\cite{BALL03,BELL93}.

By making explicit use of the time-tag model (see Eq.~(\ref{sg3b})) it can be shown that~\cite{ZHAO08} (i) if we ignore
the time-tag information ($W>T_0$), the two-particle probability takes the form of the hidden variable models
considered by Bell~\cite{BELL93}, and we cannot reproduce the results of quantum theory~\cite{BELL93},
(ii) if we focus on the case $W\rightarrow 0$ the single-particle averages are zero and the two-particle average
$E(\theta_1 , \theta_2)=-\cos 2(\theta_1 -\theta_2)$.

Although our simulation model and its probabilistic version Eq.~(\ref{propmod2}) involve local processes only, the filtering of the
detection events by means of the time-coincidence window $W$ can produce correlations which violate Bell-type
inequalities~\cite{FINE82,PASC86,LARS04}.
Moreover, for $W\rightarrow 0$ our classical, local and causal simulation model can produce single-particle and two-particle
averages that correspond with those of a singlet state in quantum theory.

\begin{figure}[t]
\begin{center}
\includegraphics[width=9.5cm]{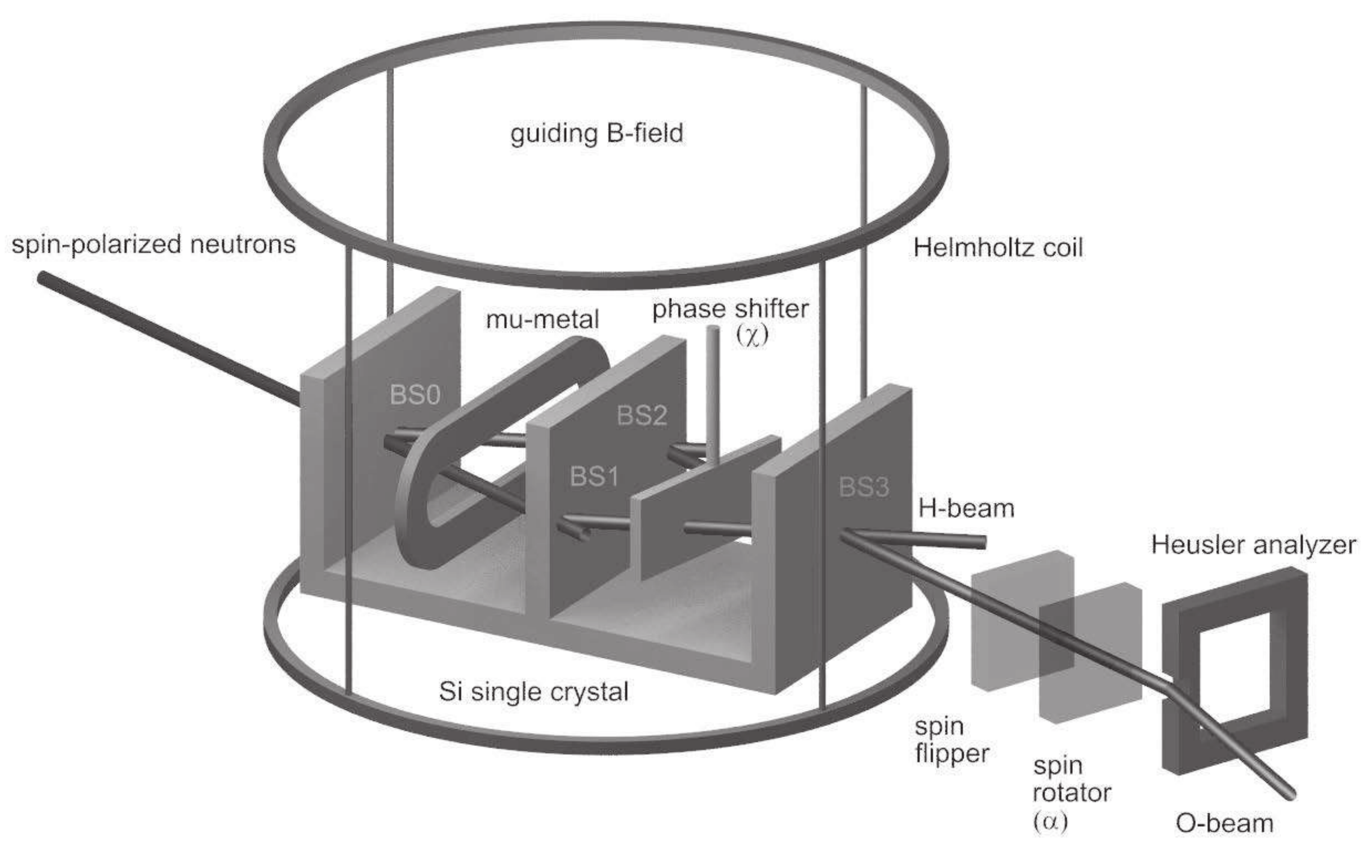}
\caption{%
Top: Schematic picture of the single-neutron interferometry experiment to test a Bell inequality violation (see also Fig.~1 in Ref.~\citen{HASE03}).
BS0, $\ldots$, BS3: beam splitters; phase shifter $\chi$: aluminum foil; neutrons that are transmitted by BS1 or BS2 leave the
interferometer and do not contribute to the interference signal. Detectors count the number of neutrons in the
O- and H-beam.
}\label{fig14}
\end{center}
\end{figure}

\begin{figure}[t]
\begin{center}
\includegraphics[width=9.5cm]{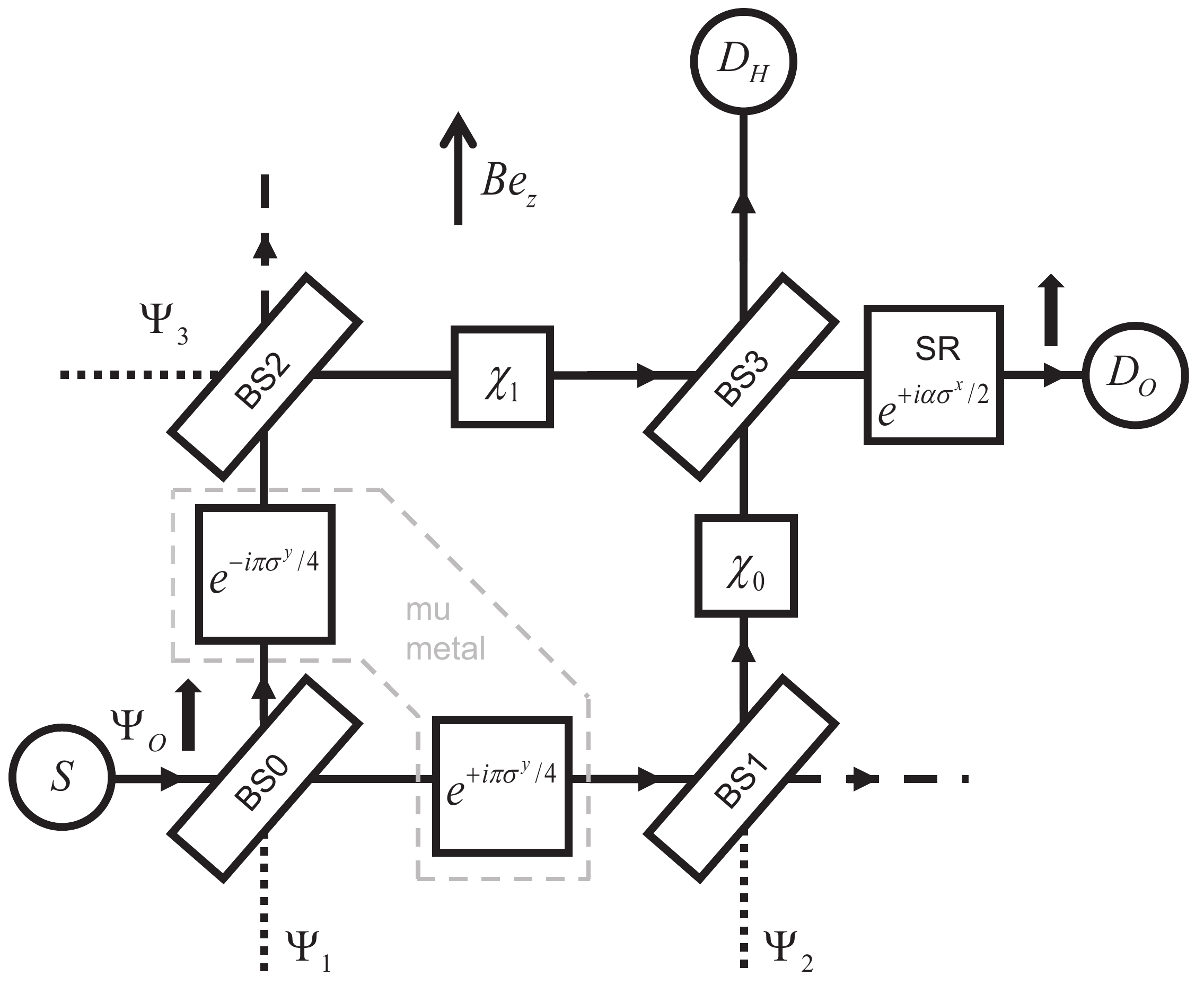}
\caption{%
Event-based network of the experimental setup shown in Fig.~\ref{fig14}.
S: single neutron source; BS0, $\ldots$ , BS3: beam splitters;
$e^{+i\pi\sigma^y/4}$, $e^{-i\pi\sigma^y/4}$: spin rotators modeling the action of a mu-metal;
$\chi_0$, $\chi_1$: phase shifters;
SR $e^{i\alpha\sigma^x/2}$: spin rotator;
$D_O$, $D_H$: detectors counting all neutrons
that leave the interferometer via the O- and H-beam, respectively.
In the experiment and in
the event-based simulation, neutrons with spin up (magnetic moment aligned parallel with respect to the
guiding magnetic field ${\mathbf B}$) enter the interferometer via the path
labeled by $\Psi_0$ only. The wave amplitudes labeled by $\Psi_1$, $\Psi_2$, and $\Psi_3$ (dotted lines) are used in the quantum
theoretical treatment only (see text). Particles leaving the interferometer via
the dashed lines are not counted.
}\label{fig14b}
\end{center}
\end{figure}

\subsection{Bell-test experiment with single neutrons}
The single-neutron interferometry experiment of Hasegawa {\it et al.}~\cite{HASE03}
demonstrates that the correlation between
the spatial and spin degree of freedom of neutrons violates a Bell-CHSH inequality.
In this section we construct an event-based model that reproduces this correlation by using detectors that count every neutron and without
using any post-selection procedure. We show that the event-based model reproduces
the exact results of quantum theory if $\gamma\rightarrow 1^-$ and that by changing $\gamma$ it can also reproduce the numerical
values of the correlations, as measured in experiments.~\cite{HASE03,BART09}
Note that this Bell-test experiment involves two degrees of freedom of one particle, while the EPRB thought experiment~\cite{BOHM51}
and EPRB experiments with single photons~\cite{WEIH98,WEIH00,HNIL02,AGUE09} involve two degrees of freedom of two particles.
Hence, the Bell-test experiment with single neutrons is not performed according to the CHSH protocol~\cite{CLAU69} because the two degrees of freedom
of one particle are not manipulated and measured independently.

Figure~\ref{fig14} shows a schematic picture of the single-neutron interferometry experiment.
Incident neutrons pass through a magnetic-prism polarizer (not shown) which produces two spatially separated beams of
neutrons with their magnetic moments aligned parallel (spin up), respectively anti-parallel (spin down) with respect
to the magnetic axis of the polarizer which is parallel to the guiding field ${\mathbf B}$. The spin-up neutrons
impinge on a silicon-perfect-crystal interferometer.~\cite{RAUC00} On leaving the first beam splitter BS0,
neutrons are transmitted or refracted.
A mu-metal spin-turner changes the orientation of the magnetic moment of the neutron from parallel to perpendicular to the guiding field ${\mathbf B}$.
Hence, the magnetic moment of the neutrons following path H (O) is rotated by $\pi/2$ ($-\pi/2$)
about the $y$ axis. Before the two paths join at the entrance plane of beam splitter BS3, a difference between the time of flights
along the two paths can be manipulated by a phase shifter. The neutrons
which experience two refraction events when passing through the interferometer form the O-beam and are analyzed by sending them through
a spin rotator and a Heusler spin analyzer. If necessary, to induce an extra spin rotation of $\pi$, a spin flipper is placed between
the interferometer and the spin rotator. The neutrons that are selected by the Heusler spin analyzer are counted with a
neutron detector (not shown) that has a very high efficiency ($\approx 99\%$).
Note that neutrons which are not refracted by the mirror plate leave the interferometer
without being detected.

The single-neutron interferometry experiment yields the count rate $N(\alpha,\chi)$ for the spin-rotation angle $\alpha$ and
the difference $\chi$ of the phase shifts of the two different paths in the interferometer~\cite{HASE03}.
The correlation $E(\alpha,\chi)$ is defined by~\cite{HASE03}

\begin{equation}
E(\alpha,\chi)=\frac{N(\alpha,\chi)+N(\alpha+\pi,\chi+\pi)-N(\alpha+\pi,\chi)-N(\alpha,\chi+\pi)}
{N(\alpha,\chi)+N(\alpha+\pi,\chi+\pi)+N(\alpha+\pi,\chi)+N(\alpha,\chi+\pi)}.
\end{equation}

\subsubsection{Event-based model}
A minimal, discrete event simulation model of the single-neutron interferometry experiment requires a specification of the
information carried by the particles, of the algorithm that simulates the source and the interferometer components
(see Fig.~\ref{fig14b}), and of the procedure to analyze the data.
Various ingredients of the simulation model have been described in Sect.~5.4.1.
In the following, we specify the action of
the remaining components, namely the magnetic-prism polarizer
(not shown), the mu-metal spin-turner, the spin-rotator and spin
analyzer.

\begin{itemize}[leftmargin=*]
\renewcommand\labelitemi{-}
\item{{\sl Magnetic-prism polarizer:}
This component takes as input a neutron
with an unknown magnetic moment and produces a neutron
with a magnetic moment that is either parallel (spin up) or antiparallel
(spin down) with respect to the $z$-axis (which by definition
is parallel to the guiding field ${\mathbf B}$). In the experiment, only a
neutron with spin up is injected into the interferometer. Therefore,
as a matter of simplification, we assume that the source $S$ only creates messengers with spin up.
Hence, we assume that $\theta =0$ in Eq.~(\ref{neutron}).
}
\item{{\sl Mu-metal spin turner:}
This component rotates the magnetic moment of a neutron that follows the H-beam (O-beam) by $\pi/2$ ($-\pi/2$) about the $y$ axis.
The processor that accomplishes this
takes as input the direction of the magnetic moment,
represented by the message $\mathbf u$ and performs
the rotation ${\mathbf u}\rightarrow e^{i\pm\pi\sigma ^y/4}{\mathbf u}$.
We emphasize that we use Pauli matrices as a convenient tool
to express rotations in three-dimensional space, not because in quantum theory
the magnetic moment of the neutron is represented by spin-$\frac{1}{2}$
operators.
}
\item{{\sl Spin-rotator and spin-flipper:}
The spin-rotator rotates the magnetic moment of a neutron by an angle $\alpha$ about the $x$ axis.
The spin flipper is a spin rotator with $\alpha=\pi$.
}
\item{{\sl Spin analyzer:}
This component selects neutrons with spin up, after which they are counted by a detector.
The model of this component projects the magnetic moment of the particle on the $z$ axis and sends the particle to the
detector if the projected value exceeds a pseudo-random number $r$.
}
\end{itemize}

\subsubsection{Simulation results}
\begin{figure}[t]
\begin{center}
\includegraphics[width=6cm]{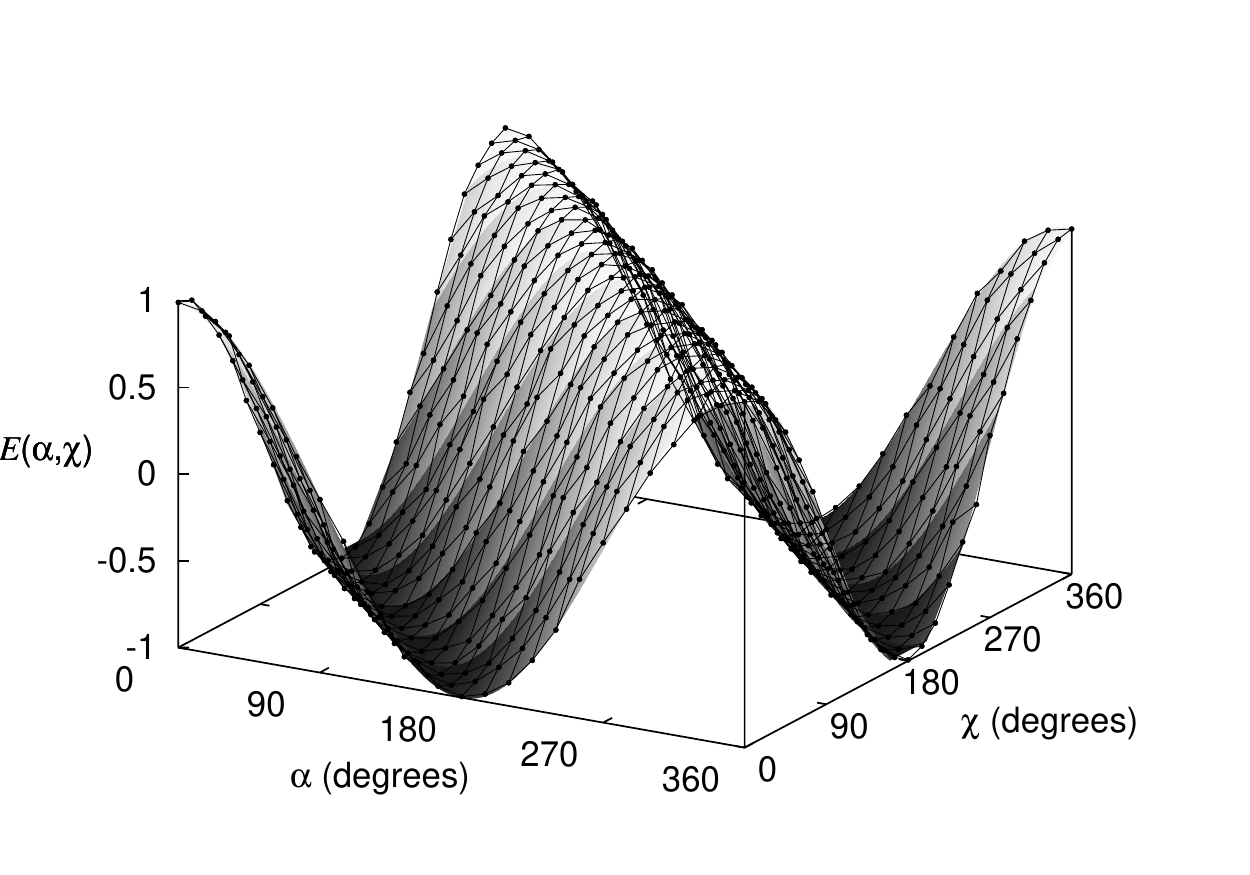}
\includegraphics[width=6cm]{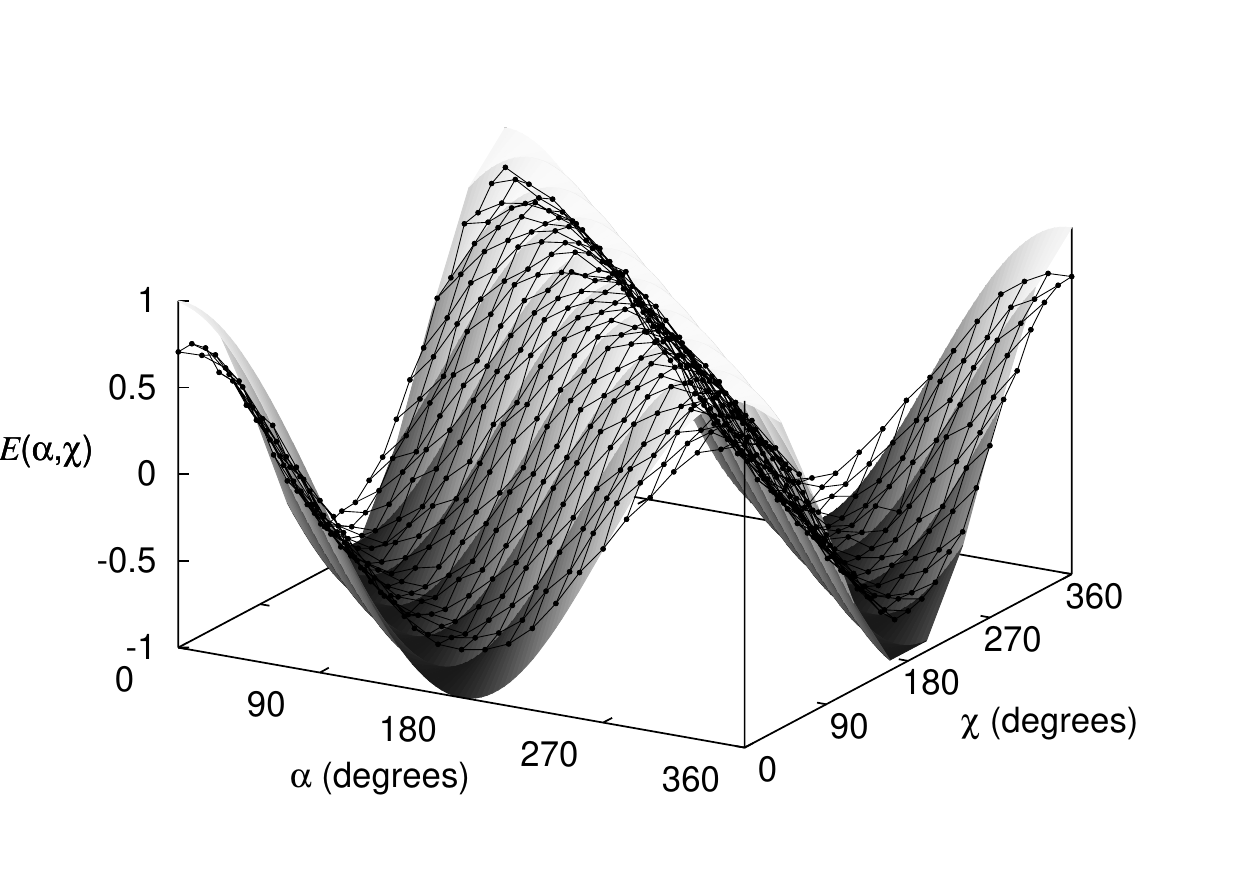}
\caption{%
Left: correlation $E(\alpha,\chi)$ between spin and path degree of freedom as obtained from an event-based simulation of the experiment
depicted in Fig.~\ref{fig14}. Solid surface: $E(\alpha,\chi)=\cos(\alpha+\chi)$ predicted by quantum theory; circles: simulation data.
The lines connecting the markers are guides to the eye only. Model parameters: reflection percentage of BS0, \ldots, BS3 is 20\% and $\gamma=0.99$.
For each pair $(\alpha,\chi)$, four times 10000 particles were used to determine the four counts $N(\alpha,\chi)$, $N(\alpha+\pi,\chi+\pi)$,
$N(\alpha,\chi+\pi)$ and $N(\alpha+\pi,\chi+\pi)$.
Right: same as figure on the left but $\gamma =0.55$.
}\label{fig15}
\end{center}
\end{figure}

In Fig.~\ref{fig15}(left) we present simulation results for the correlation $E(\alpha,\chi)$, assuming that the experimental conditions
are very close to ideal and compare them to the quantum theoretical result.

The quantum theoretical description of the experiment~\cite{HASE03} requires a four-state system for the path and another two-state system to account
for the spin-$\frac{1}{2}$ degree-of-freedom.
Thus, the statistics of the experimental data is described by the state vector Eq.~(\ref{app0}).
In the experiment~\cite{HASE03}, the neutrons that enter the interferometer
all have spin up, relative to the direction of the guiding field $\mathbf{B}$ (see Fig.~\ref{fig14}).
Thus, the state describing the incident neutrons is $|\Psi\rangle=(1,0,0,0,0,0,0,0)^T$, omitting irrelevant
phase factors.
As the state vector propagates through the interferometer and the spin rotator (see Fig.~\ref{fig14b}),
it changes according to
\begin{eqnarray}
|\Psi'\rangle
&=&
\left(\begin{array}{cc}
        \phantom{i}\cos(\alpha/2)& i\sin(\alpha/2) \\
        i\sin(\alpha/2) & \phantom{i}\cos(\alpha/2)
\end{array}\right)_{6,7}
\left(\begin{array}{cc}
        \phantom{-}t^\ast &r\\
        -r^\ast & t
\end{array}\right)_{5,7}
\left(\begin{array}{cc}
        \phantom{-}t^\ast &r \\
        -r^\ast & t
\end{array}\right)_{4,6}
\nonumber \\ &&\times
\left(\begin{array}{cc}
         e^{i\phi_1}&0 \\
         0 & e^{i\phi_1}
\end{array}\right)_{6,7}
\left(\begin{array}{cc}
         e^{i\phi_0}&0 \\
         0 & e^{i\phi_0}
\end{array}\right)_{4,5}
\left(\begin{array}{cc}
        \phantom{-}t^\ast &r\\
        -r^\ast & t
\end{array}\right)_{3,7}
\left(\begin{array}{cc}
        \phantom{-}t^\ast &r \\
        -r^\ast & t
\end{array}\right)_{2,6}
\nonumber \\ &&\times
\left(\begin{array}{cc}
        t & -r^\ast \\
        r & \phantom{-}t^\ast
\end{array}\right)_{1,5}
\left(\begin{array}{cc}
        t & -r^\ast \\
        r & \phantom{-}t^\ast
\end{array}\right)_{0,4}
\left(\begin{array}{cc}
        \phantom{-}1/\sqrt{2} & \phantom{-}1/\sqrt{2} \\
        -1/\sqrt{2} & \phantom{-}1/\sqrt{2} \\
\end{array}\right)_{2,3}
\left(\begin{array}{cc}
        \phantom{-}1/\sqrt{2} & -1/\sqrt{2} \\
        \phantom{-}1/\sqrt{2} & \phantom{-}1/\sqrt{2}
\end{array}\right)_{0,1}
\nonumber \\ &&\times
\left(\begin{array}{cc}
        t & -r^\ast \\
        r & \phantom{-}t^\ast
\end{array}\right)_{1,3}
\left(\begin{array}{cc}
        t & -r^\ast \\
        r & \phantom{-}t^\ast
\end{array}\right)_{0,2}
|\Psi\rangle
,
\label{app5}
\end{eqnarray}
where the subscripts $i,j$ refer to the pair of elements of the
eight-dimensional vector on which the matrix acts.
Reading backwards, the first pair of matrices in Eq.~(\ref{app5}) represents beam splitter BS0,
the second pair the mu-metal (a spin rotation about the $y$-axis by $\pi/4$ and $-\pi/4$, respectively),
the third and fourth pair beam splitters BS1 and BS2, respectively,
the fifth pair the phase shifters,
the sixth pair beam splitter BS3,
and the last matrix represents the spin rotator SR.

From Eq.~(\ref{app5}), it follows that the probability to detect a neutron with spin up in the O-beam is given by
\begin{eqnarray}
p_\mathrm{O}(\alpha,\chi)&=&|\Psi'_{3,\uparrow}|^2={\cal TR}^2  \left[1+\cos(\alpha+\chi)\right]
,
\label{app6}
\end{eqnarray}
where $\chi=\chi_0-\chi_1$.
From Eq.~(\ref{app6}) it follows that the correlation $E_\mathrm{O}(\alpha,\chi)$ is given by~\cite{HASE03}
\begin{eqnarray}
E_\mathrm{O}(\alpha,\chi)&\equiv&\frac{%
p_\mathrm{O}(\alpha,\chi)+p_\mathrm{O}(\alpha+\pi,\chi+\pi)-p_\mathrm{O}(\alpha+\pi,\chi)-p_\mathrm{O}(\alpha,\chi+\pi)
}{
p_\mathrm{O}(\alpha,\chi)+p_\mathrm{O}(\alpha+\pi,\chi+\pi)+p_\mathrm{O}(\alpha+\pi,\chi)+p_\mathrm{O}(\alpha,\chi+\pi)
}\nonumber \\
&=&\cos(\alpha+\chi)
,
\label{app7}
\end{eqnarray}
independent of the reflectivity ${\cal R}=|r|^2=1-{\cal T}$ of the beam splitters (which have been assumed to be identical).
The fact that $E_\mathrm{O}(\alpha,\chi)=\cos(\alpha+\chi)$ implies that the state of the neutron
cannot be written as a product of the state of the spin and the phase.
In other words, in quantum language, the spin- and phase-degree-of-freedom are entangled~\cite{BASU01,HASE03}.Repeating the calculation for the probability of detecting a neutron in the H-beam shows
that $E_\mathrm{H}(\alpha,\chi)=0$, independent of the direction of the spin.
If the mu-metal would rotate the spin about the $x$-axis instead of about the $y$-axis, then
we would find $E_\mathrm{O}(\alpha,\chi)=\cos\alpha\cos\chi$, a typical expression for a quantum system in a product state.

As shown by the markers in Fig.~\ref{fig15} (left), disregarding the small statistical fluctuations,
there is close-to-perfect agreement
between the event-based simulation data for nearly ideal experimental conditions ($\gamma =0.99$ and ${\cal R}=0.2$) and quantum theory.
However, the laboratory experiment suffers from unavoidable imperfections, leading to a reduction and distortion of the interference fringes~\cite{HASE03}.
In the event-based approach it is trivial to incorporate mechanisms for different sources of imperfections by modifying or adding update rules.
However, to reproduce the available data it is sufficient to use the parameter $\gamma$ to control the deviation from the quantum theoretical result.
For instance, for $\gamma=0.55$, ${\cal R}=0.2$ the simulation results for $E(\alpha , \chi )$ are shown in Fig.~\ref{fig15} (right).

In order to quantify the difference between the simulation results, the experimental results and quantum theory
it is customary to form the Bell-CHSH function~\cite{BELL93,CLAU69}
\begin{eqnarray}
S&=&S(\alpha,\chi,\alpha^{\prime},\chi^{\prime})\nonumber \\
&=& E_\mathrm{O}(\alpha,\chi)
+E_\mathrm{O}(\alpha,\chi^{\prime})
-E_\mathrm{O}(\alpha^{\prime},\chi)
+E_\mathrm{O}(\alpha^{\prime},\chi^{\prime})
,
\label{app8}
\end{eqnarray}
for some set of experimental settings $\alpha$, $\chi$, $\alpha^{\prime}$, and $\chi^{\prime}$.
If the quantum system can be described by a product state, then $|S|\le2$.
If $\alpha=0$, $\chi=\pi/4$, $\alpha^{\prime}=\pi/2$, and $\chi^{\prime}=\pi/4$, then
$S\equiv S_{max}=2\sqrt{2}$, the maximum value allowed by quantum theory~\cite{CIRE80}.

For $\gamma =0.55$, ${\cal R}=0.2$ the simulation results yield
$S_{max}=2.05$, in excellent agreement with the value $2.052\pm 0.010$
obtained in experiment~\cite{HASE03}. For $\gamma=0.67$, ${\cal R}=0.2$ the simulation yields $S_{max}=2.30$, in excellent agreement with the value $2.291\pm 0.008$
obtained in a similar, more recent experiment~\cite{Bartosik2009}.

In conclusion, since experiment shows that $|S|>2$, according to quantum theory it is impossible
to interpret the experimental result in terms of a quantum system in the product state~\cite{BALL03}.
The system must be described by an entangled state.
However, the event-based simulation which makes use of classical, Einstein-local and causal event-by-event processes
can reproduce all features of this entangled state.

\subsubsection{Why are results from quantum theory produced?}
From Ref.~\citen{MICH11a} we know that the event-based model for the beam splitter produces results corresponding to those
of classical wave or quantum theory when applied in interferometry experiments.
Important for this outcome is that the phase difference $\chi$ between the two paths in the interferometer
is constant for a relatively large number of incoming particles.
If, for each incoming neutron, we pick the angle $\chi$ randomly from the same set of predetermined values to
produce Fig.~\ref{fig15}, an event-based simulation
with $\gamma=0.99$ yields (within the usual statistical fluctuations) the correlation
$E(\alpha,\chi)\approx [\cos(\alpha+\chi)]/2$, which does not lead to a violation of the Bell-CHSH inequality (results not shown).
Thus, if the neutron interferometry experiment could be repeated with random choices for the phase shifter $\chi$ for each incident neutron,
and the experimental results would show a significant violation of the Bell-CHSH inequality, then the event-based model that we
have presented here would be ruled out.

\section{Discussion}
We have presented an event-based simulation method which allows for a mystery-free, particle-only description of interference and entanglement
phenomena observed in various single-photon experiments and single-neutron interferometry experiments.
The statistical distributions which are observed in these single-particle experiments and which are usually thought to be of quantum mechanical origin,
are shown to emerge from a time series of discrete events generated by causal adaptive systems, which in principle could be build
using macroscopic mechanical parts.

As shown in the examples, in the stationary state (after processing many
events), the event-based model reproduces the statistical
distributions of quantum theory. This might raise questions about the efficiency of the method.
Although the event-based simulation method can be used to
simulate a universal quantum
computer,~\cite{RAED05c,MICH05}
the so-called ``quantum speed-up''
cannot be obtained. This by itself is no surprise because
the quantum speed-up is the result of a mathematical
construct in which each unitary operation on the state
of the quantum computer is counted as one operation
and in which preparation and read-out of the quantum
computer are excluded. Whether or not this mathematical
construct is realized in Nature is an open question.

We hope that our simulation results will stimulate the design of new dedicated single-photon and neutron interferometry experiments
which help extending and refining our event-based approach.

\section*{Acknowledgments}
We would like to thank K. De Raedt, K. Keimpema, F. Jin, S. Miyashita, S. Yuan, and S. Zhao
for many thoughtful comments and contributions to the work
on which this review is based.


\bibliographystyle{ws-rv-van}
\bibliography{../../../all13}

\begin{thebibliography}{138}
\providecommand{\natexlab}[1]{#1}
\providecommand{\url}[1]{\texttt{#1}}
\expandafter\ifx\csname urlstyle\endcsname\relax
  \providecommand{\doi}[1]{doi: #1}\else
  \providecommand{\doi}{doi: \begingroup \urlstyle{rm}\Url}\fi

\bibitem{HOME97}
D.~Home, \emph{{Conceptual Foundations of Quantum Physics}}. (Plenum Press, New
  York, 1997).

\bibitem{BALL03}
L.~E. Ballentine, \emph{{Quantum Mechanics: A Modern Development}}. (World
  Scientific, Singapore, 2003).

\bibitem{MICH11a}
K.~{Michielsen}, F.~Jin, and H.~{De Raedt}, {Event-based corpuscular model for
  quantum optics experiments}, \emph{J. Comput. Theor. Nanosci.} {\bf 8},
  \penalty0 1052 -- 1080,  (2011).

\bibitem{RAED12a}
H.~{De Raedt} and K.~{Michielsen}, {Event-by-event simulation of quantum
  phenomena}, \emph{Ann. Phys. (Berlin)}. {\bf 524}, \penalty0 393 -- 410,
  (2012).

\bibitem{RAED12b}
H.~{De Raedt}, F.~Jin, and K.~{Michielsen}, {Event-based simulation of neutron
  interferometry experiments}, \emph{Quantum Matter}. {\bf 1}, \penalty0 1 --
  21,  (2012).

\bibitem{GRAN86}
P.~Grangier, G.~Roger, and A.~Aspect, Experimental evidence for a photon
  anticorrelation effect on a beam splitter: A new light on single-photon
  interferences, \emph{Europhys. Lett.} {\bf 1}, \penalty0 173 -- 179,  (1986).

\bibitem{RAED05d}
H.~{De Raedt}, K.~{De Raedt}, and K.~Michielsen, {Event-based simulation of
  single-photon beam splitters and {Mach-Zehnder} interferometers},
  \emph{Europhys. Lett.} {\bf 69}, \penalty0 861 -- 867,  (2005).

\bibitem{RAED05b}
K.~{De Raedt}, H.~{De Raedt}, and K.~Michielsen, {Deterministic event-based
  simulation of quantum interference}, \emph{Comp. Phys. Comm.} {\bf 171},
  \penalty0 19 -- 39,  (2005).

\bibitem{JACQ07}
V.~Jacques, E.~Wu, F.~Grosshans, F.~Treussart, P.~Grangier, A.~Aspect, and
  J.-F. Roch, {Experimental realization of Wheeler's delayed-choice gedanken
  experiment}, \emph{Science}. {\bf 315}, \penalty0 966 -- 968,  (2007).

\bibitem{ZHAO08b}
S.~{Zhao}, S.~{Yuan}, H.~{De Raedt}, and K.~Michielsen, {Computer simulation of
  Wheeler's delayed choice experiment with photons}, \emph{Europhys. Lett.}
  {\bf 82}, \penalty0 40004,  (2008).

\bibitem{MICH10a}
K.~Michielsen, S.~Yuan, S.~Zhao, F.~Jin, and H.~{De Raedt}, {Coexistence of
  full which-path information and interference in Wheeler's delayed choice
  experiment with photons}, \emph{Physica E}. {\bf 42}, \penalty0 348 -- 353,
  (2010).

\bibitem{SCHW99}
P.~D.~D. Schwindt, P.~G. Kwiat, and B.-G. Englert, Quantitative wave-particle
  duality and nonerasing quantum erasure, \emph{Phys. Rev. A}. {\bf 60},
  \penalty0 4285 -- 4290,  (1999).

\bibitem{JIN10c}
F.~{Jin}, S.~{Zhao}, S.~{Yuan}, H.~{De Raedt}, and K.~{Michielsen},
  {Event-by-event simulation of a quantum eraser experiment}, \emph{J. Comput.
  Theor. Nanosci.} {\bf 7}, \penalty0 1771 -- 1782,  (2010).

\bibitem{RAED12e}
H.~{De Raedt}, M.~Delina, F.~Jin, and K.~Michielsen, Corpuscular event-by-event
  simulation of quantum optics experiments: application to a quantum-controlled
  delayed-choice experiment, \emph{Phys. Scr.} {\bf T151}, \penalty0 014004,
  (2012).

\bibitem{JACQ05}
V.~Jacques, E.~Wu, T.~Toury, F.~Treussart, A.~Aspect, P.~Grangier, and J.-F.
  Roch, {Single-photon wavefront-splitting interference -- An illustration of
  the light quantum in action}, \emph{Eur. Phys. J. D}. {\bf 35}, \penalty0 561
  -- 565,  (2005).

\bibitem{JIN10b}
F.~{Jin}, S.~{Yuan}, H.~{De Raedt}, K.~{Michielsen}, and S.~Miyashita,
  {Particle-only model of two-beam interference and double-slit experiments
  with single photons}, \emph{J. Phys. Soc. Jpn.} {\bf 79}, \penalty0 074401,
  (2010).

\bibitem{ZHAO08a}
S.~{Zhao} and H.~{De Raedt}, {Event-by-event simulation of quantum cryptography
  protocols}, \emph{J. Comput. Theor. Nanosci.} {\bf 5}, \penalty0 490 -- 504,
  (2008).

\bibitem{AGAF08}
I.~N. Agafonov, M.~V. Chekhova, T.~S. Iskhakov, and A.~N. Penin,
  {High-visibility multiphoton interference of Hanbury Brown-Twiss type for
  classical light}, \emph{Phys. Rev. A}. {\bf 77}, \penalty0 053801,  (2008).

\bibitem{JIN10a}
F.~{Jin}, H.~{De Raedt}, and K.~{Michielsen}, {Event-by-event simulation of the
  Hanbury Brown-Twiss experiment with coherent light}, \emph{Commun. Comput.
  Phys.} {\bf 7}, \penalty0 813 -- 830,  (2010).

\bibitem{MICH12b}
K.~Michielsen, F.~Jin, M.~Delina, and H.~{De Raedt}, Event-by-event simulation
  of nonclassical effects in two-photon interference experiments, \emph{Phys.
  Scr.} {\bf T151}, \penalty0 014005,  (2012).

\bibitem{RAED05c}
H.~{De Raedt}, K.~{De Raedt}, and K.~Michielsen, {New method to simulate
  quantum interference using deterministic processes and application to
  event-based simulation of quantum computation}, \emph{J. Phys. Soc. Jpn.
  Suppl.} {\bf 76}, \penalty0 16 -- 25,  (2005).

\bibitem{MICH05}
K.~Michielsen, K.~{De Raedt}, and H.~{De Raedt}, {Simulation of quantum
  computation: a deterministic event-based approach}, \emph{J. Comput. Theor.
  Nanosci.} {\bf 2}, \penalty0 227 -- 239,  (2005).

\bibitem{ASPE82a}
A.~Aspect, P.~Grangier, and G.~Roger, Experimental realization of
  {Einstein-Podolsky-Rosen-Bohm} gedankenexperiment: A new violation of
  {Bell}'s inequalities, \emph{Phys. Rev. Lett.} {\bf 49}, \penalty0 91 -- 94,
  (1982).

\bibitem{ASPE82b}
A.~Aspect, J.~Dalibard, and G.~Roger, Experimental test of {Bell}'s
  inequalities using time-varying analyzers, \emph{Phys. Rev. Lett.} {\bf 49},
  \penalty0 1804 -- 1807,  (1982).

\bibitem{WEIH98}
G.~Weihs, T.~Jennewein, C.~Simon, H.~Weinfurther, and A.~Zeilinger, {Violation
  of {Bell}'s inequality under strict {Einstein} locality conditions},
  \emph{Phys. Rev. Lett.} {\bf 81}, \penalty0 5039 -- 5043,  (1998).

\bibitem{RAED06c}
K.~{De Raedt}, K.~Keimpema, H.~{De Raedt}, K.~Michielsen, and S.~Miyashita, {A
  local realist model for correlations of the singlet state}, \emph{Eur. Phys.
  J. B}. {\bf 53}, \penalty0 139 -- 142,  (2006).

\bibitem{RAED07a}
H.~{De Raedt}, K.~{De Raedt}, K.~Michielsen, K.~Keimpema, and S.~Miyashita,
  {Event-based computer simulation model of Aspect-type experiments strictly
  satisfying Einstein's locality conditions}, \emph{J. Phys. Soc. Jpn.} {\bf
  76}, \penalty0 104005,  (2007).

\bibitem{RAED07b}
K.~{De Raedt}, H.~{De Raedt}, and K.~Michielsen, {A computer program to
  simulate Einstein-Podolsky-Rosen-Bohm experiments with photons}, \emph{Comp.
  Phys. Comm.} {\bf 176}, \penalty0 642 -- 651,  (2007).

\bibitem{RAED07c}
H.~{De Raedt}, K.~{De Raedt}, K.~Michielsen, K.~Keimpema, and S.~Miyashita,
  {Event-by-event simulation of quantum phenomena: Application to
  Einstein-Podolosky-Rosen-Bohm experiments}, \emph{J. Comput. Theor. Nanosci.}
  {\bf 4}, \penalty0 957 -- 991,  (2007).

\bibitem{RAED07d}
H.~{De Raedt}, K.~Michielsen, S.~Miyashita, and K.~Keimpema, {Reply to Comment
  on ``A local realist model for correlations of the singlet state''},
  \emph{Eur. Phys. J. B}. {\bf 58}, \penalty0 55 -- 59,  (2007).

\bibitem{ZHAO08}
S.~{Zhao}, H.~{De Raedt}, and K.~Michielsen, {Event-by-event simulation model
  of Einstein-Podolsky-Rosen-Bohm experiments}, \emph{Found. Phys.} {\bf 38},
  \penalty0 322 -- 347,  (2008).

\bibitem{TRIE11}
B.~{Trieu}, K.~{Michielsen}, and H.~{De Raedt}, Event-based simulation of light
  propagation in lossless dielectric media, \emph{Comp. Phys. Comm.} {\bf 182},
  \penalty0 726 -- 734,  (2011).

\bibitem{FEYN65}
R.~P. Feynman, R.~B. Leighton, and M.~Sands, \emph{The Feynman Lectures on
  Physics, Vol. 3}. (Addison-Wesley, Reading MA, 1965).

\bibitem{YOUN02}
T.~Young, On the theory of light and colors, \emph{Phil. Trans. R. Soc. Lond.}
  {\bf 92}, \penalty0 12,  (1802).

\bibitem{ZEIL88}
A.~Zeilinger, R.~G{\"{a}}hler, C.~G. Shull, W.~Treimer, and W.~Mampe, Single
  and double slit diffraction of neutrons, \emph{Rev. Mod. Phys.} {\bf 60},
  \penalty0 1067 -- 1073,  (1988).

\bibitem{RAUC00}
H.~Rauch and S.~A. Werner, \emph{Neutron Interferometry: Lessons in
  Experimental Quantum Mechanics}. (Clarendon, London, 2000).

\bibitem{KEIT91}
D.~W. Keith, C.~R. Ekstrom, Q.~A. Turchette, and D.~E. Pritchard, An
  interferometer for atoms, \emph{Phys. Rev. Lett.} {\bf 66}, \penalty0 2693 --
  2696,  (1991).

\bibitem{CARN91}
O.~Carnal and J.~Mlynek, Young's double-slit experiment with atoms: a simple
  atom interferometer, \emph{Phys. Rev. Lett.} {\bf 66}, \penalty0 2689 --
  2692,  (1991).

\bibitem{ARND99}
M.~Arndt, O.~Nairz, J.~Vos-Andreae, C.~Keller, G.~{van der Zouw}, and
  A.~Zeilinger, {Wave-particle duality of C60 molecules}, \emph{Nature}. {\bf
  401}, \penalty0 680 -- 682,  (1999).

\bibitem{BREZ02}
B.~Brezger, L.~Hackerm{\"{u}}ller, S.~Uttenthaler, J.~Petschinka, M.~Arndt, and
  A.~Zeilinger, Matter-wave interferometer for large molecules, \emph{Phys.
  Rev. Lett.} {\bf 88}, \penalty0 100404,  (2002).

\bibitem{JUFF12}
T.~Juffmann, A.~Milic, {M. M\"ullneritsch}, P.~Asenbaum, A.~Tsukernik,
  J.~{T\"uxen}, M.~Mayor, O.~Cheshnovsky, and M.~Arndt, Real-time single
  molecule imaging of quantum interference, \emph{Nature Nanotechnology}. {\bf
  7}, \penalty0 297 -- 300,  (2012).

\bibitem{BORN64}
M.~Born and E.~Wolf, \emph{{Principles of Optics}}. (Pergamon, Oxford, 1964).

\bibitem{FEYN63}
R.~Feynman, \emph{Lectures in Physics, Vol. 1}. (Addison Wesley Publishing
  Company Reading, Mass, 1963).

\bibitem{MOEL55}
{G. M\"ollenstedt and H. D\"uker}, {Fresnelscher Interferenzversuch mit einem
  Biprisma f\"ur Elektronenwellen}, \emph{Naturwissenschaften}. {\bf 42},
  \penalty0 41,  (1955).

\bibitem{MOEL56}
{G. M\"ollenstedt and H. D\"uker}, {Beobachtungen und Messungen an
  Biprisma-Interferenzen mit Elektronenwellen}, \emph{Z. Phys.} {\bf 145},
  \penalty0 377 -- 397,  (1956).

\bibitem{JONS61}
C.~J{\"{o}}nsson, {Elektroneninterferenzen an mehreren k{\"{u}}nstlich
  hergestellten Feinspalten}, \emph{Z. Phys.} {\bf 161}, \penalty0 454 -- 474,
  (1961).

\bibitem{HASS10}
F.~Hasselbach, Progress in electron- and ion-interferometry, \emph{Rep. Prog.
  Phys.} {\bf 73}, \penalty0 016101,  (2010).

\bibitem{ROSA12}
{R. Rosa}, {The Merli-Missiroli-Pozzi two-slit electron-interference
  experiment}, \emph{Phys. Perspect.} {\bf 14}, \penalty0 178 -- 195,  (2012).

\bibitem{DONA73}
O.~Donati, G.~P. Missiroli, and G.~Pozzi, An experiment on electron
  interference, \emph{Am. J. Phys.} {\bf 41}, \penalty0 639 -- 644,  (1973).

\bibitem{MERL76}
P.~G. Merli, G.~F. Missiroli, and G.~Pozzi, On the statistical aspect of
  electron interference phenomena, \emph{Am. J. Phys.} {\bf 44}, \penalty0 306
  -- 307,  (1976).

\bibitem{TONO89}
A.~Tonomura, J.~Endo, T.~Matsuda, T.~Kawasaki, and H.~Ezawa, Demonstration of
  single-electron buildup of an interference pattern, \emph{Am. J. Phys.} {\bf
  57}, \penalty0 117 -- 120,  (1989).

\bibitem{FRAB08}
S.~Frabboni, G.~C. Gazzadi, and G.~Pozzi, {Nanofabrication and the realization
  of Feynman's two-slit experiment}, \emph{Appl. Phys. Lett.} {\bf 93},
  \penalty0 073108,  (2008).

\bibitem{FRAB10}
S.~Frabboni, C.~Frigeri, G.~C. Gazzadi, and G.~Pozzi, Four slits interference
  and diffraction experiments, \emph{Ultramicroscopy}. {\bf 110}, \penalty0 483
  -- 487,  (2010).

\bibitem{FRAB11}
S.~Frabboni, C.~Frigeri, G.~C. Gazzadi, and G.~Pozzi, Two and three slit
  electron interference and diffraction experiments, \emph{Am. J. Phys.} {\bf
  79}, \penalty0 615 -- 618,  (2011).

\bibitem{FRAB12}
S.~Frabboni, A.~Gabrielli, G.~C. Gazzadi, F.~Giorgi, G.~Matteucci, G.~Pozzi,
  N.~S. Cesari, M.~Villa, and A.~Zoccoli, The {Young-Feynman} two-slits
  experiment with single electrons: Build-up of the interference pattern and
  arrival-time distribution using a fast-readout pixel detector,
  \emph{Ultramicroscopy}. {\bf 116}, \penalty0 73 -- 76,  (2012).

\bibitem{BACH13}
R.~Bach, D.~Pope, S.~Liu, and H.~Batelaan, Controlled double-slit electron
  diffraction, \emph{New J. Phys.} {\bf 15}, \penalty0 033018,  (2013).

\bibitem{THOM08}
J.~J. Thomson, On the ionization of gases by ultra- violet light and on the
  evidence as to the structure of light afforded by its electrical effects.,
  \emph{Proc. Camb. Phil. Soc.} {\bf 14}, \penalty0 417 -- 424,  (1908).

\bibitem{TAYL09}
G.~I. Taylor, Interference fringes with feeble light, \emph{Proc. Cambridge
  Phil. Soc.} {\bf 15}, \penalty0 114 -- 115,  (1909).

\bibitem{TSUC85}
Y.~Tsuchiya, E.~Inuzuka, T.~Kurono, and M.~Hosoda, Photon-counting imaging and
  its application, \emph{Adv. Electron. Electron Phys.} {\bf 64A}, \penalty0 21
  -- 31,  (1985).

\bibitem{PARK71}
S.~Parker, A single-photon double slit interference experiment, \emph{Am. J.
  Phys.} {\bf 39}, \penalty0 420 -- 424,  (1971).

\bibitem{WEIS03}
A.~Weis and R.~Wynands, Three demonstration experiments on the wave and
  particle nature of light, \emph{PhyDid}. {\bf 1/2}, \penalty0 S67 -- S73,
  (2003).

\bibitem{DIMI08}
T.~L. Dimitrova and A.~Weis, {The wave-particle duality of light: A
  demonstration experiment}, \emph{Am. J. Phys.} {\bf 76}, \penalty0 137 --
  142,  (2008).

\bibitem{SAVE02}
I.~G. Saveliev, M.~Sanz, and N.~Garcia, {Time-resolved Young's interference and
  decoherence}, \emph{J. Opt. B: Quantum Semiclass. Opt.} {\bf 4}, \penalty0
  S477 -- S481,  (2002).

\bibitem{GARC02}
N.~Garcia, I.~G. Saveliev, and M.~Sharonov, Time-resolved diffraction and
  interference: Young's interference with photons of different energy as
  revealed by time resolution, \emph{Phil. Trans. R. Soc. Lond. A}. {\bf 360},
  \penalty0 1039 -- 1059,  (2002).

\bibitem{KOLE13}
P.~Kolenderski, C.~Scarcella, K.~Johnsen, D.~Hamel, C.~Holloway, L.~Shalm,
  S.~Tisa, A.~Tosi, K.~Resch, and T.~Jennewein.
\newblock Time-resolved double-slit experiment with entangled photons.
\newblock arXiv:1304.4943.

\bibitem{NIEU13}
A.~E. Allahverdyan, R.~Balian, and T.~M. Nieuwenhuizen, {Understanding quantum
  measurement from the solution of dynamical models}, \emph{Phys. Rep.} {\bf
  525}, \penalty0 1 -- 166,  (2013).

\bibitem{WOLF02}
S.~Wolfram, \emph{{A New Kind of Science}}. (Wolfram Media Inc., 2002).

\bibitem{PETE63}
A.~Petersen, {The philosophy of Niels Bohr}, \emph{Bulletin of the Atomic
  Scientists}. {\bf 19}, \penalty0 8 -- 14,  (1963).

\bibitem{PLOT10}
A.~Plotnitsky, {The Art and Science of Experimentation in Quantum Physics},
  \emph{AIP Conf. Proc.} {\bf 1232}, \penalty0 128 -- 142,  (2010).

\bibitem{STER22}
W.~Gerlach and O.~Stern, {Der experimentelle Nachweis der Richtungsquantelung
  im Magnetfeld}, \emph{Z. Phys.} {\bf 9}, \penalty0 349 -- 352,  (1922).

\bibitem{GERL24}
W.~Gerlach and O.~Stern, {Uber die Richtungsquantelung im Magnetfeld},
  \emph{Ann. Phys.} {\bf 74}, \penalty0 673 -- 699,  (1924).

\bibitem{RAED06a}
H.~{De Raedt}, K.~{De Raedt}, K.~Michielsen, and S.~Miyashita, {Efficient data
  processing and quantum phenomena: Single-particle systems}, \emph{Comp. Phys.
  Comm.} {\bf 174}, \penalty0 803 -- 817,  (2006).

\bibitem{RAED06}
H.~{De Raedt} and K.~Michielsen.
\newblock {Computational Methods for Simulating Quantum Computers}.
\newblock In eds. M.~Rieth and W.~Schommers, \emph{Handbook of Theoretical and
  Computational Nanotechnology}, pp. 2 -- 48. American Scientific Publishers,
  Los Angeles,  (2006).

\bibitem{BROG25}
L.~de~Broglie, Recherches sur la th{\'{e}}orie des quanta, \emph{Annales de
  Physique}. {\bf 3}, \penalty0 22,  (1925).

\bibitem{WHEE83}
J.~A. Wheeler,  (1983).
\newblock {in: Mathematical foundations of quantum theory, Proc. New Orleans
  Conf. on The mathematical foundations of quantum theory, ed. A.R. Marlow
  (Academic, New York, 1978) [reprinted in Quantum theory and measurements,
  eds. J.A. Wheeler and W.H. Zurek (Princeton Univ. Press, Princeton, NJ, 1983)
  pp. 182-213]}.

\bibitem{JACQ08}
V.~Jacques, E.~Wu, F.~Grosshans, F.~Treussart, P.~Grangier, A.~Aspect, and
  J.-F. Roch, {Delayed-choice test of quantum complementarity with interfering
  single photons}, \emph{Phys. Rev. Lett.} {\bf 100}, \penalty0 220402,
  (2008).

\bibitem{PFLE67}
R.~L. Pfleegor and L.~Mandel, {Interference of independent photon beams},
  \emph{Phys. Rev.} {\bf 159}, \penalty0 1084 -- 1088,  (1967).

\bibitem{DS08}
\url{http://demonstrations.wolfram.com/EventByEventSimulationOfDoubleSlitExperimentsWithSinglePhoto/}.

\bibitem{FEYN85}
R.~P. Feynman, \emph{QED - The Strange Theory of Light and Matter}. (Princeton
  University Press, 1985).

\bibitem{RAED12}
H.~{De Raedt}, K.~Michielsen, and F.~Jin, {Einstein-Podolsky-Rosen-Bohm
  laboratory experiments: Data analysis and simulation}, \emph{AIP Conf. Proc.}
  {\bf 1424}, \penalty0 55 -- 66,  (2012).

\bibitem{TAFL05}
A.~Taflove and S.~Hagness, \emph{Computational Electrodynamics: The
  Finite-Difference Time-Domain Method}. (Artech House, Boston, 2005).

\bibitem{HADF09}
R.~H. Hadfield, Single-photon detectors for optical quantum information
  applications, \emph{Nature Photonics}. {\bf 3}, \penalty0 696 -- 705,
  (2009).

\bibitem{ALKO01}
D.~L. Alkon, {"Either-Or"}{Two-Slit} interference: Stable coherent propagation
  of individual photons through separate slits, \emph{Biophys. J.} {\bf 80},
  \penalty0 2056--2061,  (2001).

\bibitem{MZI08}
\url{http://demonstrations.wolfram.com/EventByEventSimulationOfTheMachZehnderInterferometer/}.

\bibitem{MZIdemo}
{Sample Fortran and Java programs and interactive programs that perform
  event-based simulations of a beam splitter, one Mach-Zehnder interferometer,
  and two chained Mach-Zehnder interferometers can be found at
  \url{http://www.compphys.net/}}.

\bibitem{QuantumTheory}
{We make a distinction between quantum theory and quantum physics. We use the
  term {\sl quantum theory} when we refer to the mathematical formalism, i.e.,
  the postulates of quantum mechanics (with or without the wave function
  collapse postulate)~\cite{BALL03} and the rules (algorithms) to compute the
  wave function. The term {\sl quantum physics} is used for microscopic,
  experimentally observable phenomena that do not find an explanation within
  the mathematical framework of classical mechanics.}

\bibitem{BAYM74}
G.~Baym, \emph{Lectures on Quantum Mechanics}. (W.A. Benjamin, Reading MA,
  1974).

\bibitem{HELL87}
T.~Hellmuth, H.~Walther, A.~Zajonc, and W.~Schleich, Delayed-choice experiments
  in quantum interference, \emph{Phys. Rev. A}. {\bf 72}, \penalty0 2532 --
  2541,  (1987).

\bibitem{RAUC74a}
H.~Rauch, W.~Treimer, and U.~Bonse, Test of a single crystal neutron
  interferometer, \emph{Phys. Lett. A}. {\bf 47}, \penalty0 369 -- 371,
  (1974).

\bibitem{HASE11}
Y.~Hasegawa and H.~Rauch, Quantum phenomena explored with neutrons, \emph{New
  J. Phys.} {\bf 13}, \penalty0 115010,  (2011).

\bibitem{KROU00}
G.~Kroupa, G.~Bruckner, O.~Bolik, M.~Zawisky, M.~Hainbuchner, G.~Badurek, R.~J.
  Buchelt, A.~Schricker, and H.~Rauch, {Basic features of the upgraded S18
  neutron interferometer set-up at ILL}, \emph{Nucl. Instrum. Methods Phys.
  Res. A.} {\bf 440}, \penalty0 604 -- 608,  (2000).

\bibitem{LEMM10}
H.~Lemmel and A.~G. Wagh, Phase shifts and wave-packet displacements in neutron
  interferometry and a nondispersive, nondefocusing phase shifter, \emph{Phys.
  Rev. A}. {\bf 82}, \penalty0 033626,  (2010).

\bibitem{RAUC74b}
H.~Rauch and M.~Suda, {Intensit{\"a}tsberechnung f{\"u}r ein
  Neutronen-Interferometer}, \emph{Phys. Stat. Sol. A}. {\bf 25}\penalty0 (2),
  \penalty0 495 -- 505,  (1974).

\bibitem{BOHM51}
D.~Bohm, \emph{Quantum Theory}. (Prentice-Hall, New York, 1951).

\bibitem{EPR35}
A.~Einstein, A.~Podolsky, and N.~Rosen, Can quantum-mechanical description of
  physical reality be considered complete?, \emph{Phys. Rev.} {\bf 47},
  \penalty0 777 -- 780,  (1935).

\bibitem{CIRE80}
B.~S. Cirel'son, Quantum generalizations of {Bell}'s inequality, \emph{Lett.
  Math. Phys.} {\bf 4}, \penalty0 93 -- 100,  (1980).

\bibitem{BELL64}
J.~S. Bell, {On the Einstein-Podolsky-Rosen paradox}, \emph{Physics}. {\bf 1},
  \penalty0 195 -- 200,  (1964).

\bibitem{LARS04}
J.-{\AA}. Larsson and R.~D. Gill, {Bell}'s inequality and the coincidence-time
  loophole, \emph{Europhys. Lett.} {\bf 67}, \penalty0 707 -- 713,  (2004).

\bibitem{PENA72}
L.~{de la Pe\~na}, A.~M. Cetto, and T.~A. Brody, On hidden-variable theories
  and {Bell's} inequality, \emph{Lett. Nuovo Cim.} {\bf 5}, \penalty0 177 --
  181,  (1972).

\bibitem{FINE74}
A.~Fine, {On the completeness of quantum theory}, \emph{Synthese}. {\bf 29},
  \penalty0 257 -- 289,  (1974).

\bibitem{FINE82}
A.~Fine, {Some local models for correlation experiments}, \emph{Synthese}. {\bf
  50}, \penalty0 279 -- 294,  (1982).

\bibitem{BROD93}
T.~Brody, \emph{{The Philosphy Behind Physics}}. (Springer, Berlin, 1993).

\bibitem{KUPC86}
M.~{Kupczy\'nski}, On some tests of completeness of quantum mechanics,
  \emph{Phys. Lett. A}. {\bf 116}, \penalty0 417 -- 419,  (1986).

\bibitem{JAYN89}
E.~T. Jaynes.
\newblock {Clearing up mysteries - The original goal}.
\newblock In ed. J.~Skilling, \emph{Maximum Entropy and Bayesian Methods},
  vol.~36, pp. 1 -- 27, Dordrecht,  (1989). Kluwer Academic Publishers.

\bibitem{SICA99}
L.~Sica, {{Bell's inequalities I}: An explanation for their experimental
  violation}, \emph{Opt. Comm.} {\bf 170}, \penalty0 55 -- 60,  (1999).

\bibitem{HESS01a}
K.~Hess and W.~Philipp, {Bell's theorem and the problem of decidability between
  the views of Einstein and Bohr}, \emph{Proc. Natl. Acad. Sci. USA}. {\bf 98},
  \penalty0 14228 -- 14233,  (2001).

\bibitem{HESS05}
K.~Hess and W.~Philipp, {Bell's} theorem: Critique of proofs with and without
  inequalities, \emph{AIP Conf. Proc.} {\bf 750}, \penalty0 150 -- 157,
  (2005).

\bibitem{KRAC05}
A.~F. Kracklauer, {Bell's} inequalities and {EPR-B} experiments: Are they
  disjoint?, \emph{AIP Conf. Proc.} {\bf 750}, \penalty0 219 -- 227,  (2005).

\bibitem{SANT05}
E.~Santos, {Bell's theorem and the experiments: Increasing empirical support to
  local realism?}, \emph{Stud. Hist. Phil. Mod. Phys.} {\bf 36}, \penalty0 544
  -- 565,  (2005).

\bibitem{NIEU09}
T.~M. {Nieuwenhuizen}, Where {Bell} went wrong, \emph{AIP Conf. Proc.} {\bf
  1101}, \penalty0 127 -- 133,  (2009).

\bibitem{KARL09}
K.~{Hess}, K.~{Michielsen}, and H.~{De Raedt}, {Possible experience: from Boole
  to Bell}, \emph{Europhys. Lett.} {\bf 87}, \penalty0 60007,  (2009).

\bibitem{NIEU11}
T.~M. Nieuwenhuizen, Is the contextuality loophole fatal for the derivation of
  {Bell} inequalities?, \emph{Found. Phys.} {\bf 41}, \penalty0 580 -- 591,
  (2011).

\bibitem{RAED11a}
H.~{De Raedt}, K.~{Hess}, and K.~{Michielsen}, {Extended Boole-Bell
  inequalities applicable to quantum theory}, \emph{J. Comput. Theor. Nanosci.}
  {\bf 8}, \penalty0 1011 -- 1039,  (2011).

\bibitem{KARL12}
K.~{Hess}, H.~{De Raedt}, and K.~{Michielsen}, Hidden assumptions in the
  derivation of the theorem of {Bell}, \emph{Phys. Scr.} {\bf T151}, \penalty0
  014002,  (2012).

\bibitem{BO1862}
G.~Boole, On the theory of probabilities, \emph{Phil. Trans. R. Soc. Lond.}
  {\bf 152}, \penalty0 225 -- 252,  (1862).

\bibitem{CLAU69}
J.~F. Clauser, M.~A. Horne, A.~Shimony, and R.~A. Holt, Proposed experiment to
  test local hidden-variable theories, \emph{Phys. Rev. Lett.} {\bf 23},
  \penalty0 880 -- 884,  (1969).

\bibitem{WEIH00}
G.~Weihs.
\newblock \emph{{Ein Experiment zum Test der Bellschen Ungleichung unter
  Einsteinscher Lokalit\"at}}.
\newblock PhD thesis, University of Vienna,  (2000).
\newblock {\url{http://www.uibk.ac.at/exphys/photonik/people/gwdiss.pdf}}.

\bibitem{CLAU74}
J.~F. Clauser and M.~A. Horne, Experimental consequences of objective local
  theories, \emph{Phys. Rev. D}. {\bf 10}, \penalty0 526 -- 535,  (1974).

\bibitem{BELL93}
J.~S. Bell, \emph{{Speakable and Unspeakable in Quantum Mechanics}}. (Cambridge
  University Press, Cambridge, 1993).

\bibitem{HNIL02}
A.~Hnilo, A.~Peuriot, and G.~Santiago, Local realistic models tested by the
  {EPRB} experiment with variable analyzers, \emph{Found. Phys. Lett.} {\bf
  15}, \penalty0 359 -- 371,  (2002).

\bibitem{HNIL07}
A.~A. Hnilo, M.~D. Kovalsky, and G.~Santiago, Low dimension dynamics in the
  {EPRB} experiment with variable analyzers, \emph{Found. Phys.} {\bf 37},
  \penalty0 80 -- 102,  (2007).

\bibitem{ADEN07}
G.~Adenier and A.~Y. Khrennikov, {Is the fair sampling assumption supported by
  EPR experiments}, \emph{J. Phys. B: At. Mol. Opt. Phys.} {\bf 40}, \penalty0
  131 -- 141,  (2007).

\bibitem{BIGE09}
J.~H. Bigelow.
\newblock A close look at the {EPR} data of {Weihs} et al.
\newblock arXiv: 0906.5093v1.

\bibitem{AGUE09}
M.~B. {Ag\"uero}, A.~A. Hnilo, M.~G. Kovalsksy, and M.~A. Larotonda, {Time
  stamping in EPRB experiments: application on the test of non-ergodic
  theories}, \emph{Eur. Phys. J. D}. {\bf 55}, \penalty0 705 --709,  (2009).

\bibitem{RAED10a}
H.~{De Raedt}, S.~Zhao, S.~Yuan, F.~Jin, K.~Michielsen, and S.~Miyashita,
  Event-by-event simulation of quantum phenomena, \emph{Physica E}. {\bf 42},
  \penalty0 298 -- 302,  (2010).

\bibitem{BIGE11}
J.~H. Bigelow.
\newblock Explaining counts from {EPRB} experiments: Are they consistent with
  quantum theory?
\newblock arXiv: 1112.3399.

\bibitem{RAED13a}
H.~{De Raedt}, F.~Jin, and K.~Michielsen, Data analysis of
  {Einstein-Podolsky-Rosen-Bohm} laboratory experiments, \emph{Proc. SPIE}.
  {\bf 8832}, \penalty0 88321N1--11,  (2013).

\bibitem{Rowe01}
M.~A. Rowe, D.~Kielpinski, V.~Meyer, C.~A. Sackett, W.~M. Itano, C.~Monroe, and
  D.~J. Wineland, Experimental violation of a {Bell}'s inequality with
  efficient detection, \emph{Nature}. {\bf 401}, \penalty0 791 -- 794,  (2001).

\bibitem{HASE03}
Y.~Hasegawa, R.~Loidl, G.~Badurek, M.~Baron, and H.~Rauch, {Violation of a
  Bell-like inequality in single-neutron interferometry}, \emph{Nature}. {\bf
  425}, \penalty0 45 -- 48,  (2003).

\bibitem{ANSM10}
M.~Ansmann, H.~Wang, R.~C. Bialczak, M.~Hofheinz, E.~Lucero, M.~Neeley, A.~D.
  O'Connell, D.~Sank, M.~Weides, J.~Wenner, A.~N. Cleland, and J.~M. Martinis,
  Violation of {Bell's} inequality in {Josephson} phase qubits, \emph{Nature}.
  {\bf 461}, \penalty0 504 -- 506,  (2009).

\bibitem{PEAR70}
P.~M. Pearle, Hidden-variable example based upon data rejection, \emph{Phys.
  Rev. D}. {\bf 2}, \penalty0 1418 -- 1425,  (1970).

\bibitem{PASC86}
S.~Pascazio, Time and {Bell}-type inequalities, \emph{Phys. Lett. A}. {\bf
  118}, \penalty0 47 -- 53,  (1986).

\bibitem{KARL10}
K.~{Hess}, K.~{Michielsen}, and H.~{De Raedt}, {Reply to Comment by A.J.
  Leggett and Anupam Garg}, \emph{Europhys. Lett.} {\bf 91}, \penalty0 40002,
  (2010).

\bibitem{KHRE09}
A.~Y. Khrennikov, \emph{{Contextual Approach to Quantum Formalism}}. (Springer,
  Berlin, 2009).

\bibitem{KHRE11}
A.~Y. Khrennikov, {On the role of probabilistic models in quantum physics:
  Bell's inequality and probabilistic incompatibility}, \emph{J. Comput. Theor.
  Nanosci.} {\bf 8}, \penalty0 1006 -- 1010,  (2011).

\bibitem{BART09}
C.~Bartsch and J.~Gemmer, Dynamical typicality of quantum expectation values,
  \emph{Phys. Rev. Lett}. {\bf 102}, \penalty0 110403,  (2009).

\bibitem{BASU01}
S.~Basu, S.~Bandyopadhyay, G.~Kar, and D.~Home, Bell's inequality for a single
  spin-1/2 particle and quantum contextuality, \emph{Phys. Lett. A}. {\bf 279},
  \penalty0 281 -- 286,  (2001).

\bibitem{Bartosik2009}
H.~Bartosik, J.~Klepp, C.~Schmitzer, S.~Sponar, A.~Cabello, H.~Rauch, and
  Y.~Hasegawa, Experimental test of quantum contextuality in neutron
  interferometry, \emph{Phys. Rev. Lett.} {\bf 103}, \penalty0 040403,  (2009).

\end{thebibliography}

\end{document}